\documentclass[12pt]{article}
\usepackage{graphicx, subfigure}
\usepackage{varwidth}
\usepackage{graphics,latexsym,amssymb,epsfig,bm,
amsmath,enumerate,setspace,float,placeins}
\usepackage[usenames]{color}
\usepackage{dsfont}
\usepackage[utf8]{inputenc}
\usepackage[english]{babel}
\usepackage{amsthm}
\usepackage{algpseudocode}
\usepackage{multirow}
\usepackage{cases}
\usepackage{hhline}
\usepackage{tabularx}
\usepackage[boxed, lined, ruled, linesnumbered]{algorithm2e} 

\newcommand{\blind}{0}

\newtheorem{theorem}{Theorem}
\newtheorem{lemma}{Lemma}

\newtheorem{pro}{Proposition}

\usepackage[round]{natbib}
\usepackage{enumitem}
\usepackage{hyperref}
\hypersetup{colorlinks,citecolor=blue,
	linkcolor=blue}

	\renewcommand\labelenumi{(\roman{enumi})}
	\renewcommand\theenumi\labelenumi
	\newcommand{\comment}[1]{}
	\usepackage{mathtools}
	\newcommand{\defeq}{\vcentcolon=}
	\newcommand{\eqdef}{=\vcentcolon}

        \def\z{\mathbf{z}}  

	\DeclareMathOperator*{\minimize}{minimize\,}

\begin{document}

\newfloat{procedure}{htbp}{loa}
\floatname{procedure}{Procedure} 

\pagenumbering{arabic}

\setcounter{page}{1}

\title{\Large \bf Cross validation for penalized quantile regression with a case-weight adjusted solution path}

\if0\blind
{
\author{Shanshan Tu\thanks{Latitude AI, \href{mailto:shanshantu23@gmail.com}{shanshantu23@gmail.com}} 
\and Yunzhang Zhu\thanks{Amazon, \href{mailto:ryzhux@gmail.com}{ryzhux@gmail.com}}  
\and Yoonkyung Lee\thanks{Department of Statistics, The Ohio State University, \href{mailto:yklee@stat.osu.edu}{yklee@stat.osu.edu}} 
\and Qiuyu Gu\thanks{Meta Platforms, Inc., \href{mailto:qiuyugu15@gmail.com}{qiuyugu15@gmail.com}} 
\and Haozhen Yu\thanks{Department of Statistics, The Ohio State University, \href{mailto:yu.2823@osu.edu}{yu.2823@osu.edu}}
}

\date{}
\maketitle
} \fi

\if1\blind
{
  \bigskip
  \bigskip
  \bigskip
  \begin{center}
 {\Large\bf Cross validation for penalized quantile regression with a case-weight adjusted solution path}
\end{center}
\date{}
  \medskip
} \fi

\bigskip

\begin{abstract}
	Cross validation is widely used for selecting tuning parameters in regularization methods, but it is computationally intensive in general. To lessen its
    computational burden, approximation schemes such as
    generalized approximate cross validation (GACV) are often
    employed. However, such approximations may not work well when non-smooth loss functions are involved. As a case in point, approximate cross validation schemes for penalized quantile regression do not work well for extreme quantiles. In this paper, we propose a new algorithm to compute the leave-one-out cross validation scores exactly for quantile regression with ridge penalty through a case-weight adjusted solution path. Resorting to the homotopy technique in optimization, we introduce a case-weight for each individual data point as a continuous embedding parameter and decrease the weight gradually from one to zero to link the estimators based on the full data and those with a case deleted. This allows us to design a solution path algorithm to compute all leave-one-out estimators very efficiently from the full-data solution. We show that the case-weight adjusted solution path is piecewise linear in the weight parameter and using the solution path, we comprehensively examine case influences and observe that different modes of case influences emerge, depending on the specified quantiles, data dimensions and penalty parameter. We further illustrate the utility of the proposed algorithm in real-world applications.
\end{abstract}

\doublespace 

\section{Introduction}\label{Sec:intro}

   With the rapid growth of data dimensionality, regularization is
   widely used in model estimation and prediction. In penalized
   regression methods such as LASSO and ridge regression, the penalty parameter plays an essential role in determining the trade-off  between bias and variance of the corresponding regression  estimator. Too large a penalty could lead to undesirably large bias
   while too small a penalty would lead to instability in the
   estimator. The  penalty parameter can be chosen to minimize the
   prediction error associated with the estimator. Cross validation
   (CV) \citep{cv-origin} is the most commonly used technique for
   choosing the penalty parameter based on data-driven estimates of  the prediction error, especially when there is not enough data available. 

	Typically, fold-wise CV is employed in practice. When the
        number of folds is the same as the sample size, it is known as
        leave-one-out (LOO) CV. For small data sets, LOO CV provides
        approximately unbiased estimates of the prediction error while
        the general $k$-fold CV may produce substantial bias due to
        the difference in sample size for the fold-wise training data
        and the original data \citep{cv_study}. Moreover, for linear modeling procedures such as smoothing splines, the fitted values from the full data can be explicitly related to the predicted values for LOO CV \citep{gcv-article}. Thus, the LOO CV scores are readily available from the full data fit. The linearity of a modeling procedure that enables exact LOO CV  is strongly tied to squared error loss employed for the procedure and the simplicity of the corresponding optimality condition for the solution. 

	 However,  loss functions for general modeling procedures may
         not yield such simple optimality conditions as squared error loss
         does, and result in more complex relation between the fitted
         values and the observed responses. In general, the LOO
         predicted values may not be related to the full data
         fit in closed form. Consequently, the computation needed for LOO CV becomes
         generally intensive as LOO prediction has to be made for each
         of $n$ cases separately given each candidate penalty parameter.

	  In this paper we focus on LOO CV for penalized M-estimation
          with nonsmooth loss functions, in particular, quantile
          regression  with ridge penalty. Quantile regression
          \citep{Koenker1978} can provide a comprehensive description
          of the conditional distribution of the response variable
          given a set of covariates, and it has become an increasingly
          popular tool to explore the data heterogeneity
          \citep{Koenker40ys}.  Extreme quantiles can also be used for
          outlier detection \citep{quantile_outliers}. Penalized
          quantile regression is particularly useful for analyzing
          high-dimensional  heterogeneous data. 

\begin{figure}[h]
 	\centering
 	\includegraphics[height=2in]{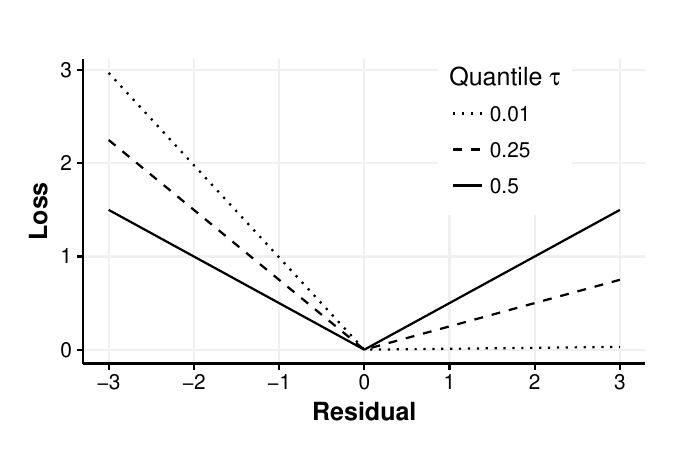}
 	\caption{The check loss for quantile regression  with quantile parameter $\tau =0 .01, ~0.25\text{ and } 0.5$}
 	\label{fig:check_loss}
 	\label{ql_plot}
 \end{figure}

         The check loss for quantile regression with a pre-specified quantile parameter $\tau \in (0,1)$ is defined as
 \begin{equation}\label{ql}
 \rho_\tau(r) = \tau r_+ + (1 - \tau) (-r)_+, \text{ where } r_+ = \max(r, 0).
 \end{equation}
  Unlike squared error loss, the check loss is nondifferentiable at 0
  as is shown in Figure \ref{ql_plot}.

\begin{figure}[h]
 	\centering
 	\includegraphics[height = 2.5in]{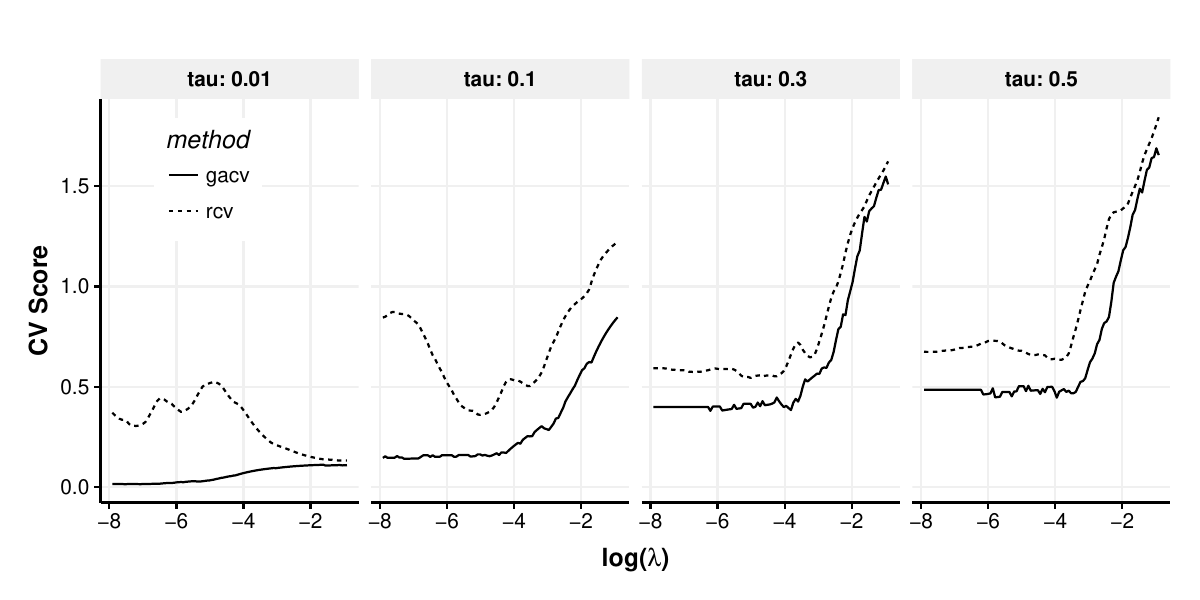}
 	\caption{Comparison of the exact LOO CV and GACV scores for
          penalized quantile regression ($\tau = 0.01, ~0.1, ~0.3,\text{ and }0.5$). The exact LOO CV score is defined as $RCV (\lambda) = \sum_{i=1}^{n} \rho_\tau\big(y_i - \hat{f}_\lambda^{[-i]} (x_i) \big)/n $, and GACV score from \cite{Li2007lampath} is defined as $GACV(\lambda) =\sum_{i=1}^{n} \rho_\tau(y_i - \hat{f} (x_i))/  (n - |\mathcal{E}_\lambda|)$. }
 	\label{fig_gacv}
 \end{figure}

  To lessen the computational cost of the exact LOO CV in this
  setting, \cite{acv1995} and \cite{yuan2006} proposed Approximate CV
  (ACV) and Generalized Approximate CV (GACV). Using a smooth
  approximation of the check loss, they applied similar arguments used
  in mean regression for
  the derivation of ordinary cross validation (OCV) \citep{ocv} and
  generalized cross validation (GCV) \citep{gcv-Golub} to quantile regression.
  The key ingredients for the arguments are the leave-one-out lemma 
  and the first-order
  Taylor expansion of the smoothed check loss. The linearization error
  from the first-order Taylor expansion may not be ignorable for
  extreme quantiles due to the increasing skewness of the distribution
  of the LOO residuals that is at odds with the increase in the slope
  of the check loss with 
  $\tau$ (see Section \ref{sec:gacv_loo} for
  details). This phenomenon can be easily illustrated. Figure
  \ref{fig_gacv} compares the exact LOO CV and GACV scores as
  a function of the penalty parameter $\lambda$
  for various quantiles in a simulation setting
		(see Section \ref{sec:empirical} for details). For instance,
	the approximate CV scores in the figure could produce penalty
		 parameter values that are very different from the
		  exact LOO CV when $\tau = 0.01$ and  $0.1$. The empirical 
      studies in \cite{Li2007lampath} and \cite{badgacv} also confirm 
      the inaccuracy of the approximation for extreme quantiles. This result 
  motivates us to explore other computationally efficient schemes for exact LOO CV.

 Instead of treating $n$ LOO problems separately, we exploit  the homotopy strategy to relate them to the full-data problem. The $n$ LOO problems can be viewed as perturbations of the full-data problem. The key idea of homotopy is to start from a problem with known solution and gradually adjust the problem with respect to a continuous homotopy parameter until we reach the desired target problem and its solution. In our approach, we leverage the full-data solution as a starting point for the LOO problems.
 In optimization, homotopy techniques have been used in many algorithms including the interior point algorithm derived from perturbed KKT conditions \citep{homotopy-interiorpoint} and parametric active set programming \citep{path-origin}.
  In statistical learning community, the latter has been widely used
  in the form of path-following algorithms. For instance,
  \cite{Osborne1992} and \cite{homo_osborne2000} apply the homotopy
  technique to generate piecewise linear trajectories in quantile
  regression and LASSO problems, respectively. Later
  \cite{leastangle}, \cite{svmpath} and \cite{Rosset07genericpath}
  exploit the homotopy path-following methods to generate an entire
  solution path for a family of regularization problems
  indexed by the penalty parameter.

In this paper, we propose an exact path-following algorithm for LOO cross validation in penalized quantile regression by introducing a case-weight $\omega$ for the held-out case as a continuous homotopy parameter.
We vary the case-weight $\omega$ from 1 to 0 to link the full-data
setting to the LOO setting. Let $\{(x_i, y_i)\}_{i=1}^n$ be the full
data with covariates $x_i\in \mathbb{R}^p$ and response $y_i\in
\mathbb{R}$. Given a fixed quantile $\tau$ and penalty parameter $\lambda$,
for each case $i^\star  \in \{1,\cdots,n\}$, consider the following
case-weight adjusted quantile regression problem with linear regression quantiles:
\begin{equation}\label{w-set-up}
\minimize_{\beta_0\in \mathbb{R}, {\beta}\in \mathbb{R}^p} \sum_{i \ne i^\star } \rho_\tau(y_i - \beta_0 - x_i^\top {\beta}) + \omega \rho_\tau(y_{i^\star } - \beta_0 - x_{i^\star }^\top {\beta}) +  \frac{\lambda}{2} \|\beta\|_2^2.
\end{equation}
 The problem in \eqref{w-set-up} with $\omega = 1$ involves the full data while $\omega = 0$ leaves out the case $i^\star $. By decreasing the case-weight $\omega$ from 1 to 0, we successfully link the two separate but intrinsically related problems. Notice that the full data solution needs to be computed only once and can be used repeatedly as a starting point for $n$ LOO problems.
  We provide an efficient homotopy algorithm to generate the solution path indexed by $\omega$, which results in the LOO solution. Hence, with the LOO solutions, we can compute CV scores exactly, circumventing the issues with approximate CV especially for extreme quantiles.

 There have been many works on computation of the solution paths for
 penalized quantile regression. In spirit of \cite{svmpath},
 \cite{l1qr} and \cite{Li2007lampath} proposed algorithms for solution
 paths in $\lambda$ given quantile $\tau$ in $l_1$-penalized quantile
 regression and kernel quantile regression, respectively. By varying
 the quantile parameter $\tau$, \cite{Takeuchi2009taupath}  examined the  solution path as a function of  $\tau$ for fixed $\lambda$ in kernel quantile regression. Further, \cite{Rosset2009bilevelpath} developed an algorithm for a generalized bi-level solution path as a function of both $\lambda$ and $\tau$. These algorithms are driven by a set of optimality conditions that imply piecewise linearity of the solution paths. Due to the linear structure in the additional term with a case-weight $\omega
 $ in \eqref{w-set-up}, it can be shown that the case-weight adjusted
 solution path is also piecewise linear in $\omega$.
This piecewise linearity allows us to devise a new path-following
 algorithm, which starts from the full-data solution and reaches the
 LOO solution at the end. We derive the optimality conditions for the case-weight adjusted solution and provide a formal proof that solutions from the algorithm satisfy the KKT conditions at every $\omega \in [0, 1]$.

 The proposed path-following algorithm with a varying case-weight
 $\omega$ does not only offer the LOO solutions efficiently, but also provides
 case-influence measures. 
We demonstrate numerically and analytically 
that the computational cost 
of the proposed algorithm 
in evaluation of LOO CV scores could be much lower 
than that of a simple competing method.  
 This also allows an efficient evaluation of
  the influence of the case on the
 fitted model as a function of $\omega$. Different from case-deletion
 diagnostics \citep{cook1977, belsley1980},  \cite{cook1986localassess} 
 proposed analogous case-influence graphs to assess local influence of 
 a statistical model. Using the case-weight adjusted solution path, 
 we can generate case-influence graphs efficiently for penalized quantile regression 
 and examine the influence of small perturbations of data on regression 
 quantiles. 
In contrast to mean regression, 
 it is observed that cases with almost identical  
case deletion statistics could have quite different case-influence graphs in quantile regression.  
Numerically, we observe that the data dimension and the value of the penalty parameter can influence the computational time of the algorithm.

The paper is organized as follows. Section \ref{Sec: w_path} proposes
a path-following algorithm for case-weight adjusted quantile 
regression with ridge penalty for cross validation.  
A formal validation of the algorithm is provided on the 
basis of the optimality conditions. 
Section \ref{sec_caseinf} presents another application of 
the case-weight adjusted solutions for measuring case influence on regression quantiles. 
In Section \ref{sec:empirical}, some numerical studies are presented to 
illustrate the applications of the proposed case-weight adjusted solution 
path algorithm and its favorable computational efficiency for 
computing LOO CV scores. In Section~\ref{sec:data}, we demonstrate the utility of the proposed method for analyzing real-world data by assessing the case influence in quantile regression.
We conclude with some remarks in Section \ref{sec:discussion}.
Technical proofs are provided in Appendix.

\section{Case-weight Adjusted Solution Path in
			Quantile Regression with Ridge Penalty}
\label{Sec: w_path}

In this section, we present a path-following algorithm for solving
the penalized quantile
regression problem in \eqref{w-set-up} with case-weight $\omega$.
We illustrate in detail how to construct a solution path
from the full-data solution as the case-weight decreases from $1$ to $0$.
As with many existing solution path algorithms, the key to our
  derivations is the optimality conditions for \eqref{w-set-up}. We
  analyze the Karush-Kuhn-Tucker (KKT) conditions for the problem after reformulating it as a constrained optimization problem.
We formally prove that the path generated by the proposed algorithm
solves the problem \eqref{w-set-up}, and is piecewise linear in
$\omega$.

\subsection{Optimality Conditions}
In the path-following algorithm, we start from the full-data solution at $\omega = 1$, and specify a scheme to update the solution as $\omega$ decreases from $1$ to $0$. The updating scheme is designed so that the path generated satisfies the KKT conditions for every
$\omega$ in $[0, 1]$.
The KKT conditions for the optimization problem \eqref{w-set-up} 
are 
\begin{align}
\label{w_kkt1}
& X^\top \theta_\omega = \lambda \beta_\omega \, , \text{ and } \,
 \theta_\omega^\top {\mathbf 1}_n = 0 &\\
\label{w_kkt3}
& \theta_{i,\omega} =
\begin{cases}
\tau - 1 & \text{ for } i \ne i^\star \\
\omega(\tau - 1) & \text{ for } i = i^\star
\end{cases}
& \text{if }   y_i - \beta_{0,\omega} - x_i^\top \beta_\omega < 0\\
\label{w_kkt4}
& \theta_{i,\omega}
\in \begin{cases} [\tau - 1, \tau] & \text{ for } i \ne i^\star \\
[\omega(\tau - 1), \omega \tau]  & \text{ for } i = i^\star
\end{cases}
& \text{if }
y_i - \beta_{0,\omega} - x_i^\top \beta_\omega =  0 \\
\label{w_kkt5}
& \theta_{i,\omega} =
\begin{cases}
\tau   & \text{ for } i \ne i^\star \\
\omega\tau & \text{ for } i = i^\star
\end{cases}
& \text{if }  y_i - \beta_{0,\omega} - x_i^\top \beta_\omega >  0
\end{align}
where 
$(\beta_{0,\omega}, \beta_{\omega})$ denotes the solution of \eqref{w-set-up}, 
$\theta_\omega = (\theta_{1,\omega},\ldots,
\theta_{n,\omega})\in \mathbb{R}^n$ is the set of dual variables
associated with the residual bounds, 
$X =(x_1, \ldots, x_n)^\top \in \Bbb{R}^{n \times p}$
is the $n\times p$ design matrix, and ${\mathbf 1}_n$ is the vector of $n$ ones.
A detailed derivation of the KKT conditions 
is included in the Appendix \ref{appx:kkt}.
The solution $(\beta_{0, \omega}, \beta_{\omega})$ and $\theta_{\omega}$ can thus be determined by the equality conditions in \eqref{w_kkt1}--\eqref{w_kkt5}.

\subsection{Outline of the Solution Path Algorithm}
			Let $r_{i,\omega} = y_i - \beta_{0,\omega} - x_i^\top \beta_\omega$ denote
			the residual for the $i$th case with
			 $(\beta_{0,\omega}, \beta_\omega)$.
			According to the sign of each residual, we can partition the $n$ cases
			into three sets. Depending on which side of $0$
			each residual falls on,
			the three sets are called the elbow set, $\mathcal{E_\omega} = \{i: r_{i,\omega} = 0 \}$, the left
			set of the elbow, $\mathcal{L_\omega} = \{i: r_{i,\omega} < 0 \}$ and the right
			set of the elbow, $\mathcal{R_\omega} = \{i: r_{i,\omega} > 0 \}$.
			The three sets may
			evolve as $\omega$ decreases. We call $\omega_m$ a breakpoint if
			the three sets change at $\omega_m$. The following rules specify
			how and when we should update the three sets at each breakpoint:
			\begin{enumerate}[label = (\alph*)]
				\item 
				if $\theta_{i,\omega}$ =
				$\tau [\omega+ (1 - \omega)  \mathbb{I}(i \ne i^\star)]$ for some
				$i \in \mathcal{E_\omega}$, then
                                move case $i$ from $\mathcal{E_\omega}$
                                to  the right set of the elbow
                                $\mathcal{R_\omega}$.
                                Here $\mathbb{I}(\cdot)$ is the
                                  indicator function for a condition,
                                  which takes the value 1 if the condition is met, and 0 otherwise.
				\item 
				if $\theta_{i,\omega}$
				= $ (\tau - 1)[\omega + (1 - \omega) \mathbb{I}(i \ne i^\star)]$
				for some $i \in \mathcal{E_\omega}$,
				then move case $i$ from
                                $\mathcal{E_\omega}$ to the
				left set of the elbow $\mathcal{L_\omega}$.
				\item
				if $r_{i,\omega} = 0$
				for some $i \in \mathcal{L_\omega} \cup \mathcal{R_\omega}$,
				then move case $i$ from
                                $\mathcal{L_\omega} \cup
                                \mathcal{R_\omega}$ to  the elbow set $\mathcal{E_\omega}$.
			\end{enumerate}
			Given the three sets, we next analyze how the solution should evolve between two breakpoints.
			Toward this end, we let
			$\{\omega_m, \text{ for } m = 0, 1,\ldots, M \mid
			0 \leq \omega_M < \cdots < \omega_1 < \omega_0 = 1\}$ be the set of breakpoints, and
			denote by $\mathcal{E}_m$, $\mathcal{L}_m$ and $\mathcal{R}_m$
			the three sets between $\omega_{m + 1}$ and $\omega_{m}$ for $m = 1,\ldots, M$. 
                        Note that we have abbreviated $\mathcal{E}_{\omega_m}$,
                          $\mathcal{L}_{\omega_m}$ and $\mathcal{R}_{\omega_m}$ to
                $\mathcal{E}_m$, $\mathcal{L}_m$ and $\mathcal{R}_m$, respectively.
			Now, when $\omega_{m+1} < \omega < \omega_{m}$,
			the KKT conditions determine how
			$(\beta_{0,\omega}, \beta_{\omega})$ and $\theta_\omega$ should change
			as functions of $\omega$ and we can show that they
			satisfy the following:
			\begin{equation}\label{eq:linear}
			\begin{array}{clc}
			 - \sum_{i \in \mathcal E_m} \theta_{i, \omega}
			& = & \sum_{i \notin \mathcal E_m} \theta_{i, \omega} \\[1ex]
			 \lambda \beta_\omega -
			\sum_{i \in \mathcal E_m} \theta_{i, \omega} x_i
			& = &  \sum_{i \notin \mathcal E_m} \theta_{i, \omega} x_i   \\[1ex]
			 \beta_{0, \omega}  + x_i^\top \beta_{\omega} & = & y_i
			 \quad \text{ for }
			i \in \mathcal E_m
			\end{array}
			\end{equation}
using \eqref{w_kkt1} and \eqref{w_kkt4}, and the fact that
$\theta_{i, \omega} = \{\tau - \mathbb I(i \in \mathcal L_m)\}
\{\omega + (1 - \omega) \mathbb{I}(i \ne i^\star)\}$ for $i \notin \mathcal E_m$ from \eqref{w_kkt3} and \eqref{w_kkt5}.

Next, we show that $(\beta_{0,\omega}, \beta_{\omega}, \theta_{\omega})$ satisfying  \eqref{eq:linear} must be linear in $\omega$.
Before proceeding, we introduce some notations.
For any vector $v = (v_1,\ldots, v_n)^\top \in \mathbb R^n$ and
any index set $A = \{i_1, \ldots, i_{k}\} \subseteq \{1,2, \ldots,
n\}$, define $v_A = (v_{i_1}, \ldots, v_{i_{k}} )$
 be a sub-vector of
$v$.
Similarly, for any matrix $M = (m_{1}, \ldots, m_{n})^\top \in \mathbb R^{n \times L}$,
let $M_{A} = (m_{i_1}, \ldots, m_{i_{k} })^\top$ be a submatrix of
$M$, where $m_i^\top$ is the $i$th row of $M$ for $i = 1, \ldots, n$.
Let $\mathbf{0}_k$ be the vector of $k$ zeros and
  $\mathbf{O}_k$ be the $k \times k$ matrix of zeros.
Now we can rewrite
\eqref{eq:linear} into a matrix form:
			\begin{equation*}
			\begin{pmatrix}
			0 & \mathbf 0^\top_p & - {\mathbf 1}^\top_{\mathcal E_m}\\[1ex]
			0 & \lambda I_{p} & -X^\top_{\mathcal E_m}\\
			{\mathbf 1}_{\mathcal E_m} & X_{\mathcal E_m} & {\mathbf O_{|\mathcal E_m|}}
			\end{pmatrix}
			\begin{pmatrix}
			\beta_{0,\omega}\\[1ex]
			\beta_{\omega}\\
			\theta_{{\mathcal E_m},\omega}
			\end{pmatrix} =
			\begin{pmatrix}
			{\mathbf 1}^\top_{\mathcal L_m}
                        \theta_{\mathcal L_m} +{\mathbf 1}^\top_{\mathcal R_m}\theta_{\mathcal R_m} \\[1ex]
			X^\top_{\mathcal L_m} \theta_{\mathcal L_m} +
			X^\top_{\mathcal R_m} \theta_{\mathcal R_m}\\
			y_{\mathcal E_m}
			\end{pmatrix},
            \end{equation*}
 which is a system of linear equations of dimension
 $1 + p + |\mathcal E_m|$.
${\mathbf 1}_{\mathcal E_m}$ is defined from
${\mathbf 1}_n$, and $I_p$ is the $p\times p$ identity matrix. 
Note that the left hand side of the
above linear equation does not depend on $\omega$, while
the right hand side is a linear function of $\omega$. This
implies that its solution must be a linear function of $\omega$.
The following lemma summarizes the properties of the solution path
described thus far.
\begin{lemma}\label{w_piecelinear}
				The solution path $(\beta_{0,\omega},\beta_{\omega})$
				 satisfying the KKT conditions
				\eqref{w_kkt1}--\eqref{w_kkt5} is piecewise linear in $\omega$.
			\end{lemma}
Using Lemma \ref{w_piecelinear}, we propose a solution path algorithm that	updates the three sets following the aforementioned rules at each breakpoint and linearly updates the solutions between two consecutive breakpoints. First we provide an outline of our algorithm:
			\begin{itemize}
				\item Start with the full-data solution at $\omega = 1$.
				\item While $\omega > 0$,
				\begin{enumerate}
					\item Decrease $\omega$ and update $(\beta_{0,\omega}, \beta_\omega)$ and $\theta_\omega$ until one of the inequalities in the KKT conditions is violated.
					\item When the violation happens,
					update the three sets according to
					the rules (a)--(c). Then go back to Step (i).
				\end{enumerate}
			\end{itemize}

\subsection{Determining Breakpoints}
For implementation of Step (i), we need to derive a formula for the
next breakpoint $\omega_{m+1}$ among $\omega \le \omega_m$. From the
KKT conditions \eqref{w_kkt1}--\eqref{w_kkt5},  we can see that as
$\omega$ decreases, the conditions \eqref{w_kkt3}--\eqref{w_kkt5}
 will be violated when 
 $\theta_{i, \omega} = \tau [\omega+ (1 - \omega)  \mathbb{I}(i \ne i^\star)] \text{ or } (\tau - 1) 
 [\omega + (1 - \omega)  \mathbb{I}(i \ne i^\star)]$
  for some $i \in \mathcal E_m$, or  $r_{i, \omega} = 0$ for some $i
 \in \mathcal L_m \cup \mathcal R_m$.
Thus, to find the next breakpoint $\omega_{m+1}$,
 we need to derive how $\theta_{{\mathcal E_m}, \omega}$ and $r_\omega = (r_{1,\omega}, \ldots, r_{n,\omega})$ change as functions of $\omega$.
This is established in the following
proposition, for which we need to impose an assumption that
 any $\min(p+2, n)$ {points} of 
 $\{(\tilde x_i, y_i)\}^{n}_{i=1}$ 
 are linearly independent, where 
 $\tilde x_i = (1, x_i^\top)^\top$.
We call this condition the \textit{general position condition}. 
A similar condition is also imposed in \cite{Li2007lampath}. 
\begin{pro}
	\label{prop:sol_path_evolution}
	Suppose that the data points
	$\{(\tilde x_i, y_i)\}_{i=1}^{n}$ satisfy the
	 \textit{general position condition}.
	 Then {the} solution path for \eqref{w-set-up} satisfies the following properties: 
         \begin{itemize}
         \item[I.]
	 When $i^\star
	 	\in \mathcal R_m \cup \mathcal L_m$, we have that
	 	\begin{equation}
	 		\label{eq:theta_evol}
	 	 \theta_{\mathcal{E}_m, \omega} -
	 	\theta_{\mathcal{E}_m, \omega_m}
	 	= b_{m}(\omega  - \omega_m)  \, ,
	 \end{equation}
	 	where
	 	\begin{eqnarray}
	 		\label{eq:theta_slopes}
	 	b_{m} &=& - (\tilde{X}_{\mathcal{E}_m} \tilde{X}_{\mathcal{E}_m}^\top)^{-1}
	 	\left[ b_{0,m} 1_{\mathcal E_m}
	 	+
	 	\tilde{X}_{\mathcal{E}_m}\tilde{x}_{i^\star}  (\tau -  \mathbb I(i^\star \in \mathcal L_m))
	 	\right]  \\
		& \text{ with } &
		b_{0, m} = \frac{ 1 - 1_{\mathcal{E}_m}^\top (\tilde{X}_{\mathcal{E}_m} \tilde{X}_{\mathcal{E}_m}^\top)^{-1}\tilde{X}_{\mathcal{E}_m}\tilde{x}_{i^\star}     }{  1_{\mathcal{E}_m}^\top (\tilde{X}_{\mathcal{E}_m} \tilde{X}_{\mathcal{E}_m}^\top)^{-1}1_{\mathcal{E}_m} } (\tau -  \mathbb I(i^\star \in \mathcal L_m)) \label{eq:slope_beta0} \, ,
	\end{eqnarray}
	 and
	 	\begin{equation}
	 		\label{eq:resid_evol}
	 	\lambda(r_{\omega} -  r_{\omega_m})
	 = h_{m}(\omega - \omega_m) \, ,
	 \end{equation}
	 	where
	 	\begin{equation}
	 		\label{eq:resid_slopes}
	 	h_{m} = - b_{0,m}{1} - \tilde{X}\left[
	 	\tilde{X}_{\mathcal{E}_m}^\top b_{m} +
	 	\{\tau - \mathbb I( i^\star \in \mathcal{L}_m )\}
	 	\tilde{x}_{i^\star}\right] \, .
	 	\end{equation}
\item[II.]
Moreover, $i^\star \in \mathcal E_m$ can only happen
	 when $m = 0$, and if that happens,
	 both $r_\omega$ and $\theta_{\omega}$
	 are constant vectors {for $\omega \in [\omega_1, 1]$}, and
	 $i^\star$ will move from $\mathcal E_0$ to
	  $\mathcal L_1 \cup \mathcal R_1$ at the next breakpoint
		 $\omega_1 = \frac{{\theta}_{i^\star,\omega_0}}{\tau - \mathbb I({\theta}_{i^\star,\omega_0} < 0)}$, and
	 stay in $\mathcal L_m \cup \mathcal R_m$ for all $m = 1,\ldots, M$.	
         \end{itemize}
\end{pro}
The proof of Proposition
 \ref{prop:sol_path_evolution} is
provided in the Appendix.
Using Proposition \ref{prop:sol_path_evolution}, we can easily determine the
next breakpoint if $i^\star \in \mathcal L_m \cup \mathcal R_m$.
Specifically,
the next breakpoint is determined by
	the largest $\omega < \omega_m$ such that
$\theta_{i, \omega} = \tau \text{ or } \tau - 1$
			for some $i \in \mathcal E_m$, or
			$r_{i, \omega} = 0$ for some $i \in \mathcal L_m \cup \mathcal R_m$.
Hence, the next breakpoint is
			\begin{equation}
				\label{eq:next_breakpoint}
			\omega_{m+1} = \max(\omega_{1, m+1}, \omega_{2, m+1}) \, ,
			\end{equation}
			where
			$\omega_{1, m+1}$ is the
largest $\omega < \omega_m$,
at which $\theta_{i,\omega}$,
for some $i \in \mathcal E_m$,  hits either of the boundaries $\tau$ or $\tau - 1$, and $\omega_{2, m+1}$ is the largest $\omega < \omega_m$, at which $r_{i,\omega}$ hits $0$ for some $i \in \mathcal L_m \cup \mathcal R_m$.
Moreover, we know that $\theta_{\mathcal E_m, \omega}$ and $r_{\omega}$ evolve
as linear functions of $\omega$ according to \eqref{eq:theta_evol} and \eqref{eq:resid_evol},
 from which we obtain the following for $\omega_{1, m+1}$
			and $\omega_{2, m+1}$:
			\begin{subequations}
				\begin{align}
				\omega_{1, m+1} & = \underset{\theta \in \{\tau, \tau - 1\}}{\max}
				\left(\underset{i \in \mathcal E_m \text{ and }
					-\omega_m \le \frac{\theta - \theta_{i, \omega_m}}{b_{i,m}} < 0 }{\max}
				\frac{\theta - \theta_{i, \omega_m}}{b_{i,m}}   + \omega_m\right)
				\label{eq: w_next_bk1_2_1}\\
				\omega_{2, m+1} & =  \underset{i \in \mathcal{L}_m \cup\mathcal{R}_m \text{ and } 0 < \frac{\lambda r_{i, \omega_m}}{h_{i,m}} \le \omega_m }{\max}
				-\frac{\lambda r_{i, \omega_m}}{h_{i,m}} + \omega_m \, ,
				\label{eq: w_next_bk1_2_2}
				\end{align}
			\end{subequations}
			where $b_{i,m}$ and $h_{i,m}$  are the $i$th component of
			slopes $b_m$ and $h_m$ defined in
			\eqref{eq:theta_slopes} and \eqref{eq:resid_slopes}, respectively.
			From these two formulas, we can see that the
			next breakpoint can be determined
			without evaluating the solutions
			between two breakpoints.
		
   \subsection{A Path-Following Algorithm}
			We summarize the detailed
			description of our proposed solution-path 
      algorithm in Algorithm \ref{alg_w} in the Appendix. 
   The following theorem states that the path generated by Algorithm \ref{alg_w} is indeed a solution path to
	the optimization problem \eqref{w-set-up}, provided that the data points
	satisfy the general position condition.

		\begin{theorem}\label{w_rigor}
			Assume that the set of data points $\{(\tilde x_i,y_i)\}^{n}_{i=1}$
			satisfies the  general position condition.
			The case-weight adjusted path generated by
                        Algorithm \ref{alg_w}
                        solves
			the optimization problem \eqref{w-set-up}
                        indexed by $\omega\in[0,1]$.
		\end{theorem}

\section {Case Influence Assessment in Quantile Regression} 
\label{sec_caseinf}
In addition to model validation through LOO CV, the case-weight
  adjusted solution path for quantile regression
can be used for assessing case influences on regression quantiles.
 In this section, we further explore the use of the  case-weight
 adjusted solutions for measuring case influence. 

\begin{figure}[h]
	\centering
	\includegraphics[height=2in]{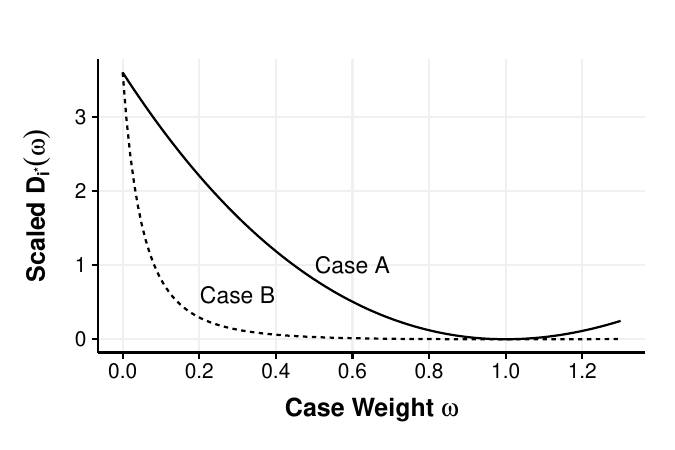}
	\caption{An illustrative example of case-influence graphs in least squares regression based on Figure 1 in \cite{cook1986localassess}.}
	\label{fig: cook1986}
\end{figure}

In statistical modeling, assessing case influence and identifying  influential cases is crucial for model diagnostics. Assessment of case influence on a statistical model has been extensively studied in robust statistics literature. Seminal works on case-influence assessment include \cite{cook1977}, \cite{cook1979} and \cite{cook1982}. As a primary example, Cook proposed the following measure, known as Cook's distance for case $i^\star $:
\begin{equation*}
D_{i^\star } = \frac{\sum_{i=1}^{n} (\hat{f}(x_i) - \hat{f}^{[-i^\star ]}(x_i))^2}{p \hat{\sigma}^2},
\end{equation*}
where $\hat{\sigma}^2$ is an estimate of the error variance in a regression model $y_i=f(x_i) +\epsilon_i$ with $\epsilon_i \sim N(0, \sigma^2)$.
Cook's distance measures an aggregated effect of one single case on $n$
fitted values after that case is deleted. In other words, it compares
two sets of fitted values when case $i^\star $ has weight $\omega = 1$
and $\omega = 0$. In contrast to  the standard Cook's distance,
\cite{cook1986localassess} also introduced the notion of a
case-influence graph to get a broad view of case influence as a
function of the case-weight $\omega$. As a general version of $D_{i^\star}$, a case-weight adjusted Cook's distance is defined as
\begin{equation}\label{eq:general_cookdist}
D_{i^\star }(\omega) = \frac{\sum_{i=1}^{n} (\hat{f}(x_i) - \hat{f}_\omega^{i^\star }(x_i))^2}{p \hat{\sigma}^2}
\end{equation}
for each $\omega \in [0,1]$, where $\hat{f}_\omega^{i^\star }$ is the
fitted model when case $i^\star $ has weight $\omega$ and the
remaining cases have weight 1. When $\omega = 0$, $\hat{f}_0^*$
coincides with $\hat{f}^{[-i^\star ]}$ and thus $D_{i^\star }(0) = D_{i^\star }$.
With this generalized distance, we could examine more complex modes of
case influence that may not be easily detected by Cook's distance $D_{i^\star }$.
Figure \ref{fig: cook1986} provides an example  of case-influence graphs where two cases A and B have the same Cook's distance but obviously different influence on the model fit depending on $\omega$. Two cases A and B can be treated the same if merely assessed by $D_{i^\star }$, but since $D_A(\omega) \ge D_B(\omega)$ at each $\omega \in [0,1]$, case A should be treated as more influential than case B.

Case-influence graphs provide  comprehensive information on local influence of cases in general, and they can be used to assess the differences in  robustness of modeling procedures. However, generation of such graphs is  computationally more expensive than Cook's distance. To circumvent the computational issue, \cite{cook1986localassess} suggested to focus on the local influence  around $\omega = 1$ through the curvature of the graph. As evident in \eqref{eq:general_cookdist}, once $\hat{f}_\omega^{i^\star }$ is obtained, the generalized Cook's distance $D_{i^\star }(\omega)$ is readily available. Thus, using path-following algorithms that generate case-weight adjusted solutions, we can easily construct case-influence graphs without additional computational cost.

Leveraging our solution path algorithm for quantile regression with
adjusted case-weight, we specifically study the characteristics of
case-influence graphs for various quantiles. In addition, we include
case-influence graphs of ridge regression for comparison of mean
regression and quantile regression as a robust counterpart in terms of
case influences. For ridge regression with penalty parameter $\lambda$ and case-weight $\omega$ for case $i^\star  \in \{1,\cdots, n\}$, we solve the following problem:
	\begin{equation}\label{lsr}
	\minimize_{\beta_0\in \mathbb{R}, {\beta}\in \mathbb{R}^p} \sum_{i \ne i^\star } (y_i - \beta_0 - x_i^\top {\beta})^2 + \omega (y_{i^\star } - \beta_0 - x_{i^\star }^\top {\beta})^2 +  \lambda \|\beta\|_2^2.
	\end{equation}
  With squared error loss, the case-weight adjusted fit
  $\hat{f}^{i^\star }_\omega$ can be computed in closed form and,
  thus, obtaining $D_{i^\star }(\omega)$ is straightforward for ridge
  regression.  For quantile regression with the check loss, however,  $\hat{f}^{i^\star }_\omega$ cannot be obtained as easily,  but our proposed solution path algorithm readily offers the path for $D_{i^\star }(\omega)$ as $\omega$ decreases from 1 to 0.

We present the case-weight adjusted Cook's distance  $D_{i^\star }(\omega)$ for ridge regression in the following proposition.
	Let $H(\lambda) = \tilde{X} (\tilde{X}^\top \tilde{X} + \lambda \tilde{I})^{-1}\tilde{X}^\top$ denote the hat matrix for ridge regression with full data, where $\tilde{I} =
	\begin{pmatrix}
	0 &\\
	& I
	\end{pmatrix}$.  $h_{ij}(\lambda)$ denotes the $ij$th entry of $H(\lambda)$ and $h_{ii}(\lambda)$ is the leverage of case $i$ in ridge regression.
\begin{pro}\label{ridgereg_sCD}
		 For ridge regression with penalty parameter $\lambda$,
		\begin{equation}\label{eq:rr-scaledCD}
		D_{i^\star }(\omega)
		= \frac{r^2_{i^\star }  \sum_{j = 1}^n h^2_{j i^\star}(\lambda)  }{p \hat{\sigma}^2\{1/(1 - \omega) - h_{i^\star i^\star}(\lambda) \}^2},
		\end{equation}
		where $r_{i^\star}$ is the residual for case $i^\star$ from the  full data fit.
	\end{pro}
 The proposition above  shows that $D_{i^\star}(\omega)$ for ridge
 regression is smooth and convex in $\omega$. The convexity comes from
 the fact that the second derivative of $g_h(\omega) = \{ 1/(1 -
 \omega) - h\}^{-2}$ for a positive constant $h$ is $\{2 + 4h(1 -
 \omega)\}\{1 - h(1 - \omega)\}^{-4}$, which is positive for $\omega
 \in (0, 1)$. Furthermore, $g_h(\omega)$ with $h \in [0, 1]$ decreases
 monotonically in $\omega \in (0,1)$ since $1/(1 - \omega)$ increases
 in $\omega$ and $1/(1 - \omega) - h > 0$. This implies that as the case-weight $\omega$ decreases from $1$ to $0$, $D_{i^\star}(\omega)$ increases monotonically since $h_{i^\star i^\star} \in [0,1]$. When both $\omega = 0$ and $\lambda = 0$,  $D_{i^\star }(\omega)$ reduces to  standard Cook's distance $(r^2_{i^\star }h_{i^\star i^\star })/\{p\hat{\sigma}^2(1 - h_{i^\star  i^\star } )^2\}$, where $h_{i^\star i^\star}$ is the leverage of case $i^\star$ in ordinary linear regression. This can be seen from the fact that $\sum_{j = 1}^n h^2_{i^\star j}(\lambda) $ is the $i^\star$th diagonal entry of $H^2(\lambda)$ and $H(0)$ is idempotent.

For penalized quantile regression, the piecewise linear solution path that we have constructed suggests that the discrepancy between the full-data fit and the case-weight adjusted fit  at any $\omega$, $\hat{f}(x_i) - \hat{f}_\omega^{i^\star }(x_i)$, is also piecewise linear, and thus $D_{i^\star }(\omega)$ is piecewise quadratic in $\omega$. Hence, $D_{i^\star }(\omega)$ can be easily obtained by aggregating the
piecewise squared difference in the fit from $1$ to
$\omega$. Equivalently, using \eqref{eq:resid_slopes}, the squared
difference in the residual, $(r_i - r_{i,\omega})^2$, can be
aggregated to produce $D_{i^\star }(\omega)$. An explicit
  expression of $D_{i^\star }(\omega)$ is provided in the proposition below.

	\begin{pro}\label{QRRP_sCD}
		 For penalized quantile regression in \eqref{w-set-up} with penalty parameter $\lambda$, if $\omega \in (\omega_{m+1}, \omega_m]$,
		\begin{align}\label{w_sCD}
		D_{i^\star }(\omega) &= \frac{1}{p\hat{\sigma}^2}\sum_{i=1}^{n} (\hat{f}(x_i) - \hat{f}_\omega^{i^\star }(x_i))^2 
    = \frac{1}{p\hat{\sigma}^2}\sum_{i=1}^{n} (r_{i,\omega} - r_{i,\omega_0 } )^2 \nonumber\\
		&= \frac{1}{p\hat{\sigma}^2} \|\mathbf r_{\omega} -
		\mathbf r_{\omega_m} + \sum_{k = 1}^{m} (\mathbf r_{\omega_k} -
		\mathbf r_{\omega_{k-1}} )\|_2^2 \nonumber \\
   &= \frac{1}{p\hat{\sigma}^2} \|(\omega - \omega_m)
    \mathbf h_m + \sum_{k = 1}^{m}(\omega_k - \omega_{k-1})\mathbf h_{k-1} \|_2^2 \, ,
		\end{align}
where  $\mathbf h_k$ is the vector of the slopes
 of the case-weight adjusted residuals $\mathbf r$ over $ (\omega_{k+1}, \omega_k]$.
\end{pro}
Numerical examples of case-influence graphs for ridge
  regression and quantile regression are presented in Section \ref{sec:empirical}.

\section{Numerical Studies}\label{sec:empirical}
In this section, we present various numerical studies to illustrate
the applications of our proposed case-weight adjusted solution path
algorithm, including LOO CV and case-influence graphs.
{We also analyze the computational complexity of the
  proposed path-following algorithm, and demonstrate its efficiency in
  computation of LOO CV scores numerically.}
Throughout the numerical studies, the standard linear model $y_i = \beta_0 +
x_i^\top\beta + \epsilon_i$ was used. We independently generated
covariates $\{x_{ij}: i=1, \cdots, n, ~j=1,\cdots, p\}$, coefficients $\big(\beta_0, \beta_1,\cdots,\beta_p \big)$ and random errors $\{\epsilon_i:  i=1, \cdots, n\}$ from the standard normal distribution.
We also created an R package \textit{NonsmoothPath} that can be used to reproduce the numerical results in this section. The R package is available at \url{https://github.com/qiuyu1995/NonsmoothPath}. 

\subsection{Leave-One-Out CV}
We first investigate the inaccuracy of GACV in approximating
  LOO CV scores as demonstrated in the introduction for extreme
  quantiles. This necessitates exact LOO CV.
 Then we numerically show that resorting to the homotopy strategy and
 applying our proposed $\omega$ path algorithm to obtain all the LOO
 solutions directly from the full-data solution could be more scalable
 and efficient than a straightforward procedure of solving
 $n$ LOO problems separately.

\subsubsection{Comparison between GACV and exact LOO
  CV}\label{sec:gacv_loo}

The exact LOO CV score in quantile regression is defined as  $RCV (\lambda) = \frac{1}{n}{\sum_{i=1}^{n} \rho_\tau\big(y_i - \hat{f}_\lambda^{[-i]} (x_i) \big) }$ and GACV score from \cite{Li2007lampath} is defined as $GACV(\lambda) = \frac{1}{n - |\mathcal{E}_\lambda|}
	{\sum_{i=1}^{n} \rho_\tau(y_i - \hat{f} (x_i))}$. We set $n = 50$, $p = 30$, and
$N_\lambda = 100$ to compare the exact LOO CV and GACV scores at
various quantiles, $\tau =0.5, ~0.3, ~ 0.1, \text{ and }0.01$.
Figure \ref{fig_gacv} reveals that as the pre-specified
  quantile $\tau$ gets extreme, the quality of GACV
  deteriorates. Similar observations have been made in the empirical
  studies of \cite{Li2007lampath} and \cite{badgacv}.

GACV is based on the smoothed  check loss, $\rho_{\tau,
    \delta}$, with a small threshold $\delta$, which is given by
 $\rho_{\tau, \delta}(r) = \big(\tau I(r > 0) + (1 - \tau)I(r < 0) \big) r^2/\delta$.
This approximate loss differs from $\rho_\tau$ only in the region of $(-\delta, \delta)$.
In the derivation of GACV, the following first-order Taylor expansion
of the smoothed loss is used:
$
\rho_\tau(y_i - \hat{f}^{[-i]}(x_i)) - \rho_\tau(y_i - \hat{f}(x_i)) \approx \rho_{\tau, \delta}^{\prime}(y_i - \hat{f}(x_i)) (\hat{f}(x_i) - \hat{f}^{[-i]}(x_i) ),
$
which may be attributed to the issue with GACV. Letting  $r^{[-i]}_i =
y_i - \hat{f}^{[-i]}(x_i)$, the LOO prediction error, and  $r_i = y_i
- \hat{f}(x_i)$, the residual from the full data fit, we define
 the approximation error of GACV from the exact LOO CV as
\begin{equation}
\Delta_{\text{approx.}} \defeq  \rho_{\tau, \delta}^{\prime}(r_i)  (r^{[-i]}_i - r_i) - \big[\rho_\tau(r^{[-i]}_i) - \rho_\tau(r_i) \big].
\end{equation}
Apparently the approximation error only comes from points with
different signs of $r^{[-i]}_i$ and $r_i$. We categorize all the
possible scenarios for the approximation error in Table \ref{tab:approx_error}
in the Appendix except the case when $r^{[-i]}_i = 0$. In the case of
  $r^{[-i]}_i = 0$, the approximation error is negligible.

\begin{figure}[h]
	\centering
	\includegraphics[height = 2.5in]{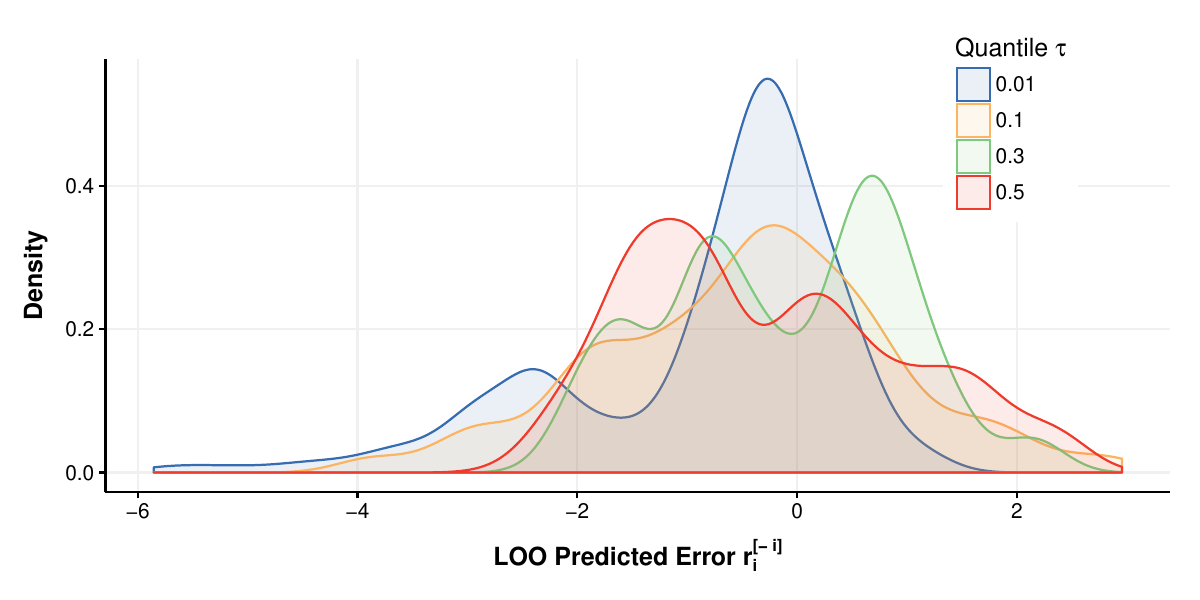}
	\caption{The distributions of LOO residual $r^{[-i]}_i$ for
          \textit{flipped} cases for various quantiles ($\tau = 0.5, ~0.3, ~0.1, \text{ and }0.01$).}
	\label{fig:gacv_issues}
\end{figure}

When the residual $r_i$ and the LOO residual $r_i^{[-i]}$ have
different signs $(+, 0, -)$, we call the case \textit{flipped} as in
scenarios (b) and (d) in Table \ref{tab:approx_error}.
Potential issues with GACV for extreme quantiles are
  summarized as follows:
\begin{enumerate}
	\item The cases in the elbow set have zero residuals. Thus, those cases are almost always flipped. In fact, in our experiment, we found that all the \textit{flipped} cases belong to the elbow set. The derivative of the smoothed check loss at $r_i = 0$ for the approximation is zero while the corresponding derivative for the true difference is either $\tau$ or $\tau - 1$. This leads to the approximation error listed in scenarios (b) and (d).

	\item For scenarios (b) and (d), the approximation error
          $\Delta_{\text{approx.}}$ depends on both $\tau$ and
          $r_i^{[-i]}$. Given $r_i^{[-i]}$, extreme values of $\tau$
          (e.g., $\tau=0.01$ in Figure \ref{fig_gacv}) lead to a larger approximation error in scenario (d), in particular. To see the effect of $\tau$ on the discrepancy between the true difference and its approximation, we examine the distribution of the LOO residuals $r_i^{[-i]}$ for flipped cases.
	Figure \ref{fig:gacv_issues} displays the distribution of $r^{[-i]}_i$ for \textit{flipped} cases for various quantiles when $log(\lambda) = -6$ from Figure \ref{fig_gacv}.  As $\tau$ becomes more extreme, the distribution tends to be more left-skewed, and scenario (d) occurs more often than scenario (b). This results in larger discrepancy between LOO CV and GACV for extreme quantiles as illustrated in Figure \ref{fig_gacv}.
\end{enumerate}

\subsubsection{Computation for Exact Leave-One-Out CV}\label{sec:lam_ome}

We compare two approaches to computing exact LOO CV scores over a set of $N_\lambda$
pre-specified grid points for the tuning parameter $\lambda$.
The first one is based on
the proposed $\omega$-path algorithm in Algorithm \ref{alg_w}.
The other one applies the
``$\lambda$-path'' algorithm proposed in \cite{Li2007lampath}
to the $n$ LOO data sets separately. We make comparisons of the two approaches
in terms of theoretical computational complexity as well as
practical runtime on simulated data sets. 

We first analyze the computational complexity of applying the
``$\lambda$-path'' algorithm proposed in \cite{Li2007lampath} $n$ times.
Note that the $\lambda$-path algorithm of \cite{Li2007lampath} generates the solution path as $\lambda$ decreases from $\infty$ to $0$.
The computation of the exact LOO CV scores involves two components:
(i) applying this algorithm to each of the $n$ LOO data sets; and (ii)
linearly interpolating the solutions between consecutive grid points.
According to \cite{Li2007lampath},
the average cost of computing one $\lambda$ path is $O(n^2p)$
and {the cost for} the linear interpolation is $O(N_\lambda p)$.
Hence, the total cost of computing the exact LOO CV scores
in this case is $O(np(n^2 + N_\lambda))$.

For the proposed $\omega$-path algorithm,
it generates each LOO solution directly from the full-data solution.
The computation consists of generating $n$ {case-weight adjusted} $\omega$-paths, whose
cost depends on the number of breakpoints for the case-weight parameter $\omega$. To simplify the analysis,
we work with the average number of $\omega$-breakpoints, denoted by $N_\omega$.
Our empirical studies show that $N_\omega$ is usually small compared to
problem dimension (see Table \ref{tab: w_steps} in the Appendix).
In fact, for large values of $\lambda$,
we can prove that $N_\omega = 1$.
Therefore, we assume that $N_\omega = O(1)$ in our analysis.
By inspecting Algorithm \ref{alg_w}, the average computational cost at
each $\omega$-breakpoint is dominated by Line 15, which computes
$b_{0,m}$, $b_m$, and $h_m$---the slopes of $\beta_{0,\omega}$, $\theta_{\mathcal E, \omega}$, and $r_\omega$.
First, the computation of $b_{0,m}$ and $b_m$
in \eqref{eq:theta_slopes} and \eqref{eq:slope_beta0}
involves inverting a $|\mathcal E_m| \times |\mathcal E_m|$ matrix
$\tilde X_{\mathcal E_m} \tilde X_{\mathcal E_m}^\top$,
which typically costs $O(|\mathcal E_m|^3)$.
This can be reduced {further} to $O(|\mathcal E_m|^2)$ by employing
a rank-one updating algorithm \citep{inversematrix}.
Moreover, the cost for computing $h_m$ in \eqref{eq:resid_slopes} is $O(n p)$.
Therefore, the average cost of generating
one $\omega$-path {at a grid point for $\lambda$} is $O(N_\omega(np + |\mathcal E_m|^2)) = O(np)$,
because $|\mathcal E_m| \leq \min(n, p+1)$
according to Lemma \ref{lemma_elbow} {in Appendix} and the assumption that
$N_\omega = O(1)$.
Consequently, the average cost of computing exact LOO CV 
{scores over $N_\lambda$ grid points} using the proposed
 $\omega$-path algorithm is
$O(N_\lambda n^2p)$, which is in contrast to the
cost of the $\lambda$-path algorithm, $O(np(n^2 + N_\lambda))$.
Note that the savings could be large when $N_\lambda \ll n$, which is corroborated by an empirical runtime comparison.

 We present numerical examples comparing the actual runtime
 performance of the two algorithms.
 Both algorithms are implemented in C++ with Armadillo package, and  were run on a Linux cluster with 24 cores and 128 GB of memory per node. We varied the quantile ($\tau = 0.1, ~0.3,~0.5$),
 sample size ($n=50$ to 300), number of covariates ($p=50$ to 300), and number of grid
 points ($N_\lambda = 20, ~50$). The grid points for $\lambda$ were
 equally spaced on the logarithmic scale over the range
 of  $\lambda$-breakpoints for the full data fit. For each setting, we
had 20 independent replicates and the results are summarized over the
replicates. To see the complexity of $\omega$-paths clearly, we
recorded the average number of $\omega$-breakpoints when $N_\lambda$ is 50 
in Table \ref{tab: w_steps}.

\begin{figure}[h]
 	\centering
 	\includegraphics[height = 2.5in]{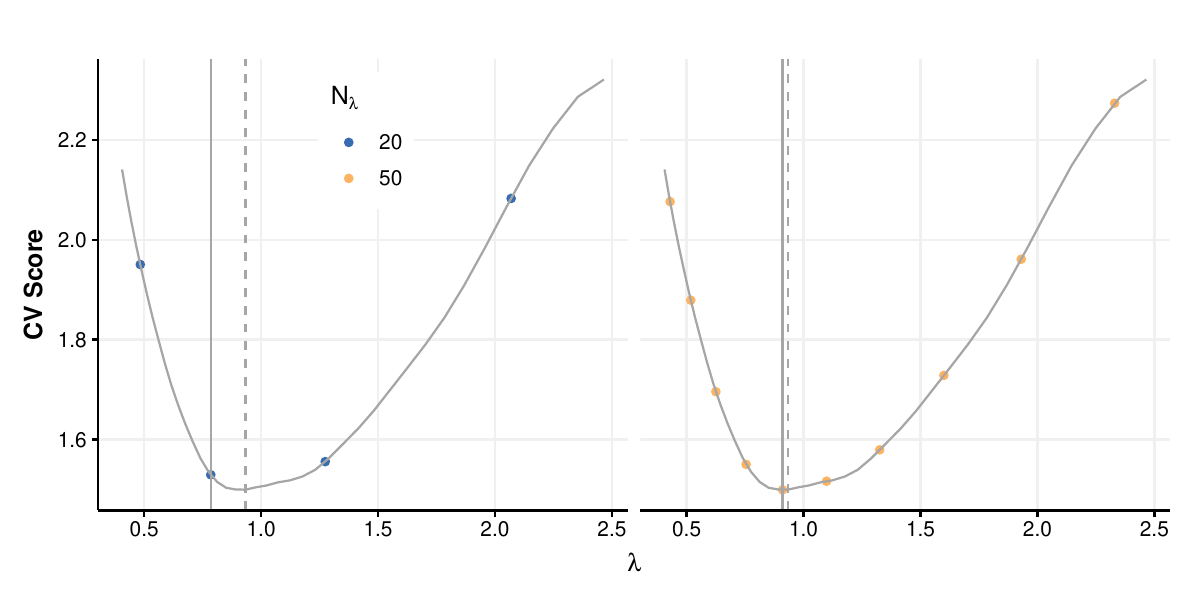}
 	\caption{Comparison of two levels of grid resolution in
          cross validation ($n = 100, p = 100, \tau = 0.5$).
          The solid curve is a complete CV
          score curve zoomed in the basin around the optimal $\lambda$,
          and the dots on the curves are CV scores at the grid points
          with $N_\lambda = 20$ in the left panel and $N_\lambda = 50$ in the right panel.  }
 	\label{fig_hv}
 \end{figure}

The runtimes for computation of CV scores
depend on the number of grid points and
generally a grid for the tuning parameter needs a sufficiently
fine resolution to locate the minimum CV score.
Figure \ref{fig_hv}
illustrates that both $N_\lambda = 20$ and $50$ are adequate
for identifying the optimal tuning parameter value for $\lambda$. The solid curves are the complete CV score curves while the dots on the curves
correspond to the CV scores at the grid points.
The minimizers of the CV scores over the grid points are
 indicated by the solid vertical lines in the two panels,
 both of which are close to the dashed vertical lines which
 correspond to the minimizers of the complete CV score curves.

\begin{figure}[h]
	\centering
	\includegraphics[height = 2.5in]{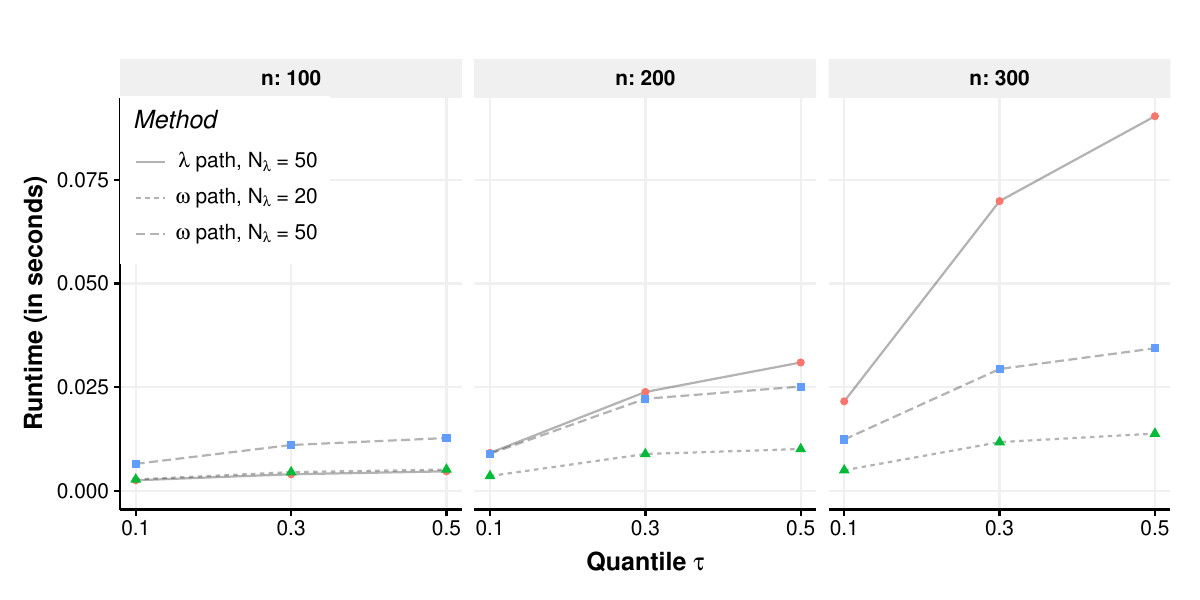}
	\caption{Comparison of the runtime per case (i.e., the total
          runtime/$n$) between the $\omega$-path and $\lambda$-path algorithms with different data dimensions, quantiles, and levels of grid resolution ($p = 50$; $n = 100, ~200,\text{ and }300$; $\tau = 0.1, ~0.3,\text{ and }0.5$; $N_\lambda = 20\text{ and }50$). }
	\label{time-plots}
\end{figure}

\begin{figure}[h]
	\centering
	\includegraphics[height = 2.5in]{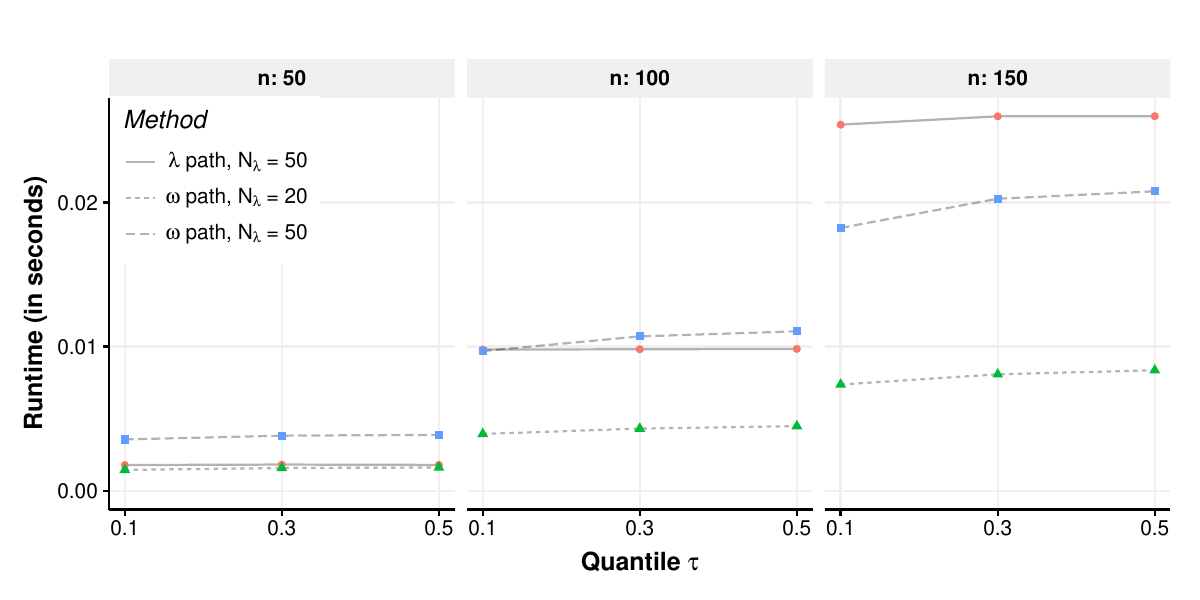}
	\caption{Comparison of the runtime per case between the $\omega$-path and $\lambda$-path algorithms with different data dimensions, quantiles and levels of grid resolution ($p = 300$; $n = 50, 100\text{ and }150$; $\tau = 0.1, ~0.3,\text{ and }0.5$; $N_\lambda = 20\text{ and }50$).}
	\label{time-plots-2}
\end{figure}

{The runtime comparisons are presented in
 Figure \ref{time-plots} for $p<n$ settings and Figure
 \ref{time-plots-2} for $p>n$ settings. The figures are based on the
 numerical summaries of the results in
 Tables \ref{table:LOOCV_p=50} and \ref{table:LOOCV_p=300} in Appendix.} 
 Overall, they show that as the sample size $n$ increases,
 our proposed $\omega$-path algorithm becomes more scalable
 than the competing $\lambda$-path algorithm.
This is consistent with our earlier analysis of computational
complexities.

\subsection{Case-Influence Graphs}\label{sec: case-inf-graphs}
This section presents case-influence graphs for ridge regression and
$\ell_2$-penalized quantile regression with the same data. As is introduced in Section \ref{sec_caseinf},
case-influence graphs show a broad view of
the influence of a case on the model as a function of the case-weight
$\omega$.
For simplicity,  we rescale the generalized Cook's distance
  $D_{i^\star}(\omega)$ in \eqref{eq:general_cookdist} by replacing  the factor
  $1/p\hat{\sigma}^2$ with $1/n$ as
\begin{equation}\label{w_hook_define}
\tilde{D}_{i^\star }(\omega) = \frac{1}{n}\sum_{i=1}^{n} (\hat{f}(x_i) - \hat{f}_\omega^{i^\star }(x_i))^2.
\end{equation}

\begin{figure}[h]
	\centering
	\includegraphics[height = 2.7in]{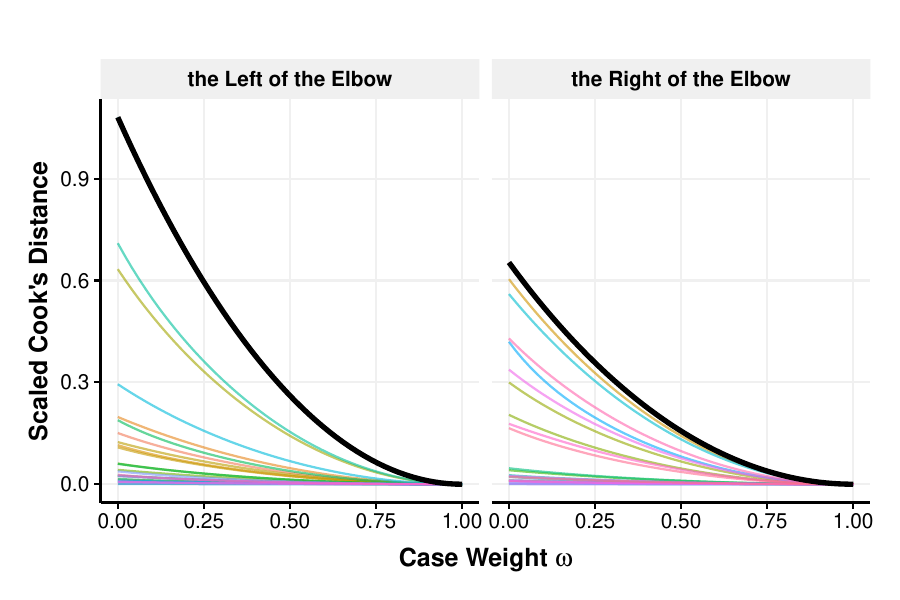}
	\caption{Case-influence graphs for ridge regression $(n = 50,
          p = 30)$. The influence curves in the left and right panels 
          correspond to cases with negative and positive full-data
          residuals, respectively. The bold curves correspond to the
          cases with the most positive or negative full-data residual.}
	\label{case-influence-plot}
\end{figure}

\begin{figure}[h]
	\centering
	\includegraphics[height = 4.5in]{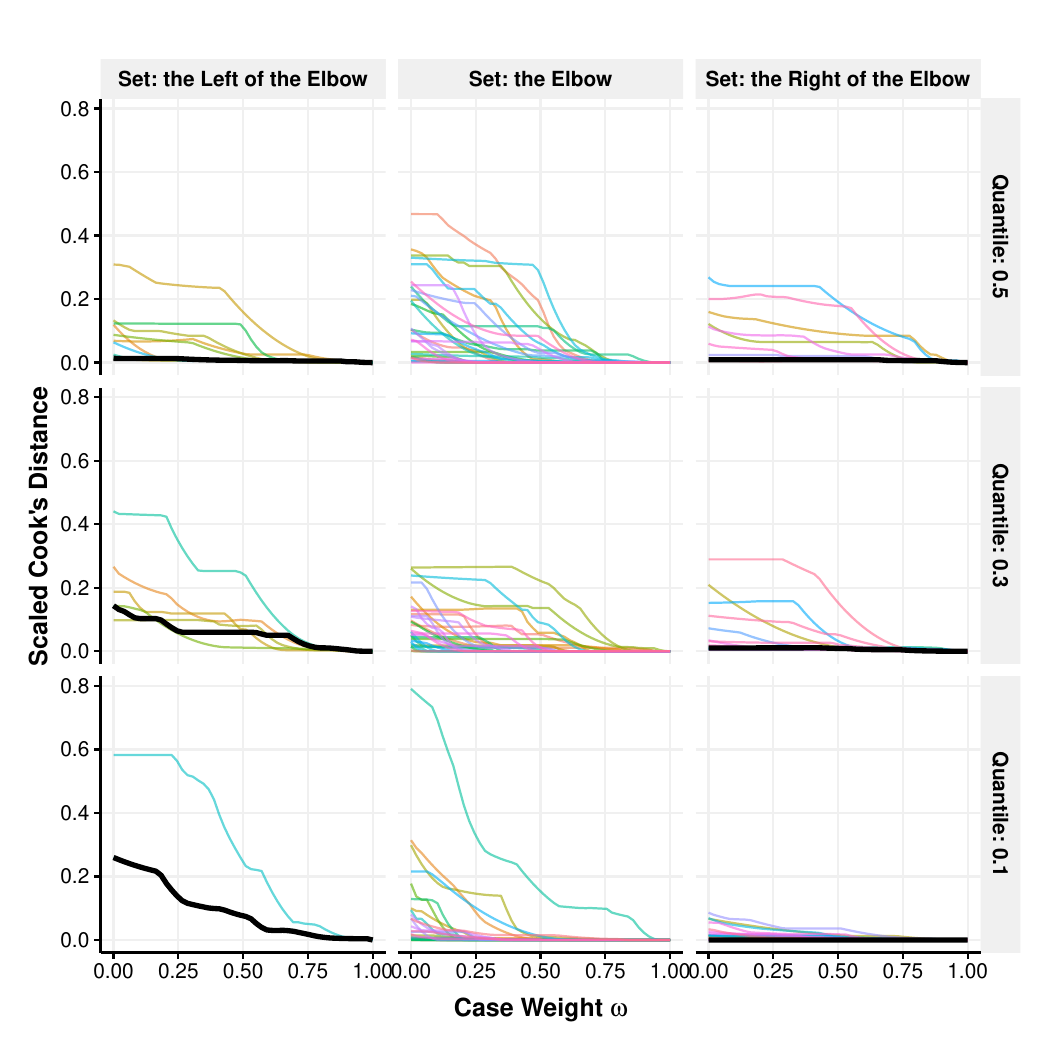}
	\caption{Case-influence graphs for penalized quantile
          regression with various quantiles $(\tau = 0.5, ~0.3,\text{
            and }0.1)$, $n = 50$ and $p = 30$. The left, middle
          and right panels correspond to cases with negative, zero and
          positive full-data residuals. The bold curves indicate the cases with the most positive or negative full-data residual.}
	\label{qr_hase-influence-plot}
\end{figure}

Figures \ref{case-influence-plot} and 
\ref{qr_hase-influence-plot} provide case-influence graphs for ridge regression and penalized quantile regression using the same
data. Here we remark some major differences in the characteristics
of the graphs. As is discussed in Section \ref{sec_caseinf},
the influence graphs for ridge regression in Figure \ref{case-influence-plot} are smooth, convex and monotonically decreasing in $\omega$, while those for quantile regression in Figure \ref{qr_hase-influence-plot} are piecewise quadratic.
Moreover, there are few crossings in the curves for ridge regression
in Figure \ref{case-influence-plot}, suggesting that the standard
Cook's distance may well be adequate for assessing case influences in
ridge regression. By contrast, Figure \ref{qr_hase-influence-plot}
reveals that for quantile regression, the relation between the
case-influence graphs and case-deletion statistics is more complex and some
cases in the elbow set with almost identical standard Cook's distance
can have quite different influence on the model fit.

 Additionally, the bold curves marked in Figures
 \ref{case-influence-plot} and \ref{qr_hase-influence-plot}
 show that for ridge regression, the cases with the most positive or
 negative full-data residuals have the greatest influence on the model
 fit, while for quantile regression that is not the case. In fact, for
 ridge regression,  \eqref{eq:rr-scaledCD} in Proposition
 \ref{ridgereg_sCD} implies strong dependence of the case influence on
 the magnitude of full-data residual.  Without much heterogeneity in
 the case leverages as in our data setting, the cases with the most
 positive or negative full-data residuals would have the greatest influence.
However, for quantile regression, the form of Cook's
  generalized distance derived in \eqref{w_sCD} does not reveal any
  specific relation between  case influence and the magnitude of
  full-data residual. It is observed that the residuals for the cases with
  the most positive or negative values tend not to change their signs
  as $\omega$ decreases from $1$ to $0$, and thus those cases are unlikely to
  enter the elbow set. They may have little influence on the model fit
  because the estimated  coefficients $(\hat \beta_{0,\omega}, \hat \beta_\omega)$ only depend on the responses in the elbow set $y_{\mathcal E_m}$ along with  $X_{\mathcal E_m}$ as shown in
  Section \ref{Sec: w_path}. This is akin to the fact that sample
  quantiles for modest $\tau$ are not affected by extreme observations. The case-influence graphs for quantile regression confirm this
  expectation, providing another perspective on the robustness of quantile regression for modest regression quantiles. Overall, the elbow set points are relatively more critical across different target quantiles while the left elbow points tend to be more influential for small quantiles ($\tau=0.3$ and $0.1$) in Figure \ref{qr_hase-influence-plot}. Similar observations will be made in the data analysis presented in the next section. 

\section{Data Analysis}\label{sec:data}

We illustrate how the proposed case-weight adjusted solution path and the corresponding case-influence graph can be utilized in real-life applications using the King County house sales data. By examining Cook's distance as a case deletion statistic (i.e., $D_{i^\star }(0)$ in \eqref{eq:general_cookdist} of Section \ref{sec_caseinf}), 
we demonstrate how the proposed methods can help identify influential observations in quantile regression. For normalization of Cook's distance, we used a robust estimate of the error variance $\sigma^2$ based on the median absolute deviation from the median regression.

The King County house sales data, publicly available on \href{https://www.kaggle.com/datasets/harlfoxem/housesalesprediction/discussion/207885}{Kaggle}, includes the price and property information of homes sold in King County, WA, between May 2014 and May 2015.  Following the exploratory analysis in \cite{luan2022measuringmodelcomplexityheteroscedastic}, we selected 12 features for an $l_2$-penalized quantile regression, and used a random sample of $n=200$ houses for illustration. The selected features are listed in Appendix \ref{sub_sec:house_data_features}. These predictors describe the quality and characteristics of a house, including the square root of the living space in square footage, the presence of a basement, the number of bathrooms, and others.

In Subsection \ref{housing_single_pred}, we consider a quantile regression with a single predictor with a negligible fixed penalty. 
In Subsection \ref{housing_multi_pred}, we utilize all 12 features for a multi-predictor quantile regression. In both subsections, the square root of the housing price is used as the response variable. A repository of the R code used in data analysis can be accessed at \url{https://github.com/hhhz-s/Case-Influence-in-Quantile-Regression}.

\subsection{Quantile Regression with a Single Predictor}
\label{housing_single_pred}

We consider a quantile regression with a single predictor for the ease of illustration. The square root of the average interior living room space in square footage from the closest 15 neighbors of a house is used as a predictor.
For various quantiles $\tau=0.1, 0.5$ and $0.95$, we set a negligible penalty of $\lambda=n*0.01$ to obtain the case-weight adjusted solution path and the corresponding case-influence graph.
This allows us to evaluate Cook's distance efficiently without refitting the model when one observation is excluded.

\begin{figure}[htp]
 \centering
    \includegraphics[width=1\textwidth]{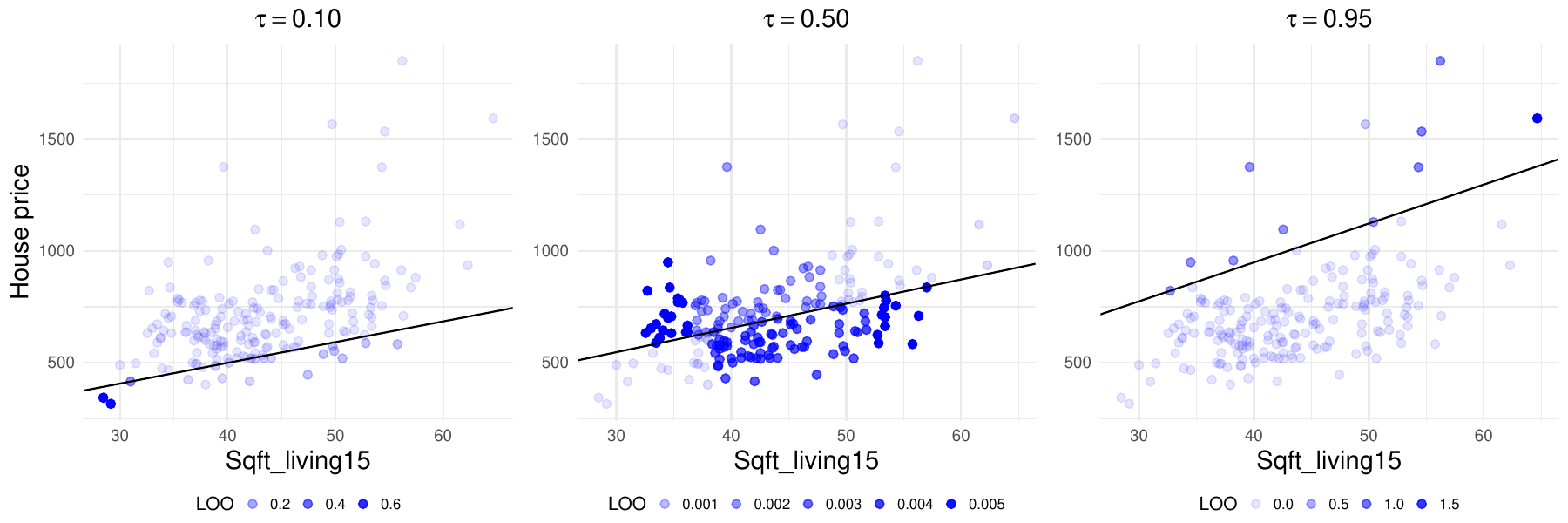}  
\caption{Scatter plot of house price versus the square root of the living room space in square footage from the nearest 15 neighbors for $\tau=0.1$ (left), $0.5$ (middle), and $0.95$ (right). The solid black line in each panel represents the fitted quantile regression line. The shade of each point is determined by Cook's distance as a LOO influence measure and is defined separately for each quantile value.}
\label{fig:qr_housing_single_pred}
\end{figure}

Figure \ref{fig:qr_housing_single_pred} illustrates how Cook's distance as a LOO influence measure is related to the relative extremeness of the predictor value and the residual in quantile regression. We observe that when 
$\tau=0.1$, influential observations tend to have negative residuals. When $\tau=0.5$, observations with either positive or negative residuals can exhibit a large influence. At $\tau=0.95$, influential observations generally have positive residuals. In terms of the predictor, influential observations tend to have more extreme predictor values.

\begin{figure}[h!]
    \centering
        \includegraphics[width=1\textwidth]{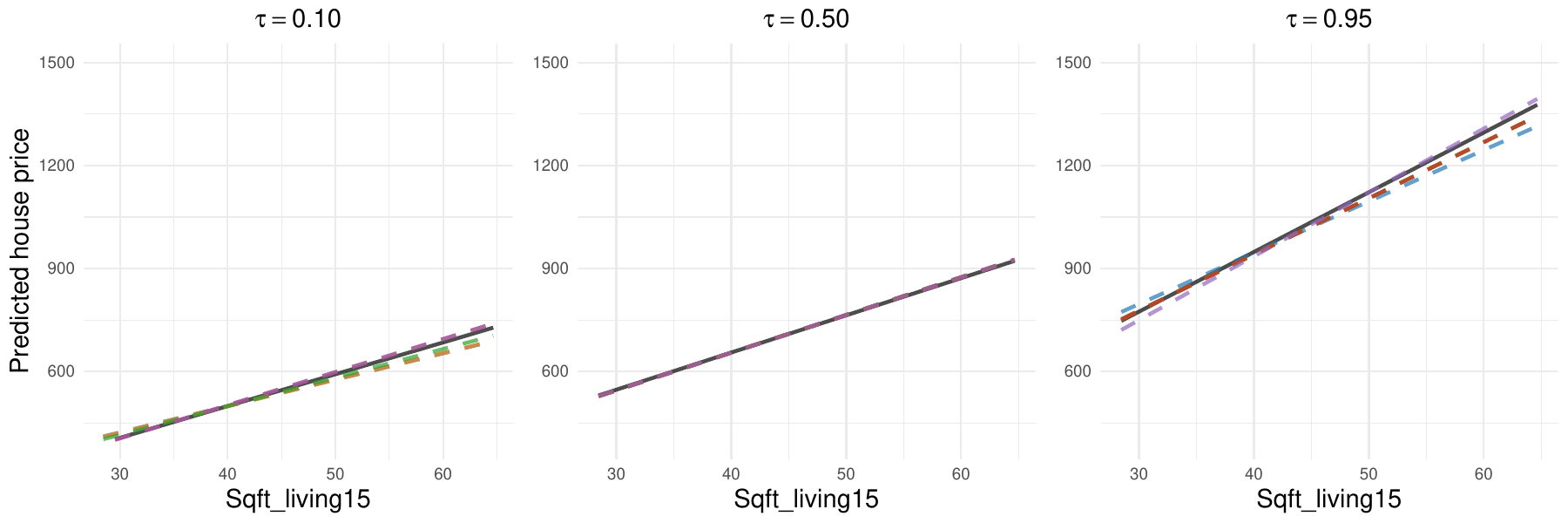} 
    \caption{Fitted single-predictor quantile regression lines for the King County house sales data when $\tau=0.1$ (left), $0.5$ (middle), and $0.95$ (right). The black solid lines are for the full-data solution, and the colored dashed lines are for the LOO solution when each of the top 5 influential observations is excluded.}
    \label{fig:qr_single_200_loo}
\end{figure}

Figure \ref{fig:qr_single_200_loo} compares the fitted regression lines for different quantiles using the full data with those when each of the top 5 influential observations is removed.  
As shown in Figure \ref{fig:qr_single_200_loo},
excluding the most influential observations from Figure \ref{fig:qr_housing_single_pred} results in various changes to the fitted quantile regression lines, depending on the specific value of the quantile $\tau$. The median regression appears to be the most robust, while the more extreme quantile regressions tend to be more sensitive to influential observations. This is also supported by a different magnitude of Cook's distance values across different quantiles in Figure \ref{fig:qr_housing_single_pred}.

Figure \ref{fig:qr_single_case_influence_graph} in Appendix \ref{sub_sec:piece_wise} displays the case-influence graphs for each observation in the single-predictor quantile regression. As in Figure \ref{fig:qr_housing_single_pred}, when $\tau = 0.1$, the influence of the cases in the right elbow set remains consistently low across all $\omega \in [0,1]$ while a couple of cases in the left elbow set have a relatively large influence. When $\tau = 0.5$, the case influence from the three sets is of the same scale. When $\tau = 0.95$, conversely, cases in the right elbow set tend to be more influential. 

The patterns of the case influence observed in Figures \ref{fig:qr_housing_single_pred} and \ref{fig:qr_single_case_influence_graph} can be explained through the property of the case-weight adjusted solution path. When adjusting the weight of a case $(x_{i^\star},y_{i^\star})$, the rate of change of the fitted value at $x_i$ can be quantified as
\begin{equation}
\label{eq_qr_case_sensitivity}
\left.\frac{\partial \hat{f}_{\omega}\left(x_i\right)}{\partial \omega}\right|_{\omega=1} =\frac{1}{\lambda} \left[\frac{\left(q_0^\top \tilde{x}_i-1\right) \left(q_0^\top \tilde{x}_{i^\star}-1\right)}{q_0^\top q_0}+\tilde{x}_i^{\top}\left(I-P_{\tilde{X}_{\mathcal{E}_0}}\right) \tilde{x}_{i^*}\right]\cdot \left(\tau-\mathbb{I}\left(i^\star\in \mathcal{L}_0\right)\right),
\end{equation}
where $q_0= \tilde{X}_{\mathcal{E}_0}^\top \left(\tilde{X}_{\mathcal{E}_0} \tilde{X}_{\mathcal{E}_0}^{\top}\right)^{-1} 1_{\mathcal{E}_0} \in \mathbb{R}^{p+1} \text { and } P_{\tilde{X}_{\mathcal{E}_0}}=\tilde{X}_{\mathcal{E}_0}^{\top}\left(\tilde{X}_{\mathcal{E}_0} \tilde{X}_{\mathcal{E}_0}^{\top}\right)^{-1} \tilde{X}_{\mathcal{E}_0}$ depend on the elbow set $\mathcal{E}_0$. Details of the derivation of the result in  \eqref{eq_qr_case_sensitivity} can be found in the proof of Proposition 1 in \cite{tu2019case}.

We refer to $\left(\tau-\mathbb{I}\left(i^\star\in \mathcal{L}_0\right)\right)$ as a generalized residual, which reflects the goodness-of-fit at $(x_{i^\star},y_{i^\star})$ similarly to the usual residual $r_{i^\star}=y_{i^\star}-\hat{f}_{i^\star}$. The generalized residual takes the value $\tau -1$ when $r_{i^\star}<0$ and $\tau$ when $r_{i^\star}>0$. 
Therefore, when $\tau<0.5$, influential observations tend to have negative residuals, as this leads to a larger generalized residual in the absolute value in quantile regression and implies more change in the fitted values when adjusting the weight of $(x_{i^\star},y_{i^\star})$.

In addition, the constant segments of the Cook's distance in Figure \ref{fig:qr_single_case_influence_graph} are due to the property of the case-weight adjusted solution path when the elbow set is of full rank. Justification of this property is provided in Appendix~\ref{sub_sec:piece_wise}. 

\subsection{Quantile Regression with Multiple Predictors}
\label{housing_multi_pred}
To demonstrate how the computational and graphical tools for case influence assessment can be utilized in the presence of multiple predictors, we fit a multi-variable quantile regression using the 12 predictors. 
 We selected the regularization parameter $\lambda$ based on the empirical loss on a separate validation set of size $200$. We may regard the term involving the predictor value in Equation \eqref{eq_qr_case_sensitivity} as the $(i, i^\star)$ entry of a generalized hat matrix akin to the hat matrix in least squares mean regression and refer to its diagonal entry as the generalized leverage:
\begin{equation}
h_{i^\star i^\star} =\frac{1}{\lambda} \left[\frac{\left(q_0^\top \tilde{x}_{i^\star}-1\right) \left(q_0^\top\tilde{x}_{i^\star}-1\right)}{q_0^\top q_0}+\tilde{x}_{i^\star}^{\top}\left(I-P_{\tilde{X}_{\mathcal{E}_0}}\right) \tilde{x}_{i^\star}\right].
\end{equation}

\begin{figure}[thp]
  \centering
  \includegraphics[width=1\textwidth]{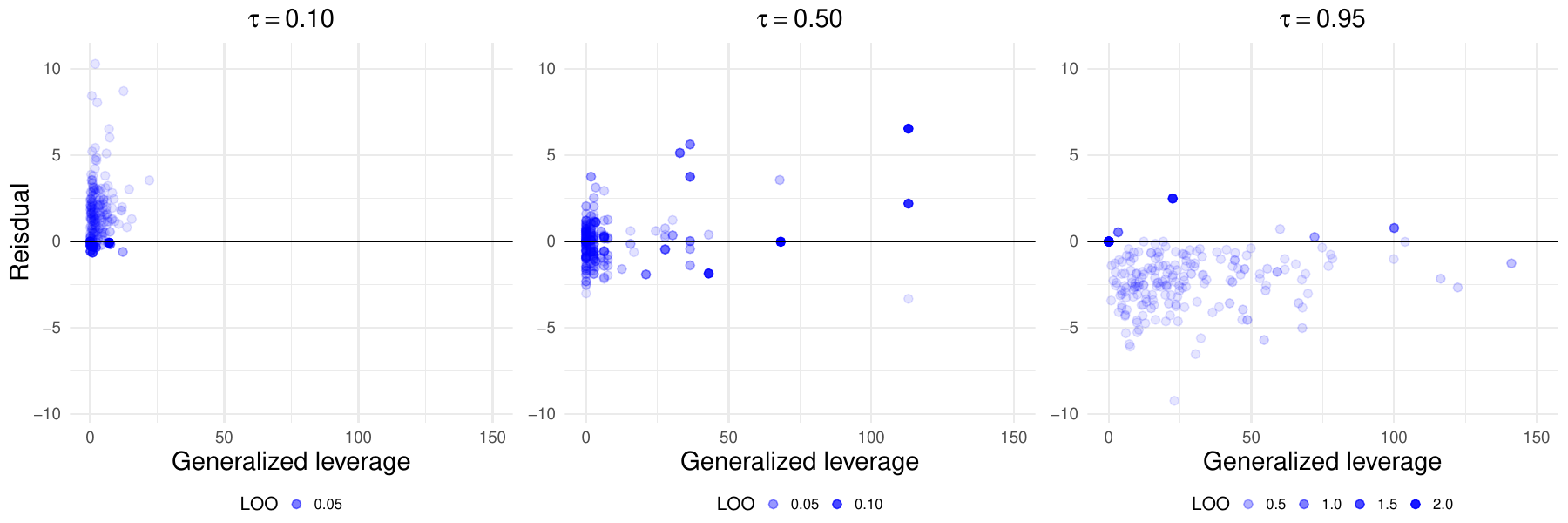}  
\caption{Scatter plot of the residual versus the generalized leverage when $\tau=0.1$ (left), $0.5$ (middle), and $0.95$ (right). The shade of each point is determined by Cook's distance as a LOO influence measure.}
 \label{fig:qr_housing_multi_pred}
\end{figure}

As illustrated in Figure \ref{fig:qr_housing_multi_pred}, similar to how Cook's distance in mean regression increases with leverage, case influence in quantile regression tends to be higher among observations with higher generalized leverage. However, the two types of leverage are distinct and may not align with each other, as generalized leverage depends on the quantile value $\tau$ through the elbow set. For instance, in Figure  \ref{fig:qr_housing_multi_pred}, the observation with the highest generalized leverage at $\tau=0.95$ has a moderate value of generalized leverage at $\tau=0.1$ unlike the leverage in least squares regression. We observe a similar pattern for the residuals as in the single-predictor quantile regression shown in Figure \ref{fig:qr_housing_single_pred}. Specifically, when $\tau<0.5$, observations with negative residuals tend to be more influential, highlighting the importance of the residual sign in determining the influence in quantile regression.

\begin{figure}[h!]
    \centering
        \includegraphics[width=0.8\textwidth]{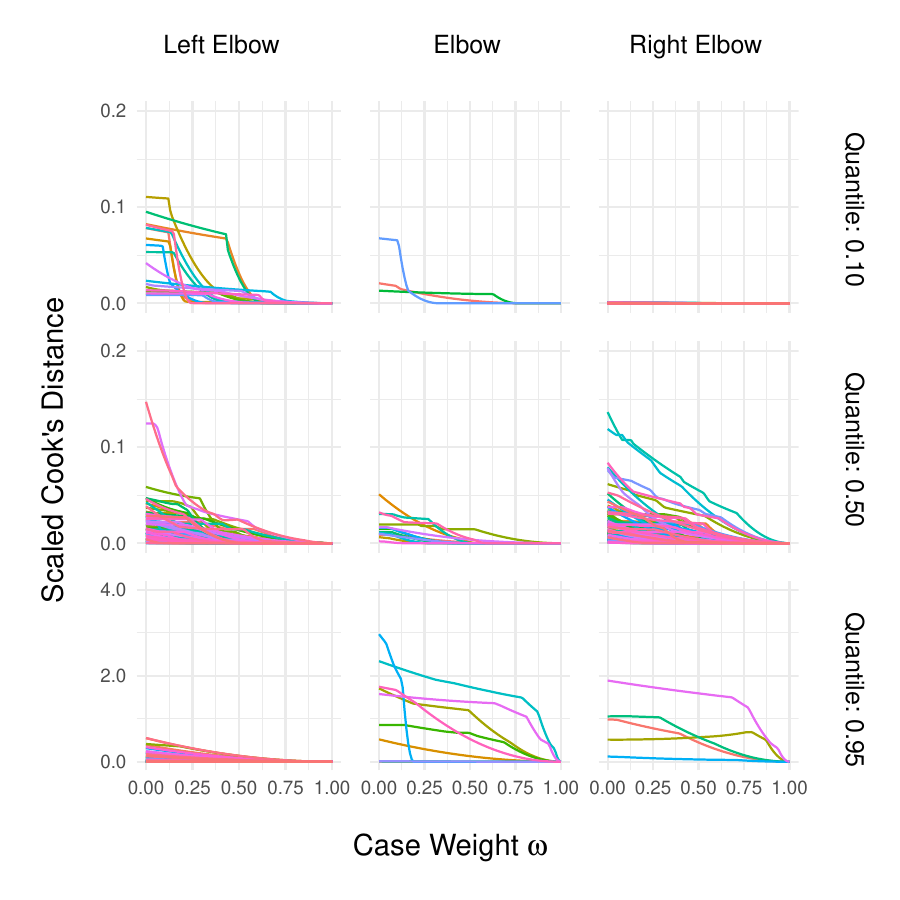} 
    \caption{Case-influence graphs from an $l_2$-penalized multi-predictor quantile regression for the King County house sales data with various quantiles: $\tau=0.1$ (top), 0.5 (middle) and 0.95 (bottom). The left, middle and right panels correspond to the left elbow, elbow and right elbow set points, respectively.}
    \label{fig:qr_multi_case_influence_graph}
\end{figure}

Figure~\ref{fig:qr_multi_case_influence_graph} shows the case-influence graphs for different quantile values based on the random sample of 200 houses. As illustrated in Figure \ref{fig:qr_multi_case_influence_graph}, when $\tau =0.1$, the houses in the left elbow set tend to be more influential than those in other sets.  When $\tau = 0.5$, the three sets are similar in terms of the magnitude of the case influence. When $\tau = 0.95$, highly influential points tend to appear from the elbow set or right elbow set.

We compare the estimated coefficients from the full-data solution and the LOO solution when the observation with the largest Cook's distance is removed for each quantile level. 
Tables of the estimated coefficients are provided in Appendix \ref{sub_sec:qr_housing_multi_loo}. As indicated in the tables, excluding the most influential observation could substantially change the quantile regression model, especially for extreme quantiles. For example, when $\tau=0.95$, removing the most influential observation resulted in the reversal of statistical significance of multiple features.

\section{Discussion}\label{sec:discussion}

In this article, we have proposed a novel path-following algorithm to compute the leave-one-out cross validation scores 
exactly for quantile regression with ridge penalty. 
Numerical analysis has demonstrated that 
the proposed algorithm compares favorably 
 to an alternative approach
in terms of computational  efficiency. 
Theoretically, we have provided a formal proof to establish the validity of the solution path 
algorithm. Moreover we have demonstrated that 
our proposed method can be used to 
efficiently compute the case-influence graph, 
which provides a more comprehensive approach to assessing
case influence.

We have primarily focused on $\ell_2$ 
penalized linear quantile regression. 
Similar case-weight adjusted path following algorithms 
can be  derived for nonparametric 
 quantile regression 
 and $\ell_1$ penalized quantile regression.
 Additionally, following the ideas proposed in 
 \cite{Rosset2009bilevelpath}, 
 it may be possible to derive 
 a bi-level solution path 
for each pair of $(\lambda, \omega)$, or even a tri-level path for each trio of $(\tau, \lambda, \omega)$. 
Furthermore, the idea of 
linking the full-data solution and the leave-one-out solution
can  be extended to classification settings. 
This will allow us to extend the notion of case influence to
 classification \citep{koh2017} 
  and to study the stability of classifiers
  using case-influence measures. How to efficiently 
  assess case influence in classification in itself is an important future direction.   

\section*{Acknowledgements}
This research was supported in part by National Science Foundation grants DMS-15-3566, DMS-17-21445, DMS-17-12580, and DMS-20-15490.

\bibliographystyle{dcu}
\bibliography{w_path_ref}

@article{homotopy-interiorpoint,
	Author = {X. Zhao and S.G. Zhang and Q.H. Liu},
	Journal = {Journal of Applied Mathematics},
	Pages = {1 - 12},
	Title = {Homotopy Interior-Point Method for a General Multiobjective Programming Problem},
	Volume = {77},
	Year = {2012}}

@InProceedings{cv_study,
	Author = {Ron Kohavi},
	Title = {A Study of Cross-Validation and Bootstrap for Accuracy Estimation and Model Selection},
	booktitle = {Proceedings of the 14th International Joint Conference on Artificial Intelligence (IJCAI)},
	pages = {1137-1145},
	Year = {1995}}

@article{quantile_outliers,
	Author = {Mohamed Chaouch and Camelia Goga},
	Issn = {0167-9473},
	Journal = {Computational Statistics and Data Analysis},
	Pages = {2214 - 2229},
	Title = {Design-based estimation for geometric quantiles with application to outlier detection},
	Volume = {54},
	Year = {2010}}

@article{inversematrix,
	Author = {Hager, W. W.},
	Journal = {SIAM Review},
	Number = {2},
	Pages = {221 - 239},
	Title = {Updating the Inverse of a Matrix},
	Volume = {31},
	Year = {1989}}

@book{belsley1980,
	Author = {Belsley, D.A. and Kuh, E. and Welsch, R.E.},
	Publisher = {New York: Wiley},
	Title = {Regression Diagnostics},
	Year = {1980}}

@techreport{ocv,
	Author = {Allen, D. M.},
	Institution = {Department of Statistics, University of Kentucky},
	Number = {23},
	Title = {The prediction sum of squares as a criterion for selecting predictor variables},
	Type = {Technical Report},
	Year = {1971}}

@inproceedings{koh2017,
  title={Understanding black-box predictions via influence functions},
  author={Koh, Pang Wei and Liang, Percy},
  booktitle={International Conference on Machine Learning},
  pages={1885--1894},
  year={2017},
  organization={PMLR}
}

@article{Osborne1992,
	Author = {MR Osborne},
	Journal = {IMA Journal of Numerical Analysis},
	Pages = {151 - 166},
	Title = {An effective method for computing regression quantiles},
	Volume = {12},
	Year = {1992}}

@article{homo_osborne2000,
	Author = {MR Osborne and B Presnell and BA Turlach},
	Journal = {IMA Journal of Numerical Analysis},
	Number = {3},
	Pages = {389 - 403},
	Title = {A new approach to variable selection in least squares problems},
	Volume = {20},
	Year = {2000}}

@article{Koenker40ys,
	Author = {Roger Koenker},
	Journal = {Annual Review of Economics},
	Pages = {155 - 176},
	Title = {Quantile Regression: 40 years On},
	Volume = {9},
	Year = {2017}}

@article{cook1979,
	Author = {Cook, D.},
	Journal = {Journal of American Statistical Association},
	Number = {365},
	Pages = {169 - 174},
	Title = {Influential observations in linear regression},
	Volume = {74},
	Year = {1979}}

@article{cook1977,
	Author = {Cook, D.},
	Journal = {Technometrics},
	Number = {1},
	Pages = {15 - 18},
	Title = {Detection of influential observations in linear regression},
	Volume = {19},
	Year = {1977}}

@article{l1qr,
	Author = {Youjuan Li and Ji Zhu},
	Journal = {Journal of Computational and Graphical Statistics},
	Number = {1},
	Pages = {163 - 185},
	Title = {{$L_1$}-Norm Quantile Regression},
	Volume = {17},
	Year = {2008}}

@article{path-origin,
	Author = {Allgower, E. and Georg, K.},
	Journal = {Acta Numerica},
	Pages = {1 - 64},
	Title = {Continuation and Path Following},
	Volume = {2},
	Year = {1993}}

@article{leastangle,
	Author = {Efron, B. and Hastie,T. and Johnstone, I. and Tishirani,R.},
	Journal = {The Annals of Statistics},
	Number = {2},
	Pages = {407 - 499},
	Title = {Least Angle Regression},
	Volume = {32},
	Year = {2004}}

@article{cv-origin,
	Author = {Stone, M.},
	Journal = {Journal of the Royal Statistical Society. Series B (Methodological)},
	Pages = {111 - 147},
	Title = {Cross-validatory choice and assessment of statistical predictions},
	Volume = {36},
	Year = {1974}}

@article{gcv-article,
	Author = {Craven, P. and Wahba, G.},
	Journal = {Numerische Mathematik},
	Pages = {377 - 403},
	Title = {Smoothing noisy data with spline functions: Estimating the correct degree of smoothing by the method of generalized cross-validation},
	Volume = {31},
	Year = {1979}}

@article{gcv-Golub,
	Author = {Wahba, G. and Golub, G. and Health, M.},
	Journal = {Technometrics},
	Pages = {215 - 223},
	Title = {Generalized Cross-Validation as a Method for Choosing a Good Ridge Parameter},
	Year = {1979}}

@article{badgacv,
	Author = {Reiss, Philip and Huang, Lei},
	Journal = {The International Journal of Biostatistics},
	Number = {1},
	Pages = {Article 10},
	Title = {Smoothness Selection for Penalized Quantile Regression Splines},
	Volume = {8},
	Year = {2012}}

@article{acv1995,
	Author = {Nychka,D. and Gray,G. and Haaland,P. and Martin,D.},
	Journal = {Journal of American Statistical Association},
	Number = {432},
	Pages = {1171 - 1178},
	Title = {A Nonparametric Regression Approach to Syringe Grading for Quality Improvement},
	Volume = {90},
	Year = {1995}}

@article{yuan2006,
	Author = {Ming Yuan},
	Journal = {Computational Statistics and Data Analysis},
	Pages = {813 - 829},
	Title = {{GACV} for quantile smoothing splines},
	Volume = {50},
	Year = {2006}}

@book{cook1982,
	Author = {Dennis Cook and Sanford Weisberg},
	Publisher = {Chapman and Hall},
	Title = {Residuals and Influence in Regression},
	Year = {1982}}

@article{Koenker1978,
	Author = {Koenker, R. and Bassett, G.},
	Journal = {Econometrica},
	Number = {1},
	Pages = {33 - 50},
	Title = {Regression Quantiles},
	Volume = {46},
	Year = {1978}}

@article{svmpath,
	Author = {Hastie,T. and Rosset,S. and Tishirani,R. and Zhu,J.},
	Journal = {Journal of Machine Learning Research},
	Pages = {1391 - 1415},
	Title = {The Entire Regularization Path for the Support Vector Machine},
	Volume = {5},
	Year = {2004}}

@article{Rosset07genericpath,
	Author = {Rosset, Saharon and Zhu, Ji},
	Journal = {The Annals of Statistics},
	Number = {3},
	Pages = {1012 - 1030},
	Title = {Piecewise Linear Regularized Solution Paths},
	Volume = {35},
	Year = {2007}}

@article{Takeuchi2009taupath,
	Author = {Takeuchi, I. and Nomura,K. and Kanamori, T.},
	Journal = {Neural Computation},
	Number = {2},
	Pages = {533 - 559},
	Title = {Nonparametric conditional density estimation using piecewise-linear solution path of kernel quantile regression},
	Volume = {21},
	Year = {2009}}

@article{Rosset2009bilevelpath,
	Author = {Saharon Rosset},
	Journal = {Journal of Machine Learning Research},
	Month = {Nov},
	Pages = {2473 - 2505},
	Title = {Bi-Level Path Following for Cross Validated Solution of Kernel Quantile Regression},
	Volume = {10},
	Year = {2009}}

@article{cook1986localassess,
	Author = {Dennis Cook},
	Journal = {Journal of the Royal Statistical Society. Series B (Methodological)},
	Number = {2},
	Pages = {133 - 169},
	Title = {Assessment of Local Influence},
	Volume = {48},
	Year = {1986}}

@article{Li2007lampath,
	Author = {Li, Youjuan and Liu, Yufeng and Zhu, Ji},
	Journal = {Journal of American Statistical Association},
	Month = {March},
	Number = {477},
	Pages = {255 - 268},
	Title = {Quantile Regression in Reproducing Kernel {Hilbert} Spaces},
	Volume = {102},
	Year = {2007}}

@misc{luan2022measuringmodelcomplexityheteroscedastic,
  title={On Measuring Model Complexity in Heteroscedastic Linear Regression}, 
  author={Bo Luan and Yoonkyung Lee and Yunzhang Zhu},
  year={2022},
  eprint={2204.07021},
  archivePrefix={arXiv},
  primaryClass={math.ST}
}

@phdthesis{tu2019case,
  author       = {Tu, Shanshan},
  title        = {Case Influence and Model Complexity in Regression and Classification},
  school       = {The Ohio State University},
  year         = {2019},
  address      = {Columbus, OH},
  type         = {Ph.D. dissertation},
}

@article{koenker2018package,
  title={Package ‘quantreg’},
  author={Koenker, Roger and Portnoy, Stephen and Ng, Pin Tian and Zeileis, Achim and Grosjean, Philip and Ripley, Brian D},
  journal={Reference manual available at R-CRAN: https://cran. rproject. org/web/packages/quantreg/quantreg. pdf},
  year={2018}
}

\newpage

\appendix

\section{Appendix}
\subsection{Derivation of KKT conditions \eqref{w_kkt1}--\eqref{w_kkt5} } 
\label{appx:kkt}

We derive the KKT conditions for the optimization problem \eqref{w-set-up}. Toward this end, let $(\beta_{0,\omega}, \beta_{\omega})$ denote the solution of \eqref{w-set-up}.
By \eqref{ql} and the fact that $\max(x, 0) = \inf_{t \geq 0, t \geq
  x} t$, we introduce auxiliary variables $\xi = (\xi_1, \cdots,
\xi_n)$ and $\zeta = (\zeta_1, \cdots, \zeta_n)$ with $\xi_i \ge
\max(y_i - \beta_0 - x_i^\top \beta, 0)$ and $\zeta_i \ge \max(-(y_i -
\beta_0 - x_i^\top \beta), 0)$ for $i = 1,\ldots, n$ to reexpress the check loss as follows:
\begin{eqnarray*}
\rho_{\tau}(y_i - \beta_0 - x_i^\top \beta) &=& \tau \max(y_i - \beta_0 - x_i^\top \beta,0) + (1 - \tau) \max(-(y_i - \beta_0 - x_i^\top \beta) , 0) \\
&=& \inf_{\xi_i, \zeta_i \geq 0 \text{ and }
-\zeta_i \le y_i -  \beta_0 - x_i^\top \beta \le \xi_i}
\tau \xi_i + (1 -\tau) \zeta_i \, .
\end{eqnarray*}
Thus, we can rewrite the optimization problem \eqref{w-set-up} as
\begin{equation}\label{w_path_opt}
\begin{array}{cl}
\underset{\beta_0 \in \mathbb{R}, \beta
\in \mathbb{R}^p, \xi \in \mathbb{R}^n, \zeta \in \mathbb{R}^n}
{\minimize}
& \tau \sum_{i \ne i^\star} \xi_i
			+ (1 - \tau) \sum_{i \ne i^\star} \zeta_i + \omega \tau \xi_{i^\star} + \omega (1 - \tau) \zeta_{i^\star} + \frac{\lambda}{2} \|\beta\|_2^2 \\
			\text{subject to  }& -\zeta_{i} \le y_{i} -  \beta_0 - x_{i}^\top \beta \le \xi_{i}, \text{and  } \zeta_i, \xi_{i} \ge 0 \text{ for } i = 1,\cdots,n \, .
			\end{array}
			\end{equation}
Note that \eqref{w_path_opt} is
in the standard form of a constrained convex optimization problem:
\begin{equation}
\label{eq:convex_opt_prob}
	\begin{array}{cl}
		\underset{\z \in \mathbb R^\ell}{\minimize} & f(\z) \\
		\text{ subject to } & g_i(\z) \leq 0 \text{ for }
		i = 1,\ldots, m \, ,
	\end{array}
\end{equation}
where $f(\cdot)$ and $g_i(\cdot)$ are
convex functions. It is well-known that
the KKT conditions for \eqref{eq:convex_opt_prob} are
\begin{equation}
  \begin{array} {ll}
    	\nabla f(\z) + \sum_{i = 1}^m \lambda_i \nabla g_i(\z) = 0 \, ,\text{ and } \\
	\lambda_i g_i(\z) = 0, \quad
	 \text{ for some real numbers } \lambda_i\geq 0, ~i = 1,\ldots, m \, .
\end{array}
\end{equation}
By letting $\z = (\beta_0, \beta, \xi, \zeta)$,
$g_i(\z) = -\zeta_i$, $g_{n+i}(\z) = -\xi_i$, $g_{i+2n}(\z) =
\beta_0 + x_i^\top \beta - y_i - \zeta_i$, and
$g_{i+3n}(\z) = y_i - \beta_0 - x_i^\top \beta - \xi_i$
for $i = 1,\ldots, n$, we next show that
equations \eqref{w_kkt1}-\eqref{w_kkt5} are the KKT conditions for \eqref{w_path_opt}. 

First, note that the Lagrangian function associated with \eqref{w_path_opt} is
\begin{eqnarray*}
&& L(\beta_0, \beta, \xi, \zeta, \alpha, \gamma, \kappa, \rho) = 
\tau \sum_{i \ne i^\star }\xi_i + (1 - \tau)\sum_{i \ne i^\star }\zeta_i + \omega \tau \xi_{i^\star } + \omega (1 - \tau) \zeta_{i^\star } + \frac{\lambda}{2} \|\beta\|_2^2 \\
 & & 
\quad + \sum_{i=1}^{n}\alpha_i(y_i - \beta_0 - x_i^\top \beta - \xi_i) - \sum_{i=1}^{n}\gamma_i(y_i - \beta_0 - x_i^\top \beta + \zeta_i)
- \sum_{i=1}^{n} \kappa_i\xi_i - \sum_{i=1}^{n}\rho_i \zeta_i \, ,
\end{eqnarray*}
where  $\alpha_i, \gamma_i, \kappa_i, \rho_i \ge 0$ are the dual
variables associated with the inequality constraints, and $\xi_i
\text{ and } \zeta_i \ge 0$ are primal variables  introduced in \eqref{w_path_opt}.
Hence, the Karush-Kuhn-Tucker (KKT)
conditions are given by
\begin{eqnarray*}
\label{w_spqr_kkt_ini}
\frac{\partial L}{\partial \beta}
        &=& \lambda \beta - \sum_{i=1}^{n}\alpha_i x_i +
    \sum_{i=1}^{n}\gamma_i x_i = 0,\\
\frac{\partial L}{\partial \beta_0}
       &=&\sum_{i=1}^{n} (\alpha_i - \gamma_i) = 0, \\
\frac{\partial L}{\partial \xi_i} &=& -\alpha_i - \kappa_i + \omega \tau + (1 - \omega) \tau \mathbf{1}_{\{i \ne i^{\star}\}}=0,\\
\frac{\partial L}{\partial \zeta_i} &=&	-\gamma_i - \rho_i + \omega (1 - \tau) +
	(1 - \omega)(1 - \tau) \mathbf{1}_{\{i \ne i^{\star}\}} =0,\\
	&&  \alpha_i(y_i - \beta_0 - x_i^\top \beta - \xi_i) = 0,
	\quad \gamma_i(y_i - \beta_0 - x_i^\top \beta + \zeta_i) = 0,
	\quad \kappa_i \xi_i = 0, \quad \rho_i \zeta_i = 0,  \\
	&&  -\zeta_i \le y_i - \beta_0 - x_i^\top \beta \le \xi_i, \quad
	\alpha_i \ge 0, ~\gamma_i \ge 0, ~\kappa_i \ge 0, ~\rho_i \ge 0,~\xi_i \ge 0, 
  ~\zeta_i \ge 0, ~ i = 1,\cdots,n.
\end{eqnarray*}
Defining $\theta_i \defeq \alpha_i - \gamma_i$ for $i =
  1,\cdots,n$, we obtain \eqref{w_kkt1} from the first two equations.

Note that when $y_i - \beta_0 - x_i^\top \beta > 0$, we must have
$\xi_i \ge y_i - \beta_0 - x_i^\top \beta > 0 $, which, together
with $\kappa_i \xi_i = 0$, implies that $\kappa_i = 0$.
Consequently, we have that
$\alpha_i = \omega \tau + (1 - \omega) \tau \mathbb{I}_{\{i \ne i^{\star}\}}$ and $\xi_i = y_i - \beta_0 - x_i^\top \beta$,
because $\alpha_i \neq 0$. Moreover, we also have that $\gamma_i = 0$, because $\gamma_i(y_i - \beta_0 - x_i^\top \beta + \zeta_i) = 0$
 and $\zeta_{i} \ge 0$. Hence, $\theta_i = \alpha_i - \gamma_i
 = \omega \tau + (1 - \omega) \tau \mathbb{I}_{\{i \ne i^{\star}\}}$, which proves \eqref{w_kkt5}. Similarly, when $y_i - \beta_0 - x_i^\top \beta < 0$, we have that
$\gamma_i = \omega (1 - \tau) + (1 - \omega)(1 - \tau) \mathbb{I}_{\{i \ne i^{\star}\}}$ and $\alpha_i = 0$. Hence,
$\theta_i = \alpha_i - \gamma_i
 = - \omega \tau - (1 - \omega) \tau \mathbb{I}_{\{i \ne i^{\star}\}}$, which proves \eqref{w_kkt3}.

Finally, when $y_i - \beta_0 - x_i^\top \beta = 0$, we must have that
$\xi_i = 0$, because $\alpha_i \xi_i = 0$, $\kappa_i \xi_i = 0$, and
$\alpha_i + \kappa_i = \omega \tau +
(1 - \omega) \tau \mathbb{I}_{\{i \ne i^{\star}\}} > 0$. Similarly, $\zeta_i = 0$. Moreover, note that
$\alpha_i \in [0,\omega \tau +
(1 - \omega) \tau \mathbb{I}_{\{i \ne i^{\star}\}}]$ and
$\gamma_i \in [0, \omega (1 - \tau) + (1 - \omega)(1 - \tau)
 \mathbb{I}_{\{i \ne i^{\star}\}}]$. Hence,
 $\theta_i = \alpha_i - \gamma_i \in [ -\omega (1 - \tau) -
 (1 - \omega)(1 - \tau)
 \mathbb{I}_{\{i \ne i^{\star}\}}
 \, ,\omega \tau +
(1 - \omega) \tau \mathbb{I}_{\{i \ne i^{\star}\}}]$, which
proves \eqref{w_kkt4}.

\subsection{Proof of Lemma \ref{lemma_elbow} }

\begin{lemma}\label{lemma_elbow}
	Let $\mathcal E_m$ be the elbow set defined in Algorithm 1. Suppose 
  that $\{(\tilde x_i, y_i)\}_{i=1}^{n}$ satisfies the general position 
        condition that 
	any {$\min(p+2, n)$} points of $\{(\tilde x_i, y_i)\}^{n}_{i=1}$  are linearly independent.  Then  we have $|\mathcal E_m| \le p+1$ and
	$\tilde{X}_{\mathcal{E}_m} \tilde{X}_{\mathcal{E}_m}^\top \succ 0$
	for each $m = 0, 1, \ldots, M$.
	\end{lemma}

\begin{proof}
  We prove $\tilde{X}_{\mathcal{E}_m} \tilde{X}_{\mathcal{E}_m}^\top
  \succ 0$ by showing that  (i) $|\mathcal E_m| \le p+1$ and
   (ii) the  rows of $\tilde X_{\mathcal E_m}$ are linearly independent. 

\textbf{ (i). } We prove $|\mathcal E_m| \le p+1$ by contradiction.
Suppose that $|\mathcal E_m| \geq p+2$. Then we must have
$p+2 \leq |\mathcal E_m| \leq n$. Moreover, by the \textit{general position condition}, we know that any $\min(n,p+2) = p+2$ points of
$\{(\tilde x_i, y_i)\}^{n}_{i=1}$  are linearly independent, which 
implies that $rank(\tilde{X}_{\mathcal E_m},  Y_{\mathcal E_m} ) \ge p+2$.
On the other hand, we can rewrite the KKT condition \eqref{w_kkt4} as
\begin{equation*}
y_{\mathcal E_m} = \beta_0 {1}_{\mathcal E_m} + X_{\mathcal E_m} \beta = \tilde{X}_{\mathcal E_m}
\begin{pmatrix}
\beta_0 \\
\beta
\end{pmatrix} \, ,
\end{equation*}
which implies that  $rank(\tilde{X}_{\mathcal E_m},  Y_{\mathcal E_m} ) 
= rank(\tilde{X}_{\mathcal E_m}) \le min(p+1, n) \le p+1$. 
This is a contradiction. Thus $|\mathcal E_m| \le p+1$.

\textbf{(ii).} Since the number of rows of $\tilde{X}_{\mathcal E_m}$,
$|\mathcal E_m| \le \min(p+1,n) \leq \min(p+2,n)$ by (i),
$rank(\tilde{X}_{\mathcal E_m},  Y_{\mathcal E_m} )=|\mathcal E_m|$
by the \textit{general position condition}, which implies that
$rank(\tilde{X}_{\mathcal E_m})=|\mathcal E_m|$. Thus, 
the rows of $\tilde{X}_{\mathcal E_m}$ must be linearly independent.
\end{proof}

\subsection{Proof of Proposition \ref{prop:sol_path_evolution} }
Since $\lambda r_\omega =
			\lambda( y - \beta_{0,\omega}{\mathbf 1} - X\beta_{\omega})
	= \lambda y - \lambda \beta_{0,\omega}{\mathbf 1} - X X^\top \theta_{\omega}$, using the fact that $\lambda \beta_{\omega} = X^\top \theta_{\omega} $ from \eqref{w_kkt1},
			we only need to derive the updating formulas for
			$\beta_{0,\omega}$ and $\theta_{{\mathcal E_m},\omega}$.



			Let $\tilde{X} = (\mathbf{1}, X)$ denote the expanded design matrix with each row
			$\tilde{x}_i^\top = (1, x_i^\top) \text{ for }
			 i = 1,\cdots, n$.
			Combining the first two equations in \eqref{eq:linear},
			we rewrite \eqref{eq:linear} as
			\begin{equation}
			\label{rewritten_eq}
			\begin{array}{cll}
			\lambda \begin{pmatrix}
			0\\
			\beta_\omega
			\end{pmatrix} - \tilde{X}^\top_{\mathcal E_m} \theta_{\mathcal E_m, \omega} &=&
			\tilde{X}^\top_{\mathcal L_m} \theta_{\mathcal L_m, \omega} + \tilde{X}^\top_{\mathcal R_m} \theta_{\mathcal R_m, \omega}, \\
			\beta_{0,\omega} {1}_{\mathcal{E}_m} + \tilde{X}_{\mathcal{E}_m}
			\begin{pmatrix}
			0\\
			\beta_\omega
			\end{pmatrix} & = &
			y_{\mathcal E_m}.
			\end{array}
			\end{equation}
By eliminating $\beta_\omega$ from \eqref{rewritten_eq},
we have that
\begin{equation}
			\label{eq:modified_KKT1}
			(\tilde{X}_{\mathcal{E}_m} \tilde{X}_{\mathcal{E}_m}^\top)	\theta_{\mathcal{E}_m, \omega} =
			\big[\lambda y_{\mathcal{E}_m} - \lambda \beta_{0,\omega} {1}_{\mathcal{E}_m}
			- \tilde{X}_{\mathcal{E}_m}
			\big(
			\tilde{X}^\top_{\mathcal L_m} \theta_{\mathcal L_m, \omega} +	\tilde{X}^\top_{\mathcal R_m} \theta_{\mathcal R_m, \omega}
			\big)
			\big] \, ,
			\end{equation}
			and
			\begin{equation}
			1^\top_{\mathcal E_m} \theta_{\mathcal E_m, \omega} =
			- 1^\top_{\mathcal L_m} \theta_{\mathcal L_m, \omega}
			- 1^\top_{\mathcal R_m} \theta_{\mathcal R_m, \omega}.
			\end{equation}
%
Under the \textit{general position condition}, 
			we have that
			\begin{equation}
			\label{eq:beta0_exp}
			\lambda \beta_{0, \omega} =
			\frac{1^\top_{\mathcal{E}_m}
				(\tilde{X}_{\mathcal{E}_m} \tilde{X}_{\mathcal{E}_m}^\top)^{-1}
				\left(
				\lambda y_{\mathcal{E}_m} -
				\tilde{X}_{\mathcal{E}_m}
				(\tilde{X}^\top_{\mathcal L_m} \theta_{\mathcal L_m, \omega} + \tilde{X}^\top_{\mathcal R_m} \theta_{\mathcal R_m, \omega})
				\right) +  1^\top_{\mathcal L_m} \theta_{\mathcal L_m, \omega}
				+ 1^\top_{\mathcal R_m} \theta_{\mathcal R_m, \omega}
			}
			{
				1^\top_{\mathcal{E}_m} (\tilde{X}_{\mathcal{E}_m}
				\tilde{X}_{\mathcal{E}_m}^\top)^{-1} 1_{\mathcal{E}_m}}
			\end{equation}
			and
			\begin{equation}
			\label{eq:modified_KKT}
			\theta_{\mathcal{E}_m, \omega} = (\tilde{X}_{\mathcal{E}_m} \tilde{X}_{\mathcal{E}_m}^\top)^{-1}
			\big[\lambda y_{\mathcal{E}_m} - \lambda \beta_{0,\omega} {1}_{\mathcal{E}_m}
			- \tilde{X}_{\mathcal{E}_m}
			\big(
			\tilde{X}^\top_{\mathcal L_m} \theta_{\mathcal L_m, \omega} +	\tilde{X}^\top_{\mathcal R_m} \theta_{\mathcal R_m, \omega}
			\big)
			\big] \, ,
			\end{equation}
  where the fact that $\tilde{X}_{\mathcal{E}_m}
	\tilde{X}_{\mathcal{E}_m}^\top$ is invertible and
  $1^\top_{\mathcal{E}_m} (\tilde{X}_{\mathcal{E}_m}
	\tilde{X}_{\mathcal{E}_m}^\top)^{-1} 1_{\mathcal{E}_m} \neq 0$
			are ensured by the
			\textit{general position condition} in view of 
      Lemma \ref{lemma_elbow}. 

From \eqref{eq:beta0_exp} and \eqref{eq:modified_KKT},
note that the dependence of $\beta_{0,\omega}$ and
 $\theta_{\mathcal E_m, \omega}$ on $\omega$ stems
 from $\theta_{\mathcal L_m, \omega}$ and
 $\theta_{\mathcal R_m, \omega}$, which may be a
 function of $\omega$ depending
			on whether the weighted case $i^\star$ is
			in $\mathcal L_m \cup \mathcal R_m$ or not.
More specifically, from \eqref{w_kkt3}--\eqref{w_kkt5},
if case $i^\star \in \mathcal E_m$, then
$\beta_{0,\omega}$ and $\theta_{\mathcal E_m, \omega}$ are
independent of $\omega$, because $\theta_{\mathcal L_m, \omega}$ and
$\theta_{\mathcal R_m, \omega}$ are both independent of $\omega$.
On the other hand, if case $i^\star \in \mathcal L_m
\cup \mathcal R_m$, then $\beta_{0,\omega}$ and $\theta_{\mathcal E_m,
  \omega}$ are linear in $\omega$ as $\theta_{i^\star} = \omega( \tau
- \mathbb{I}(i^\star \in \mathcal L_m))$ is linear in $\omega$. As a
result, we consider these two cases separately to determine
  the next breakpoint:
\begin{itemize}
	\item Case I: $i^\star \in
	\mathcal L_m \cup \mathcal R_m$
	\item Case II: $i^\star \in
	\mathcal E_m$
\end{itemize}
		\textbf{I. }
			For Case I, note that
			\begin{equation*}
			\theta_{i^\star, \omega} = \omega
			[\tau -  \mathbb I(i^\star \in \mathcal L_m)] \text{ and }
			\theta_{i, \omega} = [\tau -  \mathbb I(i \in \mathcal L_m)] \text{ for } i \neq i^\star \text{ and } i
			\in \mathcal L_m \cup
			\mathcal R_m.
			\end{equation*}
			Hence, taking the difference
			of \eqref{eq:beta0_exp} at $\omega$ and $\omega_m$,
                        and using the fact that only $\theta_{i^\star,\omega}$ changes
                          with $\omega$,
			we obtain that
			\begin{equation}
			\label{eq:beta0_slope}
			\lambda \beta_{0,\omega} - \lambda \beta_{0, \omega_m} = b_{0,m} (\omega  - \omega_m) \, ,
			\end{equation}
			where
			\begin{equation*}
			b_{0, m} = \frac{ 1 - 1_{\mathcal{E}_m}^\top (\tilde{X}_{\mathcal{E}_m} \tilde{X}_{\mathcal{E}_m}^\top)^{-1}\tilde{X}_{\mathcal{E}_m}\tilde{x}_{i^\star}     }{  1_{\mathcal{E}_m}^\top (\tilde{X}_{\mathcal{E}_m} \tilde{X}_{\mathcal{E}_m}^\top)^{-1}1_{\mathcal{E}_m} } [\tau -  \mathbb I(i^\star \in \mathcal L_m)]. 
		\end{equation*}
			Similarly,  taking the difference
			of \eqref{eq:modified_KKT} at $\omega$ and $\omega_m$,
			we obtain that
			\begin{eqnarray*}
			&& \theta_{\mathcal{E}_m, \omega} -
			\theta_{\mathcal{E}_m, \omega_m}  \nonumber \\
			&=& - (\tilde{X}_{\mathcal{E}_m} \tilde{X}_{\mathcal{E}_m}^\top)^{-1}
			\left[ (\lambda \beta_{0,\omega} - \lambda \beta_{0, \omega_m}) {1}_{\mathcal{E}_m}
			+
			\tilde{X}_{\mathcal{E}_m}\tilde{x}_{i^\star}
			\{\tau -  \mathbb I(i^\star \in \mathcal L_m)\} (\omega  - \omega_m)
			\right] \nonumber \\
			& = &
			- (\tilde{X}_{\mathcal{E}_m} \tilde{X}_{\mathcal{E}_m}^\top)^{-1}
			\left[ b_{0,m} 1_{\mathcal E_m}
			+
			\tilde{X}_{\mathcal{E}_m}\tilde{x}_{i^\star}
			\{\tau -  \mathbb I(i^\star \in \mathcal L_m)\}
			\right]  (\omega  - \omega_m) \nonumber \\
			&=& b_{m}(\omega  - \omega_m)  \, ,
		\end{eqnarray*}
			where
			\begin{equation*}
			b_{m} = - (\tilde{X}_{\mathcal{E}_m} \tilde{X}_{\mathcal{E}_m}^\top)^{-1}
			\left[ b_{0,m} 1_{\mathcal E_m}
			+
			\tilde{X}_{\mathcal{E}_m}\tilde{x}_{i^\star}  \{\tau -  \mathbb I(i^\star \in \mathcal L_m)\}
			\right]  \, .
		\end{equation*}
			This proves \eqref{eq:theta_evol} and \eqref{eq:theta_slopes}.
			Note that \eqref{eq:theta_evol} and \eqref{eq:theta_slopes}
			give how $\theta_{\mathcal E_m, \omega}$ changes
			as a function of $\omega$. Next, we derive a similar formula for $r_{\omega}$.
			To that end, multiplying both sides of the first equation
			in \eqref{rewritten_eq} by $\tilde{X}$, we have that for any
			$\omega \in [\omega_{m+1}, \omega_m]$,
			\begin{equation*}
			\lambda \tilde{X}
			\begin{pmatrix}
			0\\
			\beta_\omega
			\end{pmatrix} - \tilde{X}
			\tilde{X}^\top_{\mathcal E_m} \theta_{\mathcal E_m, \omega} =
			\tilde{X} \tilde{X}^\top_{\mathcal L_m} \theta_{\mathcal L_m, \omega}
			+ \tilde{X} \tilde{X}^\top_{\mathcal R_m} \theta_{\mathcal R_m, \omega}.
			\end{equation*}
Together with \eqref{w_kkt3}--\eqref{w_kkt5}, this
 further implies that
			\begin{equation*}
			\lambda \tilde{X}
			\begin{pmatrix}
			0\\
			\beta_\omega - \beta_{\omega_m}
			\end{pmatrix}
			= \tilde{X}
			\tilde{X}^\top_{\mathcal E_m} (\theta_{\mathcal E_m, \omega}
			- \theta_{\mathcal E_m, \omega_m})
			+ \{\tau -
			\mathbb I(i^\star \in \mathcal L_m)\}
			\tilde{X}\tilde{x}_{i^\star}   (\omega - \omega_m) \, .
			\end{equation*}
			Combining this with \eqref{eq:beta0_slope} and
			\eqref{eq:theta_slopes}, we obtain the
                        following result for residual $r_\omega$:
			\begin{eqnarray*}
			\lambda r_{\omega} - \lambda r_{\omega_m}
			&=&  \lambda \left(y - \tilde{X}
			\begin{pmatrix}
			\beta_{0, \omega} \\
			\beta_{\omega}
			\end{pmatrix} \right)
			- \lambda \left(y - \tilde{X}
			\begin{pmatrix}
			\beta_{0, \omega_m}\\
			\beta_{\omega_m}
			\end{pmatrix} \right) \nonumber\\
			& = &-\lambda \tilde{X}
			\begin{pmatrix}
			0 \\
			\beta_{\omega} - \beta_{\omega_m}
			\end{pmatrix} -
			(\lambda \beta_{0,\omega} - \lambda \beta_{0, \omega_m}) \mathbf{1} \nonumber  \\
			& = &
			-\tilde{X}
			\tilde{X}^\top_{\mathcal E_m} (\theta_{\mathcal E_m, \omega}
			- \theta_{\mathcal E_m, \omega_m})
			- \{\tau -
			\mathbb I(i^\star \in \mathcal L_m)\} \tilde{X}\tilde{x}_{i^\star}
			(\omega - \omega_m)
			- b_{0,m}(\omega - \omega_m) \mathbf{1}
			\nonumber\\
			& = &
			-\tilde{X}
			\tilde{X}^\top_{\mathcal E_m} b_m (\omega - \omega_m)
			- \{\tau -
			\mathbb I(i^\star \in \mathcal L_m)\}
			\tilde{X}\tilde{x}_{i^\star}   (\omega - \omega_m)
			- b_{0,m}(\omega - \omega_m) \mathbf{1}  \nonumber\\
			&=& h_{m}(\omega - \omega_m) \, ,
		\end{eqnarray*}
			where
			\begin{equation*}
			h_{m} = - b_{0,m}\mathbf{1} - \tilde{X}\left[
			\tilde{X}_{\mathcal{E}_m}^\top b_{m} +
			\{\tau - \mathbb I( i^\star \in \mathcal{L}_m )\}
			\tilde{x}_{i^\star}\right].
		\end{equation*}
This proves \eqref{eq:resid_evol} and \eqref{eq:resid_slopes}. 

			\textbf{II. }
			For Case II: $i^\star \in \mathcal E_m$, we will
			show that $i^\star \in \mathcal E_m$ can only happen
			when $m = 0$, and if that happens,
			$i^\star$ will move from $\mathcal E_0$ to $\mathcal L_1 \cup \mathcal R_1$ at the next breakpoint, and
			stay in $\mathcal L_m \cup \mathcal R_m$ for all $m = 1,\ldots, M$.
We show this by considering two scenarios.
			\begin{itemize}
				\item Scenario 1: if $i^\star \in \mathcal L_0 \cup \mathcal R_0$, then
				case $i^\star$ will stay in $\mathcal L_m \cup \mathcal R_m$ for $m = 1,\ldots, M$.
				\item Scenario 2: if $i^\star \in \mathcal E_0$, then $i^\star$ will
				move from $\mathcal E_0$ to $\mathcal
                                L_1\cup \mathcal R_1$ at the next breakpoint, and
				stay in $\mathcal L_m \cup \mathcal R_m$ for $m = 1,\ldots, M$.
			\end{itemize}

			For Scenario 1, we show that
			if $i^\star \in \mathcal L_m \cup \mathcal R_m$,
			then case $i^\star$
			will not move from $\mathcal L_m \cup \mathcal R_m$ to $\mathcal E_m$ at the next
			breakpoint. We prove this by showing
                        that the slope of the residual for case
                        $i^\star$ {over $(\omega_{m+1},\omega_m)$} is
			negative if $i^\star \in \mathcal R_m$; and
			positive if $i^\star \in \mathcal L_m$.
			Suppose that $i^\star \in \mathcal R_m$.
		In view of \eqref{eq:resid_evol} and \eqref{eq:resid_slopes},
			we need to show that
			$\frac{\partial r_{i^\star,\omega}}{\partial \omega} = h_{i^\star, m} < 0$, or equivalently,
			\begin{equation}\label{eq:invariant_slope1}
			h_{i^\star, m} =
			- b_{0,m} - \tilde{x}_{i^\star}^\top \left[
			\tilde{X}_{\mathcal{E}_m}^\top b_{m} +
			\tilde{x}_{i^\star}\tau\right] < 0 \, .
			\end{equation}
			By \eqref{eq:theta_slopes} and \eqref{eq:slope_beta0}, we can show that
			\begin{eqnarray*}
				h_{i^\star, m} &=&
				- \left( b_{0,m} + \tilde{x}_{i^\star}^\top \left[
				\tilde{X}_{\mathcal{E}_m}^\top b_{m} +
				\tilde{x}_{i^\star}\tau\right]\right) \\
				& = &
				- \left( b_{0,m} - \tilde{x}_{i^\star}^\top \left[
				\tilde{X}_{\mathcal{E}_m}^\top \left(
				(\tilde{X}_{\mathcal{E}_m} \tilde{X}_{\mathcal{E}_m}^\top)^{-1}
				\left[ b_{0,m} 1_{\mathcal E_m}
				+
				\tilde{X}_{\mathcal{E}_m}\tilde{x}_{i^\star}  \{\tau -  \mathbb I(i^\star \in \mathcal L_m)\}				\right]
				\right) +
				\tilde{x}_{i^\star}\tau\right]\right) \\
				&=& -\left(b_{0,m} - \tilde{x}_{i^\star}^\top \tilde{X}^\top_{\mathcal{E}_m} (\tilde{X}_{\mathcal E_m} \tilde{X}_{\mathcal E_m}^\top)^{-1} b_{0,m} {1}_{\mathcal E_m} -
				\tilde{x}_{i^\star}^\top \tilde{X}^\top_{\mathcal E_m} (\tilde{X}_{\mathcal E_m} \tilde{X}_{\mathcal E_m}^\top)^{-1} \tilde{X}_{\mathcal E_m} \tilde{x}_{i^\star} \tau + \tilde{x}_{i^\star}^\top  \tilde{x}_{i^\star} \tau\right)\\
				&=& -b_{0,m}
				\left(1 - \tilde{x}_{i^\star}^\top \tilde{X}^\top_{\mathcal{E}_m} (\tilde{X}_{\mathcal E_m} \tilde{X}_{\mathcal E_m}^\top)^{-1}
				{1}_{\mathcal E_m} \right) -
				\tau \tilde{x}_{i^\star}^\top  \big(I - \tilde{X}_{\mathcal E_m}^\top(\tilde{X}_{\mathcal E_m} \tilde{X}_{\mathcal E_m}^\top)^{-1}\tilde{X}_{\mathcal E_m} \big)\tilde{x}_{i^\star}  \\
				&=& -\tau \frac{(1 - 1^\top_{\mathcal E_m}
					(\tilde{X}_{\mathcal E_m} \tilde{X}_{\mathcal E_m}^\top)^{-1} \tilde{X}_{\mathcal E_m} \tilde{x}_{i^\star})^2}{1^\top_{\mathcal E_m}
					(\tilde{X}_{\mathcal E_m} \tilde{X}_{\mathcal E_m}^\top)^{-1}
					1_{\mathcal E_m}}
				-\tau \tilde{x}_{i^\star}^\top  \big(I - \tilde{X}_{\mathcal E_m}^\top(\tilde{X}_{\mathcal E_m} \tilde{X}_{\mathcal E_m}^\top)^{-1}\tilde{X}_{\mathcal E_m} \big)\tilde{x}_{i^\star}  <  0 \, ,
			\end{eqnarray*}
where the last inequality uses the fact that 
$\tilde{x}_{i^\star}^\top  \big(I - \tilde{X}_{\mathcal E_m}^\top(\tilde{X}_{\mathcal E_m}
      \tilde{X}_{\mathcal E_m}^\top)^{-1}\tilde{X}_{\mathcal E_m} \big)\tilde{x}_{i^\star} > 0$
      under the \textit{general position condition}, which can be shown 
      as follows. 
      Note that the rows of $\tilde{X}_{\mathcal E_m}$ and $\tilde{x}_i$
  are linearly independent since 
  $|\mathcal E_m \cup \{i\}| = |\mathcal E_{m-1}| \leq 
    \min(n, p+1) \leq \min(n, p+2)$. 
  Hence, 
  $\tilde{x}_i^\top \big(
  I - \tilde{X}^\top_{\mathcal{E}_{m}} \big(\tilde{X}_{\mathcal E_{m} }  
   \tilde{X}_{\mathcal E_{m} }^\top\big)^{-1}\tilde{X}_{\mathcal{E}_{m}} \big)
    \tilde{x}_i > 0$ since 
    $\tilde{X}^\top_{\mathcal{E}_{m}} \big(\tilde{X}_{\mathcal E_{m} } 
      \tilde{X}_{\mathcal E_{m} }^\top\big)^{-1} \tilde{X}_{\mathcal{E}_{m}}$ 
    is {the projection matrix for 
      the row space of  $\tilde{X}_{\mathcal E_m}$ and 
      $\tilde{X}^\top_{\mathcal{E}_{m}} \big(\tilde{X}_{\mathcal E_{m} } 
      \tilde{X}_{\mathcal E_{m} }^\top\big)^{-1} \tilde{X}_{\mathcal{E}_{m}} \tilde{x}_i 
      \neq \tilde{x}_i$. } 

			Similarly we can also show that
			when $i^\star \in \mathcal L_m$,
			\begin{eqnarray}\label{eq:invariance_slope2}
			h_{i^\star, m} & = &
			(1-\tau) \frac{(1 - 1^\top_{\mathcal E_m}
				(\tilde{X}_{\mathcal E_m} \tilde{X}_{\mathcal E_m}^\top)^{-1} \tilde{X}_{\mathcal E_m} \tilde{x}_{i^\star})^2}{1^\top_{\mathcal E_m}
				(\tilde{X}_{\mathcal E_m} \tilde{X}_{\mathcal E_m}^\top)^{-1}
				1_{\mathcal E_m}}
			+(1-\tau) \tilde{x}_{i^\star}^\top  \big(I - \tilde{X}_{\mathcal E_m}^\top(\tilde{X}_{\mathcal E_m} \tilde{X}_{\mathcal E_m}^\top)^{-1}\tilde{X}_{\mathcal E_m} \big)\tilde{x}_{i^\star} \nonumber \\
      & > & 0  \, .
			\end{eqnarray}
			This finishes the proof for Scenario 1. 

			For Scenario 2,
			note that when $i^\star \in \mathcal  E_0$,
			all the residuals and $\theta_{\omega}$ are constant
			as $\theta_{\omega}$ and
			$(\beta_{0,\omega}, \beta_{\omega})$ are independent of
			$\omega$ for $\omega \in [\omega_1, 1]$. As a result, the next breakpoint $\omega_1$
			can be determined by setting
			$\theta_{i^\star, \omega_0} = \omega \tau \text{ or } \omega(\tau - 1)$, that is,
			\begin{equation}
				\label{eq:w_1}
			\omega_1 =
			\begin{cases}
			\frac{{\theta}_{i^\star, \omega_0}}{\tau}
			&\text{ if } 0 < {\theta}_{i^\star, \omega_0} < \tau\\
			\frac{{\theta}_{i^\star, \omega_0}}{\tau - 1}
			&\text{ if } \tau - 1 < {\theta}_{i^\star, \omega_0} < 0\\
			0 &\text{ if } {\theta}_{i^\star, \omega_0} = 0
		\end{cases}, \, \text{ or } \,
		\omega_1 = \frac{{\theta}_{i^\star,\omega_0}}{\tau - \mathbb I({\theta}_{i^\star,\omega_0} < 0)} \, ,
			\end{equation}
			and $i^\star$ will
			move from $\mathcal E_0$ to $\mathcal L_1$ or $\mathcal R_1$ at $\omega_1$.
			After $\omega_1$, by the same argument used for Scenario 1, we can show that
			$i^\star$ will stay in $\mathcal L_m \cup \mathcal R_m$ for $m = 1,\ldots, M$. This proves
			Scenario 2.

			In summary, $i^\star \in \mathcal E_m$ can only happen when $m = 0$,
			and if that happens, $i^\star$ will
			move from $\mathcal E_0$ to $\mathcal L_1$ or $\mathcal R_1$ at the next breakpoint, and
			stay in $\mathcal L_m \cup \mathcal R_m$
			for $m = 1,\ldots, M$. 
      This completes the proof of Proposition \ref{prop:sol_path_evolution}.

			\subsection{Algorithm \ref{alg_w}}
			\begin{algorithm}[H]
				\setstretch{.8}
				\caption{The $\omega$ Path Algorithm for Case-weight Adjusted Quantile
				Regression}\label{alg_w}
								\textit{Input:} \quad $X \in \mathbb{R}^{n \times p}, y \in \mathbb{R}^n, \tau \in (0,1), {\lambda} \in \mathbb{R}^+, i^\star \in \{1,\cdots, n\}, \hat{\beta}_{0,\omega_0},  \hat{\beta}_{\omega_0}$\\
								Set $\mathcal L_0 = \{i: y_i - \hat{\beta}_{0,\omega_0} - x_i^\top \hat{\beta}_{\omega_0} < 0 \}$,
								$\mathcal E_0 = \{i: y_i - \hat{\beta}_{0,\omega_0} - x_i^\top \hat{\beta}_{\omega_0} = 0 \}$, $\mathcal R_0 = \{i: y_i - \hat{\beta}_{0,\omega_0} - x_i^\top \hat{\beta}_{\omega_0} > 0 \}$.\\
								Compute $\hat{\theta}_{\omega_0}$ by setting $\hat{\theta}_{\mathcal L_0, \omega_0} = (\tau - 1) \mathbf{1}$,
								$\hat{\theta}_{\mathcal R_0, \omega_0} = \tau \mathbf 1$, and
								$\hat \theta_{\mathcal{E}_0, \omega_0} =
								(\tilde{X}_{\mathcal{E}_0} \tilde{X}_{\mathcal{E}_0}^\top)^{-1}
								\big[\lambda y_{\mathcal{E}_0} -
								\lambda \hat \beta_{0,\omega_0} {1}_{\mathcal{E}_0}
								- \tilde{X}_{\mathcal{E}_0}
								\big( \tilde{X}^\top_{\mathcal L_0} \hat \theta_{\mathcal L_0, \omega_0} +
								\tilde{X}^\top_{\mathcal R_0} \hat \theta_{\mathcal R_0, \omega_0} \big) \big]$
								(see \eqref{eq:modified_KKT})\;
								Set $m = 0$\;
				
								\If{$i^\star \in \mathcal E_{0}$}{
									$\omega_{\text{1}} = \frac{\hat{\theta}_{i^\star,\omega_0}}{\tau - \mathbb I(\hat{\theta}_{i^\star,\omega_0} < 0)}$\;
									Set  $(\hat \beta_\omega, \hat \beta_{0,\omega}, \hat \theta_\omega) = (\hat \beta_{\omega_0}, \hat \beta_{0,\omega_0}, \hat \theta_{\omega_0})$ for any $\omega \in [\omega_1, 1]$\;
									\If{$\omega_1 > 0$}{
										Update the three sets:
										$(\mathcal L_1, \mathcal E_1, \mathcal R_1) = (\mathcal L_0, \mathcal E_0\setminus \{i^\star\}, \mathcal R_0 \cup\{i^\star\})$ if $\hat{\theta}_{i^\star,\omega_0} > 0$,\\
										otherwise  $(\mathcal L_1, \mathcal E_1, \mathcal R_1) = (\mathcal L_0 \cup\{i^\star\}, \mathcal E_0\setminus \{i^\star\}, \mathcal R_0)$\; }
									$m = m + 1$\;
								}
								\While{$\omega_{m} > 0$}{
									Compute the slopes $(b_{0, m}, b_{m},  h_{m})$ according to 
						  \eqref{eq:slope_beta0}, \eqref{eq:theta_slopes} and \eqref{eq:resid_slopes}, respectively\;
									Compute the next breakpoint $\omega_{m+1}$ and its two candidates $\omega_{1,m+1}$ and $\omega_{2,m+1}$ according to \eqref{eq:next_breakpoint}, \eqref{eq: w_next_bk1_2_1} and \eqref{eq: w_next_bk1_2_2}\;
									For each $\omega \in [\max(\omega_{m+1}, 0), \omega_m)$, \hspace{10cm}
									set
									$(\hat \theta_{\mathcal L_m  \setminus \{i^\star\}, \omega}, \hat \theta_{\mathcal R_m \setminus \{i^\star\}, \omega}, \hat \theta_{i^\star, \omega}) = \big((\tau - 1) {1}_{\mathcal L_m \setminus \{i^\star\}}, \tau {1}_{\mathcal R_m \setminus \{i^\star\}},  \omega(\tau - \mathbb I(i^\star \in \mathcal L_m)) \big)$,
									$(\lambda\hat \beta_{0, \omega}, \hat \theta_{\mathcal E_{m},\omega}) =
									(\lambda\hat \beta_{0, \omega_m}, \hat \theta_{\mathcal E_{m},\omega_m}) +
									(b_{0, m}, b_m)(\omega - \omega_m)$ and
									$( \lambda\hat \beta_\omega,
														\lambda\hat r_{\omega}) =
														(\lambda\hat \beta_{\omega_m},
														\lambda\hat r_{\omega_m}) + (
														X_{\mathcal{E}_{m}}^\top b_{m}
														 +(\tau - \mathbb I(i^\star
													 \in  \mathcal L_{m}))
									x_{i^\star},     h_m)(\omega
													  - \omega_{m})$\;
									Update the three sets:\\
									\eIf{$\omega_{m+1} = \omega_{1, m+1}$}{
										Let $i$ be the index that maximizes the objective function in \eqref{eq: w_next_bk1_2_1}. \hspace{10cm}
										Set $(\mathcal L_{m+1}, \mathcal E_{m+1}, \mathcal R_{m+1}) = (\mathcal L_m, \mathcal E_m\setminus \{i\}, \mathcal R_m \cup\{i\})$ if $\hat \theta_{i, \omega_{m+1}} = \tau$, \hspace{0.4cm} Rule (a)\\
										otherwise  $(\mathcal L_{m+1}, \mathcal E_{m+1}, \mathcal R_{m+1}) = (\mathcal L_m \cup\{i\}, \mathcal E_m\setminus \{i\}, \mathcal R_m)$; \hspace{1.8cm} Rule (b)
									}{
									Let $i$ be the index that maximizes the objective function in \eqref{eq: w_next_bk1_2_2}. \hspace{10cm}
									Set $(\mathcal L_{m+1}, \mathcal E_{m+1}, \mathcal R_{m+1}) = (\mathcal L_m\setminus \{i\}, \mathcal E_m\cup\{i\}, \mathcal R_m)$ if $i \in \mathcal L_m$,\\
									otherwise  $(\mathcal L_{m+1}, \mathcal E_{m+1}, \mathcal R_{m+1}) = (\mathcal L_m, \mathcal E_m \cup\{i\}, \mathcal R_m\setminus \{i\})$; \hspace{1.8cm} Rule (c)
								}
								$m = m + 1$\;
							}
						\end{algorithm}

\subsection{Proof of Theorem \ref{w_rigor} }
	We only need to show that the case-weight adjusted path
	$(\hat{\beta}_{0, \omega}, \hat{\beta}_{\omega}, \hat{\theta}_{\omega})$
	generated by Algorithm \ref{alg_w}
	satisfies all the KKT conditions in \eqref{w_kkt1}--\eqref{w_kkt5}.
	Our plan is to show that: (i) the initial full-data solution
	$(\hat{\beta}_{0, \omega_0}, \hat{\beta}_{\omega_0})$ together with
	$\hat{\theta}_{\omega_0}$ specified in Line 3
        satisfies the KKT conditions at $\omega = \omega_0$;
	(ii) if $i^\star \in \mathcal E_0$, then
	for each $\omega \in [\omega_1, 1]$,
	$(\hat{\beta}_{0, \omega}, \hat{\beta}_{\omega}, \hat{\theta}_{\omega})$
        in Line 7
	satisfies the KKT conditions; and
	(iii) after
        Line 13 in Algorithm \ref{alg_w},
	$i^\star \notin \mathcal E_m$ for each $m$, and
	$(\hat{\beta}_{0, \omega}, \hat{\beta}_{\omega},
        \hat{\theta}_{\omega})$  {in Line 17}
	satisfies the KKT conditions.

        {\bf (i)} Note that $(\hat{\beta}_{0, \omega_0},
	\hat{\beta}_{\omega_0})$ is the full-data solution.
	Thus, there must exist a vector $\theta \in \mathbb{R}^n$ such that
	$(\hat{\beta}_{0, \omega_0}, \hat{\beta}_{\omega_0}, \theta)$
	satisfies the KKT conditions of \eqref{w_kkt1}--\eqref{w_kkt5}.
	On the other hand, similar to the derivation of \eqref{eq:modified_KKT}, we
	can show that $\theta$ must be unique and equal to $\hat{\theta}_{\omega_0}$
	specified in Algorithm \ref{alg_w}
       {given $(\hat{\beta}_{0, \omega_0},
         \hat{\beta}_{\omega_0})$.}
       Hence, $(\hat{\beta}_{0, \omega_0}, \hat{\beta}_{\omega_0},
	\hat{\theta}_{\omega_0})$ satisfies the KKT conditions.

        {\bf (ii)} If the weighted case $i^\star \in \mathcal E_0$,
	then for each $\omega \in [\omega_1, 1]$,
	$(\hat{\beta}_{0, \omega}, \hat{\beta}_{\omega},
	\hat{\theta}_{\omega})= (\hat{\beta}_{0, \omega_0}, \hat{\beta}_{\omega_0},
	\hat{\theta}_{\omega_0})$, and thus it must also satisfy the KKT conditions \eqref{w_kkt1}--\eqref{w_kkt5},
	because the only condition in
	\eqref{w_kkt1}--\eqref{w_kkt5} that involves $\omega$ is
	$\theta_{i^\star, \omega_0} \in [\omega(\tau - 1), \omega \tau]$, which remains to be true when $\omega \geq \omega_1 = \frac{\hat{\theta}_{i^\star,\omega_0}}{\tau - \mathbb I(\hat{\theta}_{i^\star,\omega_0} < 0)}$.


{\bf (iii)}
We use induction on $m$ to show that
	$(\hat{\beta}_{0, \omega}, \hat{\beta}_{\omega}, \hat{\theta}_{\omega})$
	satisfies the KKT conditions for $\omega \in [\omega_{m+1}, \omega_{m}]$
	 and $i^\star \notin \mathcal E_m$,
after {Line 13} of Algorithm \ref{alg_w}.
First we show that $i^\star \notin \mathcal E_m$ after
{Line 13} of Algorithm \ref{alg_w}. Using similar arguments in
Part II of the proof of Proposition \ref{prop:sol_path_evolution},
we can show that if $i^\star \in \mathcal L_m \cup \mathcal R_m$,
then case $i^\star$ will not move from $\mathcal L_m \cup \mathcal R_m$ to
$\mathcal E_m$ at the next breakpoint.
This can be verified using \eqref{eq:invariant_slope1} and
\eqref{eq:invariance_slope2}, both of which are still valid here since
$b_{0,m}$ and $b_m$ in Line 15 are computed according to
\eqref{eq:theta_slopes} and \eqref{eq:slope_beta0}
in Proposition \ref{prop:sol_path_evolution}.
Hence, we must have $i^\star \in \mathcal L_m \cup \mathcal R_m$ after Line 13 of
Algorithm \ref{alg_w}.

Next, we show that
$(\hat{\beta}_{0, \omega}, \hat{\beta}_{\omega}, \hat{\theta}_{\omega})$ satisfies the
KKT conditions for $\omega \in [\omega_{m+1}, \omega_{m}]$ after
{Line 13} of Algorithm \ref{alg_w},
provided that $i^\star \notin \mathcal E_m$.
{In other words, we} show that
for each $\omega \in [\omega_{m+1}, \omega_m]$,
$(\hat{\beta}_{0, \omega}, \hat{\beta}_{\omega}, \hat{\theta}_{\omega})$
generated by Algorithm \ref{alg_w}
satisfies the KKT conditions provided that 
$(\hat{\beta}_{0, \omega_m}, \hat{\beta}_{\omega_m}, \hat{\theta}_{\omega_m})$
satisfies the KKT conditions at breakpoint $\omega_m$. 
Note that the KKT conditions consist of equality conditions
and inequality conditions. We verify them separately.

        \textbf{Equality conditions:}
	First, we verify that the following equality conditions
	are satisfied between breakpoints $\omega_m$ and $\omega_{m+1}$:
	\begin{equation}
	\label{eq:KKT_thm1}
	\begin{array}{ll}
	\hat{\theta}_{\mathcal L_m \setminus \{i^\star\}, \omega} = (\tau - 1){1}_{\mathcal L_m \setminus \{i^\star\}}, \quad
	\hat{\theta}_{\mathcal R_m \setminus \{i^\star\}, \omega} = \tau {1}_{\mathcal R_m \setminus \{i^\star\}}, \quad \hat \theta_{i^\star, \omega} = \omega (\tau - \mathbb I(i^\star \in \mathcal L_m)) \\
	\tilde{X}^\top\hat \theta_\omega  =
	\begin{pmatrix}
	0 \\
	\lambda \hat \beta_\omega
	\end{pmatrix}, \quad \tilde{X}_{\mathcal E_m}
	\begin{pmatrix}
	\hat{\beta}_{0,m}\\
	\hat{\beta}_m
	\end{pmatrix}
        =y_{\mathcal E_m}.
	\end{array}
	\end{equation}
	Since the above equality conditions are satisfied at $\omega = \omega_m$,
	it is sufficient to show that for each $\omega \in [\omega_{m+1}, \omega_m]$,
	\begin{subnumcases}{}\label{eq: eq_27a}
	\tilde{X}^\top(\hat \theta_\omega - \hat \theta_{\omega_m}) =
	\begin{pmatrix}
	0\\
	\lambda(\hat \beta_\omega - \hat \beta_{\omega_m})
	\end{pmatrix} \\
	\tilde{X}_{\mathcal{E}_m}
	\begin{pmatrix}
	\hat \beta_{0,\omega} - \hat \beta_{0, \omega_m}\\
	\hat \beta_{\omega} - \hat \beta_{\omega_m}
	\end{pmatrix} = \mathbf 0. \label{eq: eq_27b}
	\end{subnumcases}
	To prove \eqref{eq: eq_27a}, first we see that
	\begin{align*}
	\tilde X^\top (\hat \theta_\omega - \hat \theta_{\omega_m})
	&=  \tilde X_{\mathcal L_m}^\top (\hat \theta_{\mathcal L_m, \omega} -
	\hat \theta_{\mathcal L_m, \omega_m}) +
	\tilde X_{\mathcal R_m}^\top  (\hat \theta_{\mathcal R_m, \omega} - \hat \theta_{\mathcal R_m, \omega_m}) +
	\tilde X_{\mathcal E_m}^\top (\hat \theta_{\mathcal E_m, \omega} - \hat \theta_{\mathcal E_m, \omega_m})\\
	&= \{\tau - \mathbb I(i^\star \in \mathcal L_m)\}
	\tilde{x}_{i^\star} (\omega - \omega_m) +
	\tilde{X}^\top _{\mathcal E_m} b_m (\omega - \omega_m)
        \quad \quad \mbox{by \eqref{eq:theta_slopes}}\\
	&= \left(\{\tau - \mathbb I(i^\star \in \mathcal L_m)\}
	\tilde{x}_{i^\star}  +
	\tilde{X}^\top _{\mathcal E_m} b_m \right) (\omega - \omega_m)\\
	&= \begin{pmatrix}
	\left(\{\tau - \mathbb I(i^\star \in \mathcal L_m)\} +
	1_{\mathcal E_m}^\top b_m\right)(\omega - \omega_m) \\
	\left(\{\tau - \mathbb I(i^\star \in \mathcal L_m)\} x_{i^\star} + X_{\mathcal E_m}^\top b_m\right)(\omega - \omega_m)
	\end{pmatrix} \, .
	\end{align*}
	Moreover, by \eqref{eq:theta_slopes} and
	\eqref{eq:slope_beta0}, we can show that
	\begin{align*}
	-1_{\mathcal E_m}^\top b_m &=
	1_{\mathcal E_m}^\top (\tilde X_{\mathcal E_m} \tilde X_{\mathcal E_m}^\top)^{-1}\left[ b_{0,m} 1_{\mathcal E_m}
	+
	\tilde{X}_{\mathcal{E}_m}\tilde{x}_{i^\star}  \{\tau -  \mathbb I(i^\star \in \mathcal L_m)\}
	\right] \\
	&= b_{0,m} 1_{\mathcal E_m}^\top (\tilde X_{\mathcal E_m}
          \tilde X_{\mathcal E_m}^\top)^{-1}1_{\mathcal E_m}
       +1_{\mathcal E_m}^\top (\tilde X_{\mathcal E_m} \tilde
          X_{\mathcal E_m}^\top)^{-1}\tilde{X}_{\mathcal{E}_m}\tilde{x}_{i^\star}  \{\tau -  \mathbb I(i^\star \in \mathcal L_m)\}\\
	&=
	\left(1 - 1_{\mathcal{E}_m}^\top (\tilde{X}_{\mathcal{E}_m}
          \tilde{X}_{\mathcal{E}_m}^\top)^{-1}\tilde{X}_{\mathcal{E}_m}\tilde{x}_{i^\star}\right)\{\tau
          - \mathbb I(i^\star \in \mathcal L_m)\} +\
          1_{\mathcal{E}_m}^\top (\tilde{X}_{\mathcal{E}_m}
          \tilde{X}_{\mathcal{E}_m}^\top)^{-1}\tilde{X}_{\mathcal{E}_m}\tilde{x}_{i^\star})\{\tau
          - \mathbb I(i^\star \in \mathcal L_m)\} \\
       &=\tau -  \mathbb I(i^\star \in \mathcal L_m).
	\end{align*}
	Combining this with Line 17 of Algorithm \ref{alg_w}, we obtain \eqref{eq: eq_27a}.

	To prove \eqref{eq: eq_27b}, by
        \eqref{eq: eq_27a}, \eqref{eq:beta0_slope}, \eqref{eq:theta_evol}, and \eqref{eq:theta_slopes},
        it follows that 
	\begin{align*}
	\lambda \tilde{X}_{\mathcal{E}_m}
	\begin{pmatrix}
	\hat \beta_{0,\omega} - \hat \beta_{0, \omega_m}\\
	\hat \beta_{\omega} - \hat \beta_{\omega_m}
	\end{pmatrix}
	&= \tilde{X}_{\mathcal{E}_m}
	\begin{pmatrix}
	\lambda (\hat \beta_{0,\omega} - \hat \beta_{0, \omega_m})\\
	\mathbf{0}
	\end{pmatrix} + \tilde{X}_{\mathcal{E}_m}
	\begin{pmatrix}
	0\\
	\lambda (\hat \beta_{\omega} - \hat \beta_{\omega_m})
	\end{pmatrix} \\
	&=  \lambda (\hat \beta_{0,\omega} - \hat \beta_{0,
          \omega_m}){1}_{\mathcal E_m} + \tilde{X}_{\mathcal E_m}
          \tilde{X}^\top (\hat \theta_\omega - \hat \theta_{\omega_m})\\
	&= \left(b_{0,m} {1}_{\mathcal E_m} + \tilde{X}_{\mathcal E_m}\tilde{X}_{\mathcal E_m}^\top b_m +
	\tilde{X}_{\mathcal E_m}\tilde{x}_{i^\star}\{\tau - \mathbb I(i^\star \in \mathcal L_m)\}\right)(\omega - \omega_m) \\
	&= \biggl(b_{0,m} {1}_{\mathcal E_m} +
	\tilde{X}_{\mathcal E_m}\tilde{x}_{i^\star}\{\tau - \mathbb I(i^\star \in \mathcal L_m)\} -\\
	&\phantom{{}=1}
	\tilde{X}_{\mathcal E_m}\tilde{X}_{\mathcal E_m}^\top(\tilde{X}_{\mathcal E_m}\tilde{X}_{\mathcal E_m}^\top)^{-1}\left[ b_{0,m} {1}_{\mathcal E_m} +
	\tilde{X}_{\mathcal E_m}\tilde{x}_{i^\star}\{\tau - \mathbb I(i^\star \in \mathcal L_m)\} \right]\biggr) (\omega - \omega_m)\\
	&= 0.
	\end{align*}


	\textbf{Inequality conditions:}
	Next we verify the inequality conditions
	between breakpoints $\omega_m$ and $\omega_{m+1}$. For this,
	we consider two cases: $m = \mathbb I(i^\star \in \mathcal E_0)$ and
	$m \geq \mathbb I(i^\star \in \mathcal E_0) + 1$.
	In the first case, if $i^\star \notin \mathcal E_0$, then $m = 0$, and
	all inequality conditions are trivially satisfied for $\omega \in [\omega_1, 1]$.
	If $i^\star \in \mathcal E_0$, then $m = 1$.
	Now for $\omega \in [\omega_2, \omega_1]$, we need to verify that
	$r_{i^\star, \omega} > 0$ if $i^\star \in \mathcal R_1$,
	and $r_{i^\star, \omega} < 0$ if $i^\star \in \mathcal L_1$.
	In fact, by similar arguments used in the proof of Part II of Proposition
	\ref{prop:sol_path_evolution},
it can be shown that the residual of case $i^\star$ will increase
  if $i^\star \in \mathcal R_1$ and
	will decrease if $i^\star \in \mathcal L_1$.
	Moreover, since $i^\star \in \mathcal E_0$, we must have
	$r_{i^\star, \omega_1} = 0$.
	Combining, we have that
	that $r_{i^\star, \omega} > 0$ if $i^\star \in \mathcal R_1$
	and $r_{i^\star, \omega} < 0$ if $i^\star \in \mathcal L_1$.

In the second case of $m \geq \mathbb I(i^\star \in \mathcal E_0) + 1$,
	we have that $i^\star \notin \mathcal E_m$ and $i^\star \notin
        \mathcal E_{m-1}$ since $i^\star \notin \mathcal E_m$ after Line 13.
	In addition, as we have shown before, the  sign of the residual of case $i^\star$ does not change after {Line 13}.
         Under these conditions, we next show that all the inequality
         conditions are satisfied by verifying that
        the three rules in Algorithm \ref{alg_w} to update the elbow
        set and non-elbow sets at each breakpoint are consistent with
        the signs of resulting residuals.
	Specifically, we need to verify that,
	for rule (a),  if $\hat{\theta}_{i, \omega_{m}} = \tau$ for some $i \in \mathcal E_{m-1}$, then
	$r_{i,\omega} > 0$ for $\omega \in (\omega_{m+1}, \omega_{m})$; for rule (b),
	if  $\hat{\theta}_{i, \omega_{m}} = \tau - 1$ for some $i \in \mathcal E_{m-1}$,
	then $r_{i,\omega} < 0$ for $\omega \in (\omega_{m+1}, \omega_{m})$,
	and for rule (c),
	if $r_{i, \omega_{m}} = 0$ for some $i \in \mathcal L_{m-1} \cup \mathcal R_{m-1}$,
	then $\theta_{i,\omega} \in (\tau-1 , \tau)$ for
	$\omega \in (\omega_{m+1}, \omega_{m})$.

	For rule (a), if there exists $i \in \mathcal E_{m-1}$ such that
	$\hat \theta_{i,\omega_{m}} = \tau$ at $\omega_{m}$,
	the rule sets $\mathcal E_{m} = \mathcal{E}_{m-1}
	\setminus\{i\}$ and
	$\mathcal R_{m} = \mathcal{R}_{m-1} \cup\{i\}$.
	We need to show that $r_{i, \omega} > 0$ for $\omega \in (\omega_{m+1}, \omega_{m})$.
	Since $r_{i, \omega_{m}} = 0$, we need to show that $h_{i,m}$---the slope of
	$r_{i,\omega}$---is negative. In view of
	 \eqref{eq:resid_slopes} and the fact that
	$\mathcal E_{m} = \mathcal{E}_{m-1}  \setminus\{i\}$, we need to prove that
	\begin{equation} \label{w_new_slope}
	-h_{i,m} = b_{0, m} + \tilde{x}_i^\top [\tilde{X}^\top_{\mathcal{E}_{m-1} \setminus \{i\}} b_{m}
	+ \tilde{x}_{i^\star} (\tau - \mathbb I(i^\star \in \mathcal L_{m})] > 0.
	\end{equation}
	Since $i \in \mathcal{E}_{m-1}$, we have that
	$\lambda r_{i,\omega} \equiv 0$ for $\omega \in (\omega_{m}, \omega_{m-1}]$,
	which implies that its slope $h_{i,m-1}=0$. This, together with \eqref{eq:resid_slopes}, implies
	that
	\begin{eqnarray}
	-h_{i,m-1}&=& b_{0,m-1} + \tilde{x}_i^\top [\tilde{X}^\top_{\mathcal{E}_{m-1}} b_{m-1} +
	\tilde{x}_{i^\star} (\tau - \mathbb I(i^\star \in \mathcal L_{m-1}))]
	 \nonumber \\
	& =&
	b_{0,m-1} + \tilde{x}_i^\top [ \tilde{X}^\top_{\mathcal{E}_m} b_{\mathcal{E}_m, m-1} +  \tilde{x}_i b_{i,m-1}
	+
	\tilde{x}_{i^\star} (\tau - \mathbb I(i^\star \in \mathcal L_{m-1}))] = 0
	\label{w_old_slope} \, .
\end{eqnarray}
	Moreover, note that
	$\mathbb I(i^\star \in \mathcal L_{m-1}) = \mathbb I(i^\star \in \mathcal L_{m})$
	for $m \ge 1 + \mathbb I(i^\star \in \mathcal E_0)$ since
	$i^\star \notin \mathcal E_m$ and $i \neq i^\star$.
In view of \eqref{w_old_slope}, \eqref{w_new_slope} is equivalent to
	\begin{equation*}
	b_{0, m} + \tilde{x}_i^\top \tilde{X}^\top_{\mathcal{E}_{m-1}\setminus \{i\}} b_{m} >
	b_{0, m-1} + \tilde{x}_i^\top \tilde{X}^\top_{\mathcal{E}_{m-1}\setminus \{i\}} b_{\mathcal{E}_{m-1}\setminus \{i\}, m-1} + \tilde{x}_i^\top \tilde{x}_i b_{i,m-1} \, ,
	\end{equation*}
	or 
	\begin{equation}\label{w_iqn_goal}
	(b_{0, m} -b_{0, m-1}) + \tilde{x}_i^\top
	\tilde{X}^\top_{\mathcal{E}_{m}}
	(b_{m} - b_{\mathcal{E}_{m}, m-1}) - \tilde{x}_i^\top \tilde{x}_i b_{i,m-1} > 0.
	\end{equation}
	Next we plan to show that
	\begin{eqnarray}
		& &
		(b_{m} - b_{\mathcal{E}_{m}, m-1}) - \tilde{x}_i^\top \tilde{x}_i b_{i,m-1}
		\nonumber \\
		& = &
		b_{i, m-1}\left(
	 -\frac{\big({1_{\mathcal E_m} }^\top \big(\tilde{X}_{\mathcal E_m }   \tilde{X}_{\mathcal E_m }^\top\big)^{-1} \tilde{X}_{\mathcal{E}_{m}} \tilde{x}_i - 1\big)^2}{{1_{\mathcal E_m} }^\top \big(\tilde{X}_{\mathcal E_m }   \tilde{X}_{\mathcal E_m }^\top\big)^{-1} {1_{\mathcal E_m} }}   - \tilde{x}_i^\top \big(
	 I - \tilde{X}^\top_{\mathcal{E}_{m}} \big(\tilde{X}_{\mathcal E_{m} }   \tilde{X}_{\mathcal E_{m} }^\top\big)^{-1} \tilde{X}_{\mathcal{E}_{m}}
	 \big) \tilde{x}_i
	 \right). 
	 \label{eq:key_eq_thm1}
 \end{eqnarray}
	To that end,
	from the second equation in \eqref{eq:KKT_thm1}, we know that
	$1^\top \hat \theta_\omega = 0$ for $\omega \in (\omega_{m+1}, \omega_{m}]$.
	Hence, for any $\omega \in (\omega_{m+1}, \omega_{m})$,
	\begin{equation}
	\begin{array}{ll}
	\label{w_dyn1}
	1_{\mathcal{E}_{m}}^\top \hat \theta_{\mathcal{E}_{m},\omega} + 1_{\mathcal L_{m}}^\top \hat \theta_{\mathcal{L}_{m}, \omega} + 1_{\mathcal R_{m}}^\top \hat \theta_{\mathcal{R}_{m}, \omega} = 0\\
	1_{\mathcal{E}_{m}}^\top \hat \theta_{\mathcal{E}_{m}, \omega_{m}} + 1_{\mathcal L_{m}}^\top \hat \theta_{\mathcal{L}_{m},\omega_{m}} + 1_{\mathcal R_{m}}^\top \hat \theta_{\mathcal{R}_{m},\omega_{m}} = 0
	\end{array} \, .
\end{equation}
	Taking difference of the above two equations and using the updating formula for
	$\hat{\theta}_{\mathcal E_m, \omega}$, we obtain that
	\begin{equation}\label{w_dyn_sum}
	1_{\mathcal E_m}^\top   b_{m}(\omega - \omega_{m}) + (\tau - \mathbb{I}(i^\star \in \mathcal L_m)) (\omega - \omega_{m}) = 0.
	\end{equation}
	Dividing both sides by $\omega - \omega_{m}$, \eqref{w_dyn_sum} reduces to 
		\begin{equation*}
	 1_{\mathcal E_m}^\top   b_{m} + (\tau - \mathbb{I}(i^\star \in \mathcal L_m)) = 0.
		\end{equation*}
		Similarly, we also have that 
		\begin{equation*}
		1_{\mathcal E_{m-1} }^\top   b_{m-1} + 
    (\tau - \mathbb{I}(i^\star \in \mathcal L_{m-1} )) = 0 \, .
		\end{equation*}
	Taking the difference and using the fact that 
	$\mathbb I(i^\star \in \mathcal L_{m-1}) = \mathbb I(i^\star \in \mathcal L_{m})$ for $m \ge 1 + \mathbb I(i^\star \in \mathcal E_0)$,  we obtain that 
	\begin{equation}\label{w_iqn_key1}
	b_{i,m-1} = 1_{\mathcal{E}_{m}}^\top b_{m}
	- 1_{\mathcal{E}_{m}}^\top b_{\mathcal{E}_{m}, m-1} :=
	1_{\mathcal{E}_{m}}^\top \Delta_{\mathcal{E}_{m}},
	\end{equation}
	where $\Delta_{\mathcal{E}_{m}} = b_{m} - b_{\mathcal{E}_{m}, m-1}$.
	On the other hand, for any $j \in \mathcal{E}_{m}$,  $r_{j, m-1} = r_{j, m} = 0$ implies their slopes $h_{j, m-1} = h_{j, m} = 0$. Hence, we have that
	$h_{\mathcal E_{m-1} \setminus \{i\}, m-1} = 0$ and $h_{\mathcal E_{m-1} \setminus \{i\}, m} = 0$,
	which implies that
	\begin{equation*}
	\begin{array}{ll}
	b_{0, m-1} {1_{\mathcal{E}_{m}}} + \tilde{X}_{\mathcal{E}_{m}} [\tilde{X}^\top_{\mathcal{E}_{m}} b_{\mathcal{E}_{m}, m-1} + b_{i, m-1} \tilde{x}_i + \tilde{x}_{i^\star}(\tau - \mathbb I(i^\star \in \mathcal L_{m-1})) ] = 0 \, , \\
	b_{0, m} {1_{\mathcal{E}_{m}}} + \tilde{X}_{\mathcal{E}_{m}} [\tilde{X}^\top_{\mathcal{E}_{m}} b_{m} + \tilde{x}_{i^\star}(\tau - \mathbb I(i^\star \in \mathcal L_{m}))] = 0 \, .
\end{array}
\end{equation*}
	Again, because of $\mathbb I(i^\star \in \mathcal L_{m-1}) = \mathbb I(i^\star \in \mathcal L_{m})$ for $m \ge 1 + \mathbb I(i^\star \in \mathcal E_0)$, taking the difference, 
  we obtain that 
	\begin{equation}
	\Delta_0 {1_{\mathcal{E}_{m}}} + \tilde{X}_{\mathcal{E}_{m}} \tilde{X}^\top_{\mathcal{E}_{m}} \Delta_{\mathcal{E}_{m}} - \tilde{X}_{\mathcal{E}_{m}} \tilde{x}_i b_{i, m-1} = 0
	\, , \label{w_iqn_key2}
	\end{equation}
	where $\Delta_0 = b_{0, m} - b_{0, m-1}$.
	Combining \eqref{w_iqn_key1} and \eqref{w_iqn_key2} and solving for
	$\Delta_{\mathcal{E}_{m}}$ and $\Delta_0$,
	 we obtain that
	\begin{equation*}
	\begin{array}{ll}
	\Delta_{\mathcal{E}_{m}} = \big(\tilde{X}_{\mathcal E_m }   \tilde{X}_{\mathcal E_m }^\top\big)^{-1}\big(\tilde{X}_{\mathcal{E}_{m}} \tilde{x}_i b_{i, m-1} - \Delta_0 {1_{\mathcal{E}_{m}}} \big) \, , 
		\\
	\Delta_0 = \frac{
		\big( 1_{\mathcal E_m}^\top \big(\tilde{X}_{\mathcal E_m }   \tilde{X}_{\mathcal E_m }^\top\big)^{-1} \tilde{X}_{\mathcal{E}_{m}} \tilde{x}_i - 1\big) b_{i, m-1}
	}{{1}_{\mathcal E_m }^\top \big(\tilde{X}_{\mathcal E_m }   \tilde{X}_{\mathcal E_m }^\top\big)^{-1} {1}_{\mathcal E_m } } \, .
	\end{array}
\end{equation*}
	Substituting the above into the LHS of \eqref{w_iqn_goal}, we have that
	\begin{align*}
	\text{LHS of \eqref{w_iqn_goal}} &=
	(b_{0, m} - b_{0, m-1}) + \tilde{x}_i^\top \tilde{X}^\top_{\mathcal{E}_{m}}(b_{m} - b_{\mathcal{E}_{m}, m-1}) - \tilde{x}_i^\top \tilde{x}_i b_{i, m-1}
	= \Delta_0 + \tilde{x}_i^\top \tilde{X}^\top_{\mathcal{E}_{m}} \Delta_{\mathcal{E}_{m}} - \tilde{x}_i^\top \tilde{x}_i b_{i, m-1}\\
	&= \frac{\big(1_{\mathcal E_m}^\top \big(\tilde{X}_{\mathcal E_m }   \tilde{X}_{\mathcal E_m }^\top\big)^{-1} \tilde{X}_{\mathcal{E}_{m}} \tilde{x}_i - 1\big) b_{i, m-1}}{{1_{\mathcal E_m}
		}^\top \big(\tilde{X}_{\mathcal E_m }   \tilde{X}_{\mathcal E_m }^\top\big)^{-1} {
		1_{\mathcal E_m}
	}} + \tilde{x}_i^\top \tilde{X}^\top_{\mathcal{E}_{m}} \big(\big(\tilde{X}_{\mathcal E_m }
	  \tilde{X}_{\mathcal E_m }^\top\big)^{-1} \tilde{X}_{\mathcal{E}_{m}} \tilde{x}_i b_{i, m-1}  - \tilde{x}_i^\top \tilde{x}_i b_{i, m-1}\\
	& \quad -
	\frac{\big(\tilde{X}_{\mathcal E_m }   \tilde{X}_{\mathcal E_m }^\top\big)^{-1} {1_{\mathcal E_m}
		} \big({1_{\mathcal E_m} }^\top \big(\tilde{X}_{\mathcal E_m }   \tilde{X}_{\mathcal E_m }^\top\big)^{-1} \tilde{X}_{\mathcal{E}_{m}} \tilde{x}_i - 1\big)b_{i, m-1}}{{1_{\mathcal E_m} }^\top \big(\tilde{X}_{\mathcal E_m }   \tilde{X}_{\mathcal E_m }^\top\big)^{-1} {1_{\mathcal E_m} }}
	\big) \\
	&= b_{i, m-1}\left(
	-\frac{\big({1_{\mathcal E_m} }^\top \big(\tilde{X}_{\mathcal E_m }   \tilde{X}_{\mathcal E_m }^\top\big)^{-1} \tilde{X}_{\mathcal{E}_{m}} \tilde{x}_i - 1\big)^2}{{1_{\mathcal E_m} }^\top \big(\tilde{X}_{\mathcal E_m }   \tilde{X}_{\mathcal E_m }^\top\big)^{-1} {1_{\mathcal E_m} }}   - \tilde{x}_i^\top \big(
	I - \tilde{X}^\top_{\mathcal{E}_{m}} \big(\tilde{X}_{\mathcal E_{m} }   \tilde{X}_{\mathcal E_{m} }^\top\big)^{-1} \tilde{X}_{\mathcal{E}_{m}}
	\big) \tilde{x}_i
	\right). 
	\end{align*}
This proves \eqref{eq:key_eq_thm1}.

   Moreover, under the 
  \textit{general position condition}, 
  we must have the rows of $\tilde{X}_{\mathcal E_m}$ and $\tilde{x}_i$
  are linearly independent since 
  $|\mathcal E_m \cup \{i\}| = |\mathcal E_{m-1}| \leq 
    \min(n, p+1) \leq \min(n, p+2)$. 
  Hence, 
	$\tilde{x}_i^\top \big(
	I - \tilde{X}^\top_{\mathcal{E}_{m}} \big(\tilde{X}_{\mathcal E_{m} }  
   \tilde{X}_{\mathcal E_{m} }^\top\big)^{-1}\tilde{X}_{\mathcal{E}_{m}} \big)
    \tilde{x}_i > 0$ since 
    $\tilde{X}^\top_{\mathcal{E}_{m}} \big(\tilde{X}_{\mathcal E_{m} } 
      \tilde{X}_{\mathcal E_{m} }^\top\big)^{-1} \tilde{X}_{\mathcal{E}_{m}}$ 
    is {the projection matrix for 
      the row space of  $\tilde{X}_{\mathcal E_m}$ and 
      $\tilde{X}^\top_{\mathcal{E}_{m}} \big(\tilde{X}_{\mathcal E_{m} } 
      \tilde{X}_{\mathcal E_{m} }^\top\big)^{-1} \tilde{X}_{\mathcal{E}_{m}} \tilde{x}_i 
      \neq \tilde{x}_i$. } 
	Thus, in order to show that the LHS of \eqref{w_iqn_goal} is positive, 
	it remains to show that $b_{i, m-1} < 0$.
	Based on the facts that $\hat{\theta}_{i, \omega}$ is a linear 
  function of $\omega$:
	$\hat{\theta}_{i, \omega} =
	\hat{\theta}_{i, \omega_{m-1}} + b_{i, m-1} (\omega - \omega_{m-1})$
	for $\omega \in [\omega_{m}, \omega_{m-1}]$
	and $\hat \theta_{i,\omega} = \tau$ at $\omega_{m}$, we must have 
	its slope $\frac{\partial \hat \theta_{i,\omega}}{\partial \omega} < 0$ for
	$\omega \in (\omega_{m}, \omega_{m-1})$. This implies that
	\begin{equation*}
	\frac{\partial \hat \theta_{i,\omega}}{\partial \omega} = b_{i, m-1} < 0 \, ,
	\end{equation*}
	which completes the proof for rule (a).


	For rule (b), similar arguments can be applied to prove that if there exists some case $i \in \mathcal{E}_{m-1}$ such that $\hat \theta_{i, \omega_{m}}  = \tau - 1 \text{ and }
	r_{i,\omega_{m} } = 0 $ at $\omega_{m}$, then $r_{i,\omega} < 0$ for any $\omega \in (\omega_{m+1}, \omega_{m})$.

	For rule (c), without loss of generality,
	we assume that  $r_{i, \omega_{m}} = 0$ for some case $i \in \mathcal{L}_{m-1}$  and $\hat \theta_{i, \omega_{m}} = \tau - 1$ at $\omega_{m}$. Then the rule 
  updates the three sets as $\mathcal{E}_{m} = \mathcal E_{m-1} \cup \{i\}$ 
  and $\mathcal L_m = \mathcal L_{m-1} \setminus \{i\}$.
	As $\omega$ starts to decrease from $\omega_{m}$, $\hat \theta_{i,\omega}$ will 
  increase from $\tau - 1$, which implies that its slope $b_{i, m} < 0$.

	In the proof of rule (a),  we have shown that
	$h_{i, m} < 0$ given $b_{i, m-1} < 0$ and $h_{i, m-1} = 0$. Here we need to prove $b_{i, m} < 0$ given $h_{i, m-1} < 0$ and $h_{i, m} = 0$.
	Reversing the arguments used for rule (a),
	we need to show that $b_{i, m} < 0$
	given \eqref{w_iqn_goal}. Similar to the proof of rule (a),
	we can show that the LHS of \eqref{w_iqn_goal} is
	\begin{equation*}
	b_{i, m}\Big(
	-\frac{\big({1_{\mathcal E_{m-1} } }^\top
	\big(\tilde{X}_{\mathcal E_{m-1}}
	 \tilde{X}_{\mathcal E_{m-1}}^\top\big)^{-1}
	 \tilde{X}_{\mathcal{E}_{m-1}}
	 \tilde{x}_i - 1\big)^2}{{1_{\mathcal E_{m-1}} }^\top
	 \big(\tilde{X}_{\mathcal E_{m-1}}
	   \tilde{X}_{\mathcal E_{m-1}}^\top\big)^{-1}
		 {1_{\mathcal E_{m-1}} }} - \tilde{x}_i^\top \big(
	I - \tilde{X}^\top_{\mathcal{E}_{m-1}}
	\big(\tilde{X}_{\mathcal E_{m-1}}
	\tilde{X}_{\mathcal E_{m-1}}^\top\big)^{-1}
	\tilde{X}_{\mathcal{E}_{m-1}} \big) \tilde{x}_i \Big)	> 0,
\end{equation*}
	which implies that $b_{i, m} < 0$. This completes the proof.

\subsection{Proof of Proposition \ref{ridgereg_sCD}} \label{proof: ridgereg_sCD}
	The normal equation for problem \eqref{lsr} is
	\begin{equation}\label{eq1}
	-\sum_{i \ne i^\star }^{n} \tilde{x}_i(y_i - \tilde{x}_i^\top \tilde{\beta}_\omega) - \omega \tilde{x}_{i^\star }(y_{i^\star } - \tilde{x}_{i^\star }^\top \tilde{\beta}_\omega) + \lambda \tilde{I} \tilde{\beta}_\omega = 0.
	\end{equation}
When $\omega = \omega_0 = 1$, in particular, the normal equation is
	\begin{equation}\label{eq2}
	-\sum_{i \ne i^\star }^{n} \tilde{x}_i(y_i - \tilde{x}_i^\top \tilde{\beta}_{\omega_0}) - \omega \tilde{x}_{i^\star }(y_{i^\star } - \tilde{x}_{i^\star }^\top \tilde{\beta}_{\omega_0}) - (1 - \omega) \tilde{x}_{i^\star }(y_{i^\star } - \tilde{x}_{i^\star }^\top \tilde{\beta}_{\omega_0}) + \lambda \tilde{I} \tilde{\beta}_{\omega_0} = 0.
	\end{equation}
	Subtracting \eqref{eq1} from \eqref{eq2}, we have
	\begin{equation*}
	\bigg[\sum_{i \ne i^\star }^{n} \tilde{x}_i \tilde{x}_i^\top + \omega \tilde{x}_{i^\star } \tilde{x}_{i^\star }^\top
	+ \lambda \tilde{I}\bigg](\tilde{\beta}_{\omega_0}-\tilde{\beta}_{\omega} )
	= (1 - \omega) \tilde{x}_{i^\star }( y_{i^\star }-\tilde{x}_{i^\star }^\top \tilde{\beta}_{\omega_0}) =(1 - \omega) \tilde{x}_{i^\star }r_{i^\star },
	\end{equation*}
	which leads to
      \[ 	\tilde{\beta}_{\omega_0} - \tilde{\beta}_\omega
             = \big[\tilde{X}^\top \tilde{X} + \lambda \tilde{I} - (1 - \omega)\tilde{x}_{i^\star } \tilde{x}_{i^\star }^\top\big]^{-1} (1 - \omega)\tilde{x}_{i^\star } r_{i^\star }. \]
     Letting $\tilde{D}_\lambda=\tilde{X}^\top \tilde{X} + \lambda
     \tilde{I}$, we have
	\begin{eqnarray*}
	\tilde{\beta}_{\omega_0} - \tilde{\beta}_\omega
	&=& \bigg[	\tilde{D}_\lambda^{-1} + \frac{
            \tilde{D}_\lambda^{-1} \tilde{x}_{i^\star
            }\tilde{x}_{i^\star }^\top
            \tilde{D}_\lambda^{-1}}{{1}/{(1 - \omega)} -
            \tilde{x}_{i^\star }^\top
            \tilde{D}_\lambda^{-1}\tilde{x}_{i^\star }}\bigg](1 -
            \omega)\tilde{x}_{i^\star }  r_{i^\star }\\
	&=& (1 - \omega)\tilde{D}_\lambda^{-1}\bigg[\tilde{x}_{i^\star } +
	\frac{	 \tilde{x}_{i^\star }\tilde{x}_{i^\star }^\top \tilde{D}_\lambda^{-1}\tilde{x}_{i^\star }}{{1}/{(1 - \omega)} - \tilde{x}_{i^\star }^\top \tilde{D}_\lambda^{-1}\tilde{x}_{i^\star }}
	\bigg]r_{i^\star }\\
	&=&
\frac{\tilde{D}_\lambda^{-1} \tilde{x}_{i^\star }r_{i^\star }}{{1}/{(1 - \omega)} - \tilde{x}_{i^\star }^\top \tilde{D}_\lambda^{-1}\tilde{x}_{i^\star }},
	\end{eqnarray*}
      which implies that
      \begin{equation}
        \label{eq:diff_fit}
     \hat{f}(x_j) - \hat{f}_\omega^{i^\star }(x_j)=\tilde{x}_j^\top(\tilde{\beta}_{\omega_0}-\tilde{\beta}_\omega)
     = \frac{\big(\tilde{x}_j^\top\tilde{D}_\lambda^{-1} \tilde{x}_{i^\star }\big)r_{i^\star }}{{1}/{(1 - \omega)} - \tilde{x}_{i^\star }^\top \tilde{D}_\lambda^{-1}\tilde{x}_{i^\star }}
    = \frac{h_{j i^\star }(\lambda) r_{i^\star }}{{1}/{(1 - \omega)}
      - h_{i^\star i^\star }(\lambda)}.
      \end{equation}
	Hence,
\[\sum_{j=1}^{n} \big(\hat{f}(x_j) - \hat{f}_\omega^{i^\star  }(x_j) \big)^2
= \frac{r_{i^\star }^2\sum_{j=1}^{n} h_{j i^\star }^2(\lambda) }{\{{1}/{(1 - \omega)}
      - h_{i^\star i^\star }(\lambda)\}^2},\]
which completes the proof.



\begin{table}[h]
	\caption{Approximation Error of $\rho_\tau(r^{[-i]}_i) - \rho_\tau(r_i) $ }\label{tab:approx_error}
	\begin{center}
		\begin{tabular}{c c c | c |  c | c}
			\hline
			\hline
			\multicolumn{3}{c|}{Scenario} & True Difference & Approximation & Approximation Error\\
			\hline
			&$r^{[-i]}_i$ & $r_i$ & $\rho_\tau(r^{[-i]}_i) - \rho_\tau(r_i) $   & $\rho_{\tau, \delta}^{\prime}(r_i)  (r^{[-i]}_i - r_i)$ & $\Delta_{\text{approx.}}$\\[.5em]
			\hline
			(a) & $(0,\infty)$ & $(-\infty, -\delta]$	& \multirow{2}{*}{$(\tau) r^{[-i]}_i - (\tau - 1) r_i$}
			&
			$ (\tau - 1) (r^{[-i]}_i - r_i)$ & $-|r^{[-i]}_i|$ \\
			\hhline{---~--}
			(b) & $(0,\infty)$ & $( -\delta, 0]$	&
			&
			$ 2(1 - \tau)\frac{r_i}{\delta} (r^{[-i]}_i - r_i)$ & $\approx -\tau|r^{[-i]}_i|$\\
			\hline
			(c) & $(-\infty, 0)$ & $[\delta, \infty)$ & \multirow{2}{*}{$(\tau - 1) r^{[-i]}_i  - (\tau) r_i$}
			&
			$\tau (r^{[-i]}_i - r_i)$ & $-|r^{[-i]}_i|$ \\
			\hhline{---~--}
			(d) & $(-\infty, 0)$ & $[0, \delta)$ &
			&
			$2\tau \frac{r_i}{\delta}(r^{[-i]}_i - r_i)$& $\approx -(1 - \tau)|r^{[-i]}_i|$\\
			\hline
			\hline
		\end{tabular}
	\end{center}
\end{table}

\begin{table}[h]
	\caption{Average number of $\omega$-breakpoints for various
          data dimensions and quantiles.
          {The results are averaged over 20 independent
		simulations with $N_\lambda = 50$.  The 50 grid points for
		$\lambda \in [0.01, 100]$  are equally spaced on the
                logarithmic scale. The values in the parentheses
		are the corresponding standard errors.} }\label{tab: w_steps}
	\begin{center}
		\begin{tabular}{c| cc c |cc  c}
			\hline
			\hline
			$\tau$ & $n$ & $p$ & Average number of $\omega$ breakpoints & $n$ & $p$ & Average number of $\omega$ breakpoints\\
			\hline
			\multirow{2}{*}{0.1} & 100 & 50 & 4.409 (0.556) &
			50 & 300 & 0.714 (0.109)\\
			& 300 & 50 &  4.417 (0.768)	& 150 & 300 & 1.044 (0.192)\\
			\hline
			\multirow{2}{*}{0.3} & 100 & 50 & 7.177 (0.686) & 50 & 300 & 0.714 (0.109)\\
			& 300 & 50 &  7.724 (1.074) & 150 & 300 & 1.360 (0.202)\\
			\hline
			\multirow{2}{*}{0.5} & 100 & 50 & 7.427 (0.695) &
			50 & 300 & 1.098 (0.119)\\
			& 300 & 50 & 8.780 (1.119) & 150 & 300 & 1.635 (0.238) \\
			\hline
			\hline
		\end{tabular}
	\end{center}
\end{table}

\begin{table}[ht]
	\caption{The elapsed runtime per case (i.e., the total
    runtime/$n$) measured in seconds for LOO CV simulation ($p = 50$
    and $n > p$). The value in the parentheses is the standard deviation over 20 replicates.} 	\label{table:LOOCV_p=50}
    	\begin{center}
	\begin{tabular}{lcc|cccc}
		\hline \hline
		$n$ & $p$ & $\tau$ & $\lambda$-path ($N_{\lambda}=50$)  &$w$-path ($N_{\lambda}=20$)  & $w$-path ($N_{\lambda}=50$)\\
		\hline
		100 &  50 &  0.1 & 0.0025 (0.0003)  & 0.0027 (0.0005) & 0.0065 (0.0010) \\
		100 &  50 &  0.3 &  0.0040 (0.0006)  & 0.0045 (0.0005)   &  0.0110 (0.0013) \\
		100 &  50 &  0.5 & 0.0047 (0.0005)   &  0.0051 (0.0008)  &  0.0127 (0.0020) \\
		\hline
		
		200 &  50 &  0.1 & 0.0091 (0.0013)  & 0.0036 (0.0006)   & 0.0089 (0.0014)\\
		200 &  50 &  0.3 & 0.0238 (0.0034) & 0.0089 (0.0014) &  0.0222 (0.0035)\\
		200 &  50 &  0.5 &  0.0309 (0.0053)&  0.0101 (0.0013)& 0.0251 (0.0032)\\
		\hline
		
		300 &  50 &  0.1 & 0.0216 (0.0030) &  0.0050 (0.0008) & 0.0124 (0.0019) \\
		300 &  50 &  0.3 & 0.0699 (0.0061) &  0.0117 (0.0014)& 0.0294 (0.0033)\\
		300 &  50 &  0.5 &  0.0904 (0.0124)&  0.0138 (0.0019)& 0.0344 (0.0049)\\
		\hline \hline
		
	\end{tabular}
        	\end{center}
\end{table}

\begin{table}[ht]
	\caption{The runtime per case for LOO CV simulation ($p = 300$
          and $n < p$). The value in the parentheses is the standard deviation over 20 replicates.} \label{table:LOOCV_p=300}
      \begin{center}
	\begin{tabular}{lcc|cccc}
		\hline \hline
		$n$ & $p$ & $\tau$ & $\lambda$-path ($N_{\lambda}=50$)  &$w$-path ($N_{\lambda}=20$)  & $w$-path ($N_{\lambda}=50$)\\
		\hline
		50 &  300 &  0.1 &0.0018 (2e-5)  &0.0015 (3e-5)  &0.0036 (9e-5)  \\
		50 &  300 &  0.3 &0.0018 (5e-5)  &0.0016 (3e-5)  &0.0038 (6e-5) \\
		50 &  300 &  0.5 & 0.0018 (2e-5) & 0.0016 (4e-5) &0.0039 (6e-5) \\
		\hline
		
		100 & 300 &  0.1 &0.0098 (7e-5)  &0.0039 (0.00014)  & 0.0097 (0.00018)\\
		100 & 300 &  0.3 & 0.0098 (8e-5) &0.0043 (0.00011)  & 0.0107 (0.00020) \\
		100 & 300 &  0.5 &0.0099 (1e-4)  & 0.0045 (0.00017) & 0.0111 (0.00022)\\
		\hline
		
		150 & 300 &  0.1 &0.0254 (0.00020)  & 0.0074 (0.00031) & 0.0182 (0.00050)\\
		150 & 300 &  0.3 &0.0260 (0.00040)  & 0.0081 (0.00033)& 0.0203 (0.00049)\\
		150 & 300 &  0.5 &0.0260 (0.00035) & 0.0084 (0.00031)  & 0.0208 (0.00047)\\
		\hline \hline
		
	\end{tabular}
        	\end{center}
\end{table}

\subsection{Data Analysis}

\subsubsection{King County House Sales Data}
\label{sub_sec:house_data_features}
\cite{luan2022measuringmodelcomplexityheteroscedastic} considered the King County House Sales Data for hetero-scedastic linear regression. Based on the exploratory data analysis results, 12 features were selected, and appropriate transformations were applied to the selected features. In this paper, we utilize the same features for an $l_2$-penalized quantile regression. Table 
\ref{tab:rda_house_var_list} lists all 12 variables, and details of the data transformations can be found in Appendix B.2 of \cite{luan2022measuringmodelcomplexityheteroscedastic}.

\begin{table}[h]
	\centering
	\caption{List of independent variables used in King County house sales data analysis}
	\label{tab:rda_house_var_list}
	\fontsize{9}{10.8}\selectfont
    \bigskip
	\begin{tabularx}{\textwidth}{lcXc}
		\hline
		Variable & Type & Description & Transformation\\
		\hline
		\texttt{basement} & categorical & Whether the house has basement or not: yes/no  & - \\ 
		
		\texttt{bathrooms} & numeric & Number of bathrooms in the house & - \\ 
		
		\texttt{bedrooms} & categorical & Number of bedrooms in the house: 1-2/3-5/6 or more  & - \\
		
		\texttt{condition} & categorical & Condition of the house: poor/average/good &  \\
		
		\texttt{grade} & numeric & Overall grade of the house regarding the types of materials used and the quality of workmanship & - \\
		
		\texttt{grade2} & numeric & Square of \texttt{grade} & - \\
		
		\texttt{sqft\_living} & numeric & Interior living space of the houses in square feet & square root \\
		
		\texttt{sqft\_living15} & numeric & Average interior living space for the closest 15 houses in square feet &  square root\\
		
		\texttt{view} & numeric & Grade of the view around the house & -\\
		
		\texttt{waterfront} & categorical & Whether the house has a waterfront or not: yes/no & -\\
		
		\texttt{yr\_built} & categorical & Year when the house was built: 1920 or before/1921-1940/1941-1960/1961-1980/1981-2000/2001 or after & -\\
		
		\texttt{zone} & categorical & Location of the house: zone 1/zone 2/zone 3 & -\\
		\hline
	\end{tabularx}
\end{table}

\subsubsection{Exploratory Data Analysis (EDA)}
\label{sub_sec: eda_sample_data}
\begin{figure}
    \centering
    \includegraphics[width=0.9\textwidth]{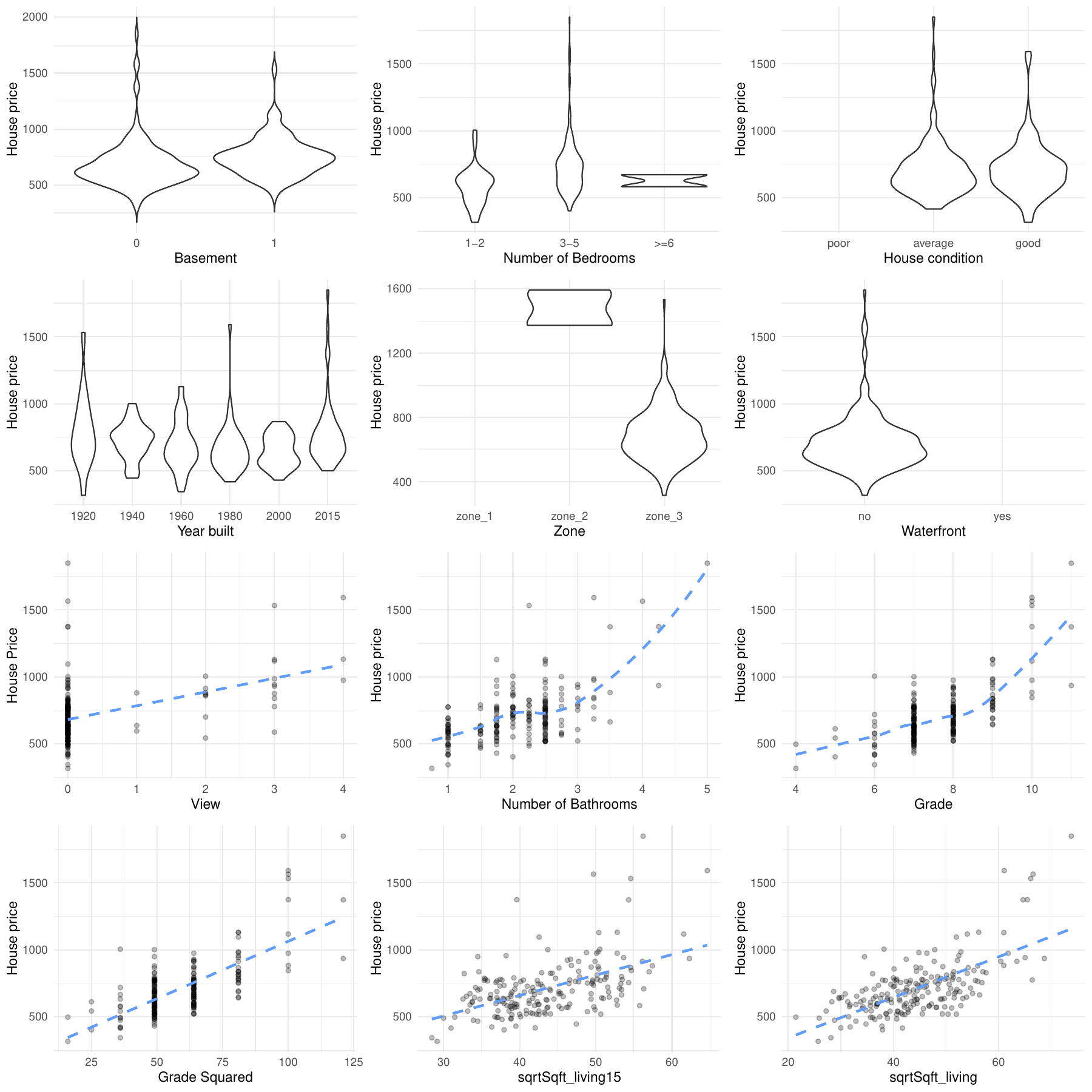}
    \caption{EDA plots for a random sample of 200 houses with the view of the house treated as a numeric variable.}
    \label{fig:eda_sample_v2}
\end{figure}

We present EDA results for the random sample of 200 houses. 
Based on the plots in Figure \ref{fig:eda_sample_v2}, we dropped some features with insufficient variability to prevent unreliable estimates. For example, the waterfront feature was excluded because only 1 out of 200 houses overlooked the waterfront. Similarly, the zone feature was dropped due to low variability. Since there are no houses in poor condition in our sample, we used average condition as the baseline. Since there are only 2 houses with more than 6 bedrooms, we decided to combine this category with the 3-5 bedrooms category. For the view of the house, since most houses have a score of 0, we decided to combine the non-zero scores into one category and treat it as a categorical variable.

\subsubsection{Proof of the Piece-wise Constant Pattern in Case-Influence Graph}
\label{sub_sec:piece_wise}
\begin{pro}
    For linear quantile regression, if the elbow set at $\omega_m$ is of full rank, i.e., $|\mathcal{E}_{m}|=p+1$, the solution path $\hat{f}_{\omega}(x)$ is constant for $\omega \in (\omega_{m+1}, \omega_m]$.
\end{pro}

\begin{proof}
From Equation \eqref{eq_qr_case_sensitivity} for $m=0$, we can conclude that when the initial elbow set is of full rank, i.e. $|\mathcal{E}_{0}|=p+1$, the solution path between two consecutive breakpoints $\omega_{1}$ and $\omega_{0}=1$ is constant. 
When $|\mathcal{E}_{0}|=p+1$, the projection matrix onto the space spanned by the elbow set points $P_{\tilde{X}_{\mathcal{E}_0}}$ is $I$, which makes the second part of the following term zero:
\begin{equation*}
    \left[\frac{\left(q_0^\top \tilde{x}_i-1\right) \left(q_0^\top \tilde{x}_{i^\star}-1\right)}{q_0^\top q_0}+\tilde{x}_i^{\top}\left(I-P_{\tilde{X}_{\mathcal{E}_0}}\right) \tilde{x}_{i^*}\right].
\end{equation*}
In addition, since $\tilde{x}_{i^\star}$ is in the row space of $\tilde{X}_{\mathcal{E}_0}$, there exists $a \in \mathbb{R}^{p+1}$ such that $\tilde{x}_{i^\star}=\tilde{X}_{\mathcal{E}_0}^\top a$, which also implies $1_{p+1}^\top a =1$. In this case, the first part of the term also becomes zero because
\[
\left(q_0^\top \tilde{x}_{i^\star}-1\right) = 1_{\mathcal{E}_0}^{\top}\left(\tilde{X}_{\mathcal{E}_0} \tilde{X}_{\mathcal{E}_0}^{\top}\right)^{-1} \tilde{X}_{\mathcal{E}_0} (\tilde{X}_{\mathcal{E}_0}^\top a) - 1 = 1_{p+1}^\top a -1 = 0.
\]
This conclusion for the solution path at $\omega_0=1$ holds true generally for any breakpoints $\omega_m$, as long as the elbow set at $\omega_m$ is of full rank since
the rate of change of the case-weight adjusted solution at $\omega \in (\omega_{m+1},\omega_m]$ is 
\begin{equation}
\frac{\partial \hat{f}_{\omega}\left(x_i\right)}{\partial \omega} =\frac{1}{\lambda} \left[\frac{\left(q_m^\top \tilde{x}_i-1\right) \left(q_m^\top \tilde{x}_{i^\star}-1\right)}{q_m^\top q_m}+\tilde{x}_i^{\top}\left(I-P_{\tilde{X}_{\mathcal{E}_m}}\right) \tilde{x}_{i^*}\right]\cdot \left(\tau-\mathbb{I}\left(i^\star\in \mathcal{L}_m\right)\right),
\end{equation}
where $q_m = \tilde{X}_{\mathcal{E}_m}^\top \left(\tilde{X}_{\mathcal{E}_m} \tilde{X}_{\mathcal{E}_m}^{\top}\right)^{-1} 1_{\mathcal{E}_m} \in \mathbb{R}^{p+1}$.
\end{proof}

\begin{figure}[h!]
    \centering
    \includegraphics[width=0.8\textwidth]{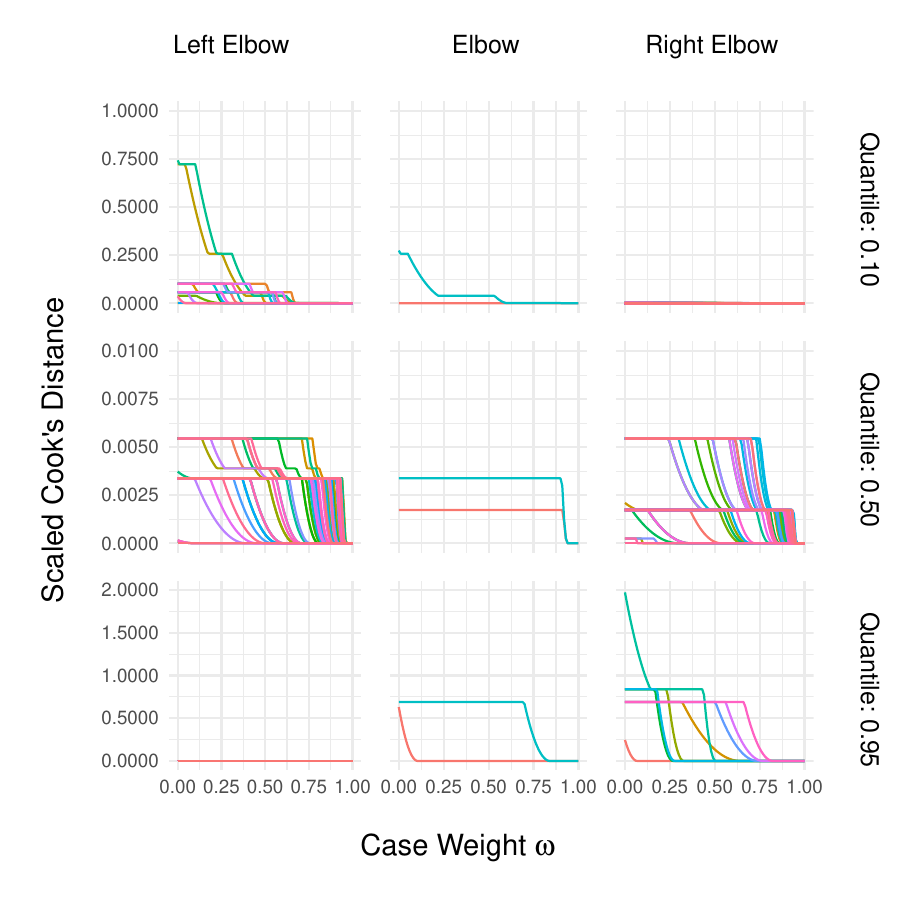} 
    \caption{Case-influence graphs from the single-predictor quantile regression for the King County house sales data with various quantiles: $\tau=0.1$ (top), 0.5 (middle) and 0.95 (bottom). The left, middle and right panels correspond to the left elbow, elbow and right elbow set points, respectively.}
    \label{fig:qr_single_case_influence_graph}
\end{figure}

\clearpage
\subsubsection{LOO Solutions for Quantile Regression with Multiple Predictors}
\label{sub_sec:qr_housing_multi_loo}
We present a comprehensive comparison of the full-data solution and the LOO solution excluding the observation with the largest Cook's distance for each quantile. In Tables \ref{tab:qr_housing_multi_tau10}, \ref{tab:qr_housing_multi_tau50} and \ref{tab:qr_housing_multi_tau95}, the first column lists the feature names after the corresponding transformations described in Table \ref{tab:rda_house_var_list}.
The second column contains the estimated coefficients based on the random sample of $200$ observations (full-data solution), and their corresponding p-values. The third column provides the estimated coefficients when excluding the observation with the largest Cook's distance (LOO solution) and corresponding p-values. The calculation  is based on the standard quantile regression function in the R package \texttt{quantreg} \citep{koenker2018package} with no regularization applied. 

Those features whose statistical significance may change between the full-data and LOO solutions (using the significance level of 0.05) are highlighted in bold in the first column. Among the three tables, at $\tau=0.5$, we observe no noticeable changes in the p-values even when excluding the most influential observation, demonstrating the robustness of median regression. In contrast, at $\tau=0.95$, statistical significance of multiple features is affected by the most influential observation.

\begin{table}[h]
    \centering
    \caption{Estimated quantile regression coefficients and their p-values when $\tau = 0.1$.}
    \label{tab:qr_housing_multi_tau10}
    \fontsize{9}{10.8}\selectfont
    \bigskip
    \begin{tabular}{lrr}
        \hline
        Feature & Full-Data Solution & LOO Solution \\
        \hline
        basement        & 28.79 (0.30) & 28.61 (0.28) \\
        bathrooms            & 63.27 (0.01) & 63.29 (0.02) \\
        condition\_good      & -18.81 (0.46) & -18.94 (0.46) \\
        yr\_built\_2000      & -29.92 (0.34) & -30.06 (0.32) \\
        yr\_built\_1980      & 40.01 (0.26) & 39.28 (0.29) \\
        \textbf{yr\_built\_1960}      & 75.98 (0.04) & 75.71 (0.06) \\
        yr\_built\_1940    & 129.21 (0.03)& 128.93 (0.04) \\
        yr\_built\_1920      & 185.44 (0.00) & 184.41 (0.00) \\
        grade               & 48.72 (0.54) & 47.90 (0.55) \\
        grade2              & 1.31 (0.80) & 1.35 (0.80) \\
        sqrtSqft\_living15   & 1.00 (0.67) & 0.95 (0.74) \\
        sqrtSqft\_living     & -0.76 (0.67) & -0.71 (0.74) \\
        bedrooms\_3        & 56.69 (0.30) & 56.72 (0.28) \\
        view         & 107.93 (0.00) & 109.32 (0.00) \\
        \hline
    \end{tabular}
\end{table}

\begin{table}[h]
    \centering
    \caption{Estimated quantile regression coefficients and their p-values when $\tau = 0.5$.}
    \label{tab:qr_housing_multi_tau50}
    \fontsize{9}{10.8}\selectfont
    \bigskip
    \begin{tabular}{lrr}
        \hline
        Feature & Full-Data Solution & LOO Solution \\
        \hline
        basement         & 16.61 (0.57) & 14.06 (0.67) \\
        bathrooms            & 28.13 (0.21) & 26.78 (0.23) \\
        condition\_good      & 34.40 (0.17) & 34.58 (0.15) \\
        yr\_built\_2000      & -32.01 (0.33) & -32.17 (0.35) \\
        yr\_built\_1980      & 25.43 (0.48) & 27.03 (0.52) \\
        yr\_built\_1960      & 88.03 (0.03) & 89.33 (0.03) \\
        yr\_built\_1940      & 160.14 (0.00) & 161.38 (0.00) \\
        yr\_built\_1920      & 143.66 (0.00) & 146.13 (0.01) \\
        grade               & -62.22 (0.69) & -83.10 (0.56) \\
        grade2              & 10.48 (0.31) & 11.92 (0.22) \\
        sqrtSqft\_living15   & 1.76 (0.49) & 1.61 (0.51) \\
        sqrtSqft\_living     & 4.11 (0.18) & 4.45 (0.15) \\
        bedrooms\_3         & -30.28 (0.39) & -28.98 (0.35) \\
        view         & 30.08 (0.51) & 24.82 (0.55) \\
        \hline
    \end{tabular}
\end{table}

\begin{table}[h]
    \centering
    \caption{Estimated quantile regression coefficients and their p-values when $\tau = 0.95$.}
    \label{tab:qr_housing_multi_tau95}
    \fontsize{9}{10.8}\selectfont
    \bigskip
    \begin{tabular}{lrr}
        \hline
        Feature & Full-data Solution & LOO Solution \\
        \hline
        basement        & -58.00 (0.18) & -53.70 (0.21) \\
        \textbf{bathrooms}            & 104.03 (0.02) & 77.47 (0.05) \\
        condition\_good      & 23.15 (0.53) & 18.67 (0.58) \\
        yr\_built\_2000      & -63.32 (0.41) & -105.47 (0.10) \\
        yr\_built\_1980      & -8.16 (0.91) & -68.91 (0.32) \\
        yr\_built\_1960      & 63.44 (0.53) & 21.55 (0.83) \\
        yr\_built\_1940      & 113.47 (0.13) & 41.39 (0.62) \\
        yr\_built\_1920      & 78.20 (0.41) & 91.33 (0.38) \\
        \textbf{grade}               & -523.42 (0.05) & -370.92 (0.12) \\
        \textbf{grade2}              & 40.36 (0.02) & 29.02 (0.06) \\
        sqrtSqft\_living15   & -0.27 (0.96) & -2.85 (0.57) \\
        \textbf{sqrtSqft\_living}     & 4.95 (0.23) & 8.10 (0.04) \\
        bedrooms\_3      & -55.37 (0.34) & -25.40 (0.64) \\
        \textbf{view}     & 121.15 (0.17) & 137.01 (0.05) \\
        \hline
    \end{tabular}
\end{table}
\end{document}